\documentclass[12pt]{article}

\usepackage{latexsym}
\usepackage{amsmath,amssymb,amsfonts}
\usepackage{bbm}

\newcommand{\be}{\begin{equation}}
\newcommand{\ee}{\end{equation}}

\def\cN{{\cal N}}
\def\dfrac{\displaystyle \frac }

\def\tr{{\rm tr}\,}

\newcommand{\bm}[1]{\mbox{\scriptsize\boldmath$#1$}}

\newcommand{\bea}{\begin{eqnarray}}
\newcommand{\eea}{\end{eqnarray}}
\newcommand{\ba}{\begin{array}}
\newcommand{\ea}{\end{array}}
\newcommand{\nn}{\nonumber}

\begin{document}

\begin{titlepage}

\phantom{a}

\vspace{0.6cm}

\begin{center}
{\Large\bf The low-energy $\cN=4$ SYM effective action
\\[3mm]in diverse harmonic superspaces} \vspace{1.5cm}
 \\[0.5cm]
 {\bf
 I.~L.~Buchbinder $^+$\footnote{joseph@tspu.edu.ru},
 E.~A.~Ivanov $^\dag$\footnote{eivanov@theor.jinr.ru}
 and
 I.~B.~Samsonov $^{*}$\footnote{samsonov@uwa.edu.au, on leave from
Tomsk Polytechnic University, 634050 Tomsk, Russia.}}
\\[3mm]
 {\it $^+$
 Department of Theoretical Physics, Tomsk State Pedagogical
 University,\\ Tomsk 634061, Russia}
\\[1mm]
 {\it $^\dag$
 Bogoliubov Laboratory of Theoretical Physics, JINR,\\
141980 Dubna, Moscow Region, Russia}
\\[1mm]
{\it $^{*}$
School of Physics M013, The University of Western Australia\\
35 Stirling Highway, Crawley, W.A. 6009, Australia
}

\vspace{0.6cm}
\bf Abstract
\end{center}
We review various superspace approaches to the
description of the low-energy effective action in $\cN=4$ super
Yang-Mills (SYM) theory. We consider the four-derivative part of the low-energy effective action in the Coulomb branch.
The typical components of this effective action are the gauge field $F^4/X^4$ and the scalar field Wess-Zumino terms. We construct $\cN=4$
supersymmetric completions of these terms in the framework of different harmonic superspaces supporting $\cN=2,3,4$ supersymmetries.
These approaches are complementary to each other in the sense that they make manifest different subgroups of the total $SU(4)$ R-symmetry group.
We show that the effective action acquires an extremely simple form in those superspaces which manifest the non-anomalous maximal subgroups of $SU(4)$.
The common characteristic feature of our construction is that we restore the superfield effective actions exclusively
by employing the
$\cN =4$ supersymmetry and/or superconformal $PSU(2,2|4)$ symmetry. \end{titlepage}
\setcounter{footnote}{0}
\tableofcontents

\numberwithin{equation}{section}

\section{Introduction}

$\cN=4$ SYM theory in four-dimensional Minkowski space is an exceptional model of quantum field theory. Originally, it
was constructed  by compactification of the $10D$ super-Yang-Mills theory \cite{BSS}. Shortly after its discovery, this theory was found to exhibit
miraculous cancelations of ultraviolet divergences, so that its beta-function is zero to all loops \cite{GRS,V,D}
and the model is UV finite and superconformal \cite{N4-conf}. This result triggered a high interest in studying other four-dimensional conformal field theories,
though $\cN=4$ SYM theory remains the key example of the UV finite field theories.

Although  $\cN=4$ SYM theory has no phenomenological applications, it plays a crucial role for the study of quantum aspects
of string theory through the so-called AdS/CFT (or ``gauge/gravity'') correspondence \cite{Maldacena,Gubser1998,Witten1998}
(see also \cite{Aharony1999} for a review).  In the original Maldacena's work \cite{Maldacena}
it was conjectured that quantum observables in IIB superstring theory on the $AdS_5\times S^5$ background can be
determined  by studying the corresponding objects in $\cN=4$ SYM theory. Since 1998, Maldacena's conjecture
has been thoroughly verified  and nowadays we have a good understanding of quantum properties of both sides of the AdS/CFT correspondence.

In quantum field theory, there are several objects exhibiting physical properties of a given model: scattering amplitudes,
correlation functions and Wilson loops. All these quantities have been investigated in $\cN=4$ SYM theory and then have been
matched with the corresponding objects in string theory. The detailed exposition of these results can be retrieved
from numerous review papers and textbooks, see, e.g., \cite{Schwarz-book}. The short summary is that
many of these quantum quantities in $\cN=4$ SYM theory can be found {\em exactly} beyond the perturbation theory.
These exact results provide a strong ground for the further studies of string theory, as well as of many other superconformal
field theories - with different amounts of supersymmetry and in diverse space-time dimensions.

An object of the crucial importance in quantum field theory is the effective action. By definition, it is the generating functional
for 1PI (``one-particle-irreducible'') Green's functions, which encodes the full information about quantum properties of given model.
It can also be viewed
as a functional reproducing the effective equations of motion which take
into account quantum corrections. Since the effective action is a very complicated object, it makes sense to study first its low-energy part,
which describes the physics below some energy scale and so serves a good approximation in this domain.

The low-energy effective action of $\cN=4$ SYM theory plays an important role in checking the AdS/CFT correspondence. According to \cite{Maldacena},
it can be matched with the effective action of a D3-brane propagating in the $AdS_5$ background. This D3-brane action can be understood
as a Born-Infeld-type
action possessing $\cN=4$ superconformal symmetry (see, e.g., \cite{BI-review}). This conjecture has been checked perturbatively,
by comparing the leading terms in the power series expansions of both these actions. We stress that the verification
of this conjecture  on the field theory side is a very non-trivial task, since it involves the computation of the quantum loop corrections
to the low-energy effective action.
To date, we have a good understanding of this issue in the one-loop approximation. Only limited
results are available beyond the one-loop order.

The significant progress in exploring quantum aspects of $\cN=4$ SYM theory has been achieved due to the property that it possesses
a reach  set of symmetries which are preserved in the quantum perturbation theory. Indeed, this model, being a non-trivial
interacting quantum field theory,
respects the highest amount of supersymmetries 
admissible in the four-dimensional Minkowski space. The supersymmetry is a part of the $PSU(2,2|4)$
superconformal group that remains unbroken on the quantum level due to the vanishing  beta-function \cite{GRS}. This symmetry imposes very strong
constraints on the quantum observables, such that some of them can be found exactly. The low-energy effective action is one of such objects.
As we will demonstrate in the present paper, its leading part is completely fixed by the underlying (super)symmetries.

Within the perturbation theory one computes the effective action as a series expansion over some small parameters, such as the coupling
constants or Planck's length. It is advantageous to use the so-called derivative expansion, which assumes that the terms with the lower number
of derivatives on fields give the leading contribution in the low-energy approximation, as compared to the terms with a larger number of derivatives.
In the present paper, we restrict our consideration only to the four-derivative terms in the low-energy effective action of $\cN=4$ SYM theory.
We will be interested in the effective action in the Coulomb branch, which describes the effective dynamics of the massless degrees of freedom.
The remaining massive degrees of freedom appearing as a result of spontaneous breaking of gauge symmetry are assumed to be integrated out.

The studies of the four-derivative part of the $\cN=4$ SYM effective action were initiated in the papers \cite{DS,S16}, where
the so-called $F^4/X^4$ term was analyzed. In these papers, it was argued that the $F^4/X^4$ term in the $\cN=4$ SYM effective action
is one-loop exact and does not receive the instanton corrections. This term was also obtained by the direct quantum computations
using different superspace methods \cite{deWit,Lindstrom,non-hol2,BFM,non-hol3}.

Another interesting term in the four-derivative part of the $\cN=4$ SYM effective action is the Wess-Zumino term for scalar fields \cite{TZ}.
Its presence  is compulsory in order to obey the anomaly-matching condition for the $SU(4)$ R-symmetry \cite{Intriligator}. Moreover,
it has a natural interpretation as the Chern-Simons term of the D3-brane action on the $AdS_5$ background \cite{TZ}.

In the  papers mentioned above only some selected terms in the four-derivative part of the $\cN=4$ SYM effective action were found.
Already in the first papers \cite{DS,S16} it was conjectured that the full four-derivative part of the effective action can be restored
as a supersymmetric completion of these particular terms. However,  the proof of this statement turned out to be  a very non-trivial exercise,
and it was accomplished only in the paper \cite{BuIv}, based on the $\cN=2$ harmonic superspace techniques \cite{GIOS-84,GIKOS}.
In the subsequent papers \cite{BelSam1,BelSam2,BISZ}, alternative descriptions of the four-derivative part of the effective
action were developed in the framework of different $\cN=3$ and $\cN=4$ harmonic superspace approaches.

The basic aim of the present
paper is to give a systematic and self-consistent review of what has been done in \cite{BuIv,BelSam1,BelSam2,BISZ}. In the course of this
consideration, we also give the appropriate account of the related issues.

We point out that the four-derivative part of the effective action constructed in \cite{BuIv,BelSam1,BelSam2,BISZ}
is the {\em exact} result which was obtained solely on the ground of symmetries of the theory, though the perturbative
checks were performed afterwards  in \cite{BIP,BBP,BP2005}
(see also \cite{BIP-review} for a review). This exposes the exceptional role of the quantum $\cN=4$ SYM theory
among other models of the quantum field theory.
We also emphasize that in the papers just mentioned not only a superfield generalization of the old results \cite{DS,S16} was obtained,
but also many important properties of the $\cN=4$ SYM low-energy effective action were explained. In particular, the following
questions were addressed:
Why is the coefficient in front of the $F^4/X^4$-term one-loop exact? What is the origin of the Wess-Zumino term in the low-energy
effective action? Why is the harmonic superspace approach so efficient for studying the effective action and which harmonic superspace
is most suitable for this purpose? All these issues are thoroughly reviewed in the present paper.

The rest of the paper is organized as follows. In section 2 we give a brief summary of basic features of the low-energy effective action
in $\cN=4$ SYM theory. A part of this effective action which is represented by the Wess-Zumino term for scalar fields
is  discussed in detail in section 3. In particular, we explain the origin of the Wess-Zumino term as the necessary consequence of the `t Hooft
anomaly-matching condition for the R-symmetry group $SU(4)$. In section
4 we review the $\cN=2$ harmonic superspace description of $\cN=4$ SYM theory and construct
its low-energy effective action possessing the full $\cN=4$ supersymmetry. Section 5 is devoted
to $\cN=3$ SYM theory in the $\cN=3$ harmonic superspace. This theory is known to be equivalent to $\cN=4$ SYM on shell
and so provides the maximally supersymmetric off-shell formulation of the latter.
For this $\cN=3$ SYM theory we construct the $\cN=3$ superconformal low-energy effective action and consider its component
field structure in the sector of bosonic fields. In sections 6 and 7 we elaborate on two different $\cN=4$ harmonic superspaces
which appear very suitable for description of the $\cN=4$ SYM low-energy effective action. We demonstrate that
the latter acquires especially simple form in these superspaces. In the last section we discuss some issues
and open problems related to the study of the low-energy effective action in $\cN=4$ SYM theory beyond the
leading low-energy approximation.

\section{Low-energy effective action in the Coulomb branch}
\label{Sect2}
\subsection[Classical action and the spontaneous gauge symmetry breaking]{Classical action and the spontaneous gauge symmetry\break
breaking}
The $\cN=4$ gauge supermultiplet consists of one vector gauge field
$A_m$, four spinor fields $\psi^I_\alpha,\; \bar\psi_{\dot\alpha I}$ and six scalar fields $\varphi^{IJ}=-\varphi^{JI}\,$, where
$I=1,2,3,4$ is the quartet index of the
R-symmetry $SU(4)$ group. The spinor fields are in the conjugated non-equivalent fundamental representations ${\bf 4}$ and ${\bf \underline{4}}$ of $SU(4)\,$,
while the scalar fields are in the real representation ${\bf 6}$, since they obey the
reality condition
\be
\overline{\varphi^{IJ}} = \bar\varphi_{IJ} = \frac12
\varepsilon_{IJKL} \varphi^{KL}\,,
\label{phi-reality}
\ee
with $\varepsilon_{IJKL}$ being the totally antisymmetric $SU(4)$ tensor,
$\varepsilon_{1234}=1$. In the non-abelian case, all these fields
are sitting in the adjoint representation of some gauge group $G$.
They can be viewed as the matrices taking values in the Lie algebra $\mathfrak g$ of the group $G\,$.

The scalars $\varphi^{IJ}$ can be equivalently represented as a
real vector in the fundamental representation of $SO(6)\sim SU(4)$
\be
X^A = (\gamma^A)_{IJ} \varphi^{IJ}\,,\quad
(X^A)^* = X^A\,,\quad
A =1,\ldots,6\,,
\ee
where $(\gamma^A)_{IJ}=-(\gamma^A)_{JI}$ are six-dimensional
gamma-matrices which provide the equivalence of the representations {\bf 6} of $SO(6)$ and $SU(4)$ groups.\footnote{The defining properties of these matrices are:
$(\gamma_A)_{IJ}(\gamma_B)^{JK} +
 (\gamma_B)_{IJ}(\gamma_A)^{JK} = -\delta_{AB} \delta_I^K$,
$(\gamma^A)_{IJ} (\gamma_A)^{KL} = \delta_I^K \delta_J^L
 - \delta_I^L \delta_J^K$, $\left((\gamma_A)^{IJ} \right)^* = (\gamma_A)_{IJ}=\frac12 \varepsilon_{IJKL}
(\gamma_A)^{KL}$. }
In the present paper we will employ both forms for the scalar fields, $X^A$ and $\varphi^{IJ}$.

The classical action of $\cN=4$ SYM theory reads
\bea
S&=&\tr \int d^4x \Big(
\frac12\nabla^{\alpha\dot\alpha} \varphi^{IJ} \nabla_{\alpha\dot\alpha}\bar\varphi_{IJ}
 -\frac12(F^{\alpha\beta}F_{\alpha\beta} + \bar F^{\dot\alpha\dot\beta}\bar F_{\dot\alpha\dot\beta})
-i\psi^{\alpha I}\nabla_{\alpha\dot\alpha} \bar\psi^{\dot\alpha}_I
\nn\\&&
+\frac{g}{2\sqrt2} \psi^{\alpha I}[\psi_\alpha^J,\bar\varphi_{IJ}]
+\frac{g}{2\sqrt2} [\varphi^{IJ},\bar\psi_{\dot\alpha J}]\bar\psi^{\dot\alpha}_I
-\frac {g^2}{16} [\varphi^{IJ},\varphi^{KL}]
 [\bar\varphi_{IJ},\bar\varphi_{KL}]
\Big).
\label{S}
\eea
Here $\nabla_{\alpha\dot\alpha} = \sigma^m_{\alpha\dot\alpha}
\nabla_m$ is the gauge-covariant derivative which acts on the fields by
the generic rule
\be
\nabla_m = \partial_m + i g [A_m ,\ \cdot\ ]\,,
\ee
$g$ is a dimensionless gauge coupling constant and $F_{\alpha\beta}\,$, $\bar F_{\dot\alpha\dot\beta}$ are the
spinorial components of the Yang-Mills field strength
\footnote{In this paper we employ the following basic conventions. The Minkowski space metric is $\eta_{mn}= {\rm diag}(1,-1,-1,-1)$.
For conversion of the vector and spinor indices we use the rules
$A^m = \frac12\sigma^m_{\alpha\dot\alpha}
A^{\alpha\dot\alpha}$,
$A_{\alpha\dot\alpha} = \sigma^m_{\alpha\dot\alpha}
A_m$. The basic properties of the sigma-matrices are
$(\sigma_m)_{\alpha\dot\alpha} (\sigma_n)^{\alpha\dot\alpha} =
2\eta_{mn}$,
$(\sigma^m)_{\alpha\dot\alpha} (\sigma_m)^{\beta\dot\beta}
=2\delta_\alpha^\beta \delta_{\dot\alpha}^{\dot\beta}
$. The convention for raising and lowering the spinorial indices is
$\psi^\alpha = \varepsilon^{\alpha\beta}\psi_\beta$,
$\psi_\alpha = \varepsilon_{\alpha\beta}\psi^\beta$,
$\varepsilon_{12} = \varepsilon^{21} = 1$,
and the same for dotted spinorial indices.
Finally, the antisymmetric tensor $F_{mn}=-F_{nm}$
is converted into its spinorial components as
$F_{mn} =\frac12(\sigma_{mn})^{\dot\alpha\dot\beta}\bar F_{\dot\alpha\dot\beta}
+\frac12(\sigma_{mn})^{\alpha\beta}F_{\alpha\beta}
$, where
$(\sigma_{mn})^{\alpha\beta} = -\frac12
 (\sigma_m{}^\alpha{}_{\dot\gamma} \sigma_n^{\beta\dot\gamma}
  -\sigma_n{}^\alpha{}_{\dot\gamma}\sigma_m^{\beta\dot\gamma})$,
$(\sigma_{mn})^{\dot\alpha\dot\beta} = \frac12(\sigma_m^{\gamma\dot\alpha}
 \sigma_{n\gamma}{}^{\dot\beta} -
 \sigma_n^{\gamma\dot\alpha}\sigma_{m\gamma}{}^{\dot\beta})$.
The basic properties of the antisymmetric products of sigma-matrices are
$
(\sigma_{mn})^{\alpha\beta}(\sigma^{mn})_{\gamma\delta} =
 4 (\delta_\gamma^\alpha \delta^\beta_\delta + \delta^\alpha_\delta
 \delta^\beta_\gamma)$,
$(\sigma_{mn})^{\dot\alpha\dot\beta}(\sigma^{mn})_{\dot\gamma\dot\delta} =
 4 (\delta_{\dot\gamma}^{\dot\alpha} \delta^{\dot\beta}_{\dot\delta} + \delta^{\dot\alpha}_{\dot\delta}
 \delta^{\dot\beta}_{\dot\gamma})$.
}
\be
F_{mn} = \partial_m A_n - \partial_n A_m +ig [A_m, A_n]\,.
\ee

The action (\ref{S}) is invariant under the
$\cN=4$ supersymmetry transformations
\bea
\delta\varphi^{IJ}&=& i \psi^{\alpha[I} \epsilon_\alpha^{J]}
 + \frac i2 \varepsilon^{IJKL} \bar\epsilon_{\dot\alpha K}
 \bar\psi^{\dot\alpha}_L \,,\nn\\
\delta\psi^{\alpha I}&=& -\sqrt2 F^{\alpha\beta} \epsilon^I_\beta
 - 2\nabla^{\alpha\dot\alpha} \varphi^{IJ} \bar\epsilon_{\dot\alpha
 J}
  + \frac{i g}{\sqrt2} [\varphi^{IJ},\bar\varphi_{JK}]\epsilon^{\alpha K}\,,\nn\\
\delta A^{\alpha\dot\alpha}&=&\frac i{2\sqrt2} \psi^{\alpha
I}\bar\epsilon^{\dot\alpha}_I
 +\frac i{2\sqrt2} \bar\psi^{\dot\alpha}_I \epsilon^{\alpha I}\,,
\label{N4susy}
\eea
with anticommuting parameters $\epsilon^I_\alpha$. These
transformations, together with the space-time translations and Lorentz transformations, form the $\cN=4$ Poincar\'e superalgebra.
The algebra of these transformations closes on shell, i.e., up to
terms proportional to the classical equations of motion.

The classical $\cN=4$ SYM action (\ref{S}) involves the non-negative potential
of scalar fields,
\be
V=\frac{g^2}{16}\tr [\varphi^{IJ},\varphi^{KL}]
 [\bar\varphi_{IJ},\bar\varphi_{KL}] \geq 0\,.
\label{scalar-potential}
\ee
This potential reaches its minimum $V=0$ for the fields valued in the
Cartan subalgebra $\mathfrak h$ of the Lie algebra $\mathfrak g$ of the gauge group
\be
V=0 \quad \Rightarrow \quad \varphi^{IJ}\equiv\varphi_{\mathfrak h}^{IJ} \in {\mathfrak h}\,.
\ee
Hence, at non-trivial vacuum expectation values (vevs) of these fields,
\be
\langle \varphi_{\mathfrak h}^{IJ} \rangle = a^{IJ} = const\,,
\ee
the spontaneous breaking of gauge symmetry becomes possible. The details of gauge
symmetry breaking in $\cN=4$ SYM theory are presented in \cite{Fayet}.
Assuming that the gauge group in $\cN=4$ SYM is
$G=SU(N)$,\footnote{Other gauge groups can be  considered as well.}
the pattern of spontaneous symmetry breaking can be summarized as
follows:
\begin{itemize}
\item In general, the  gauge group $G=SU(N)$ is broken down to
$H=[U(1)]^{N-1}\,$, which is the maximal abelian subgroup of $SU(N)$.
However, also a larger subgroup of the gauge group may remain unbroken,  when
not all of the scalars from $\mathfrak h$ acquire non-vanishing vevs. To simplify the
issue, in what follows we will basically assume that $H=[U(1)]^{N-1}\,$ and, even more, that the gauge group
$G$ is $SU(2)\,$ which can be broken down only to $H=U(1)$.
\item After the spontaneous gauge symmetry breaking,
the fields $(\varphi^{IJ}, A_m,\psi^I_\alpha, \, \bar\psi_{\dot\alpha I})_{\mathfrak h}$ from the Cartan
subalgebra $\mathfrak h$ remain massless, while the fields corresponding
to the coset space $G/H$ acquire masses specified by the vacuum values $a^{IJ}$. These $G/H$
fields realize the {\it massive} representation of $\cN=4$ superalgebra
with the {\it central charges} which are identified with some $U(1)$ generators from the subalgebra ${\mathfrak h}$,
times the parameters $a^{IJ}$. Since such central charges are vanishing on the massless fields
$(\varphi^{IJ}, A_m,\psi^I_\alpha, \, \bar\psi_{\dot\alpha I})_{\mathfrak h}$, the latter form a supermultiplet of the {\it standard}
$\cN=4$ supersymmetry.
\item
The $\cN=4$ supersymmetry itself remains unbroken whatever $G$ and $H$ are, while its R-symmetry $SU(4) \simeq SO(6)$ proves spontaneously
broken  down to some subgroup of $SU(4)$. In the case of $G=SU(2), H = U(1)$, this subgroup is $SO(5)\simeq USp(4)\,$.
The full-fledged $\cN=4$ superalgebra with central charges, because of the presence of $SU(4)$ breaking constants $a^{IJ}$ in the right-hand
sides of the basic anticommutators,  possesses the reduced R-symmetry group $SO(5)\simeq USp(4)$.
With respect to this $USp(4)$, the $\cN=4$ massive vector
multiplet comprises five complex scalars in the
representation $\bf 5$, one complex singlet massive vector and
four Dirac spinors in the representation $\bf 4$ of $USp(4)\,$.\footnote{For the simplest case of gauge group $SU(2)$ broken to $U(1)$ there is only one central charge
proportional to the $U(1)$ generator and only one set of the $SU(4)_R$ breaking parameters $a^{IJ}$, giving rise just to $SO(5) \simeq USp(4)$ as the reduced R-symmetry.
In the more general case of $G=SU(N)$ and $H = [U(1)]^{N-1}$, more central charges can appear, with different sets of $SU(4)_R$ breaking constants.
If these constant $SO(6)$ vectors are collinear, the reduced R-symmetry is still $USp(4)$ and the relevant massive supermultiplets have the
same $USp(4)$ contents, while their number is $\frac12 N(N-1)$. If the breaking constant vectors are arbitrary, the further reduction of the original $SO(6)\simeq SU(4)$
R-symmetry occurs.}
\item The R-symmetry $SO(6)\simeq SU(4)$ is spontaneously broken down to $SO(5)\simeq USp(4)$ also in the sector of massless fields, though in this case no central
charges in the $\cN=4$ superalgebra are present, and so no reduction of the R-symmetry group comes about. The effect of spontaneous breaking
consists in that the vacuum expectation values $a^{IJ}$ of the scalar fields are invariant only under the group $SO(5)$. This means that
the $SU(4)$ transformations of the physical scalars $\phi^{IJ}_{\mathfrak h}=\varphi^{IJ}_{\mathfrak h} - a^{IJ}$
acquire inhomogeneous terms (shifts), so five fields out of these massless scalars can be interpreted as the $SO(6)/SO(5)$ Goldstone fields.
It is worth pointing out  that the model is
still invariant under the full R-symmetry group $SU(4)$, but the latter is now realized on the scalar fields by the inhomogeneous transformations.
\item The original classical action (\ref{S}) is known to be
invariant under the superconformal group $PSU(2,2|4)$ involving $SU(4)$ as a subgroup. This
extended symmetry is also spontaneously broken and is realized by inhomogeneous transformations of the fields $(\varphi^{IJ},A_m,
\psi^I_\alpha, \, \bar\psi_{\dot\alpha I} )_{\mathfrak h}$. In particular, one field out of six massless scalars is a dilaton (apart from the remaining five
$SU(4)/O(5)$ Goldstone fields). Also,
the conformal $\cN=4$ supersymmetry is spontaneously
broken,  with $(\psi^I_\alpha, \, \bar\psi_{\dot\alpha I} )_{\mathfrak h}$ as the corresponding goldstini. To avoid a possible confusion, we note that
$PSU(2,2|4)$ is in fact the symmetry group of the whole effective action, including its part spanned by the massive $G/H$ fields, and this is preserved at the quantum level
due to the vanishing beta-function. However, the realization of the superconformal symmetry on the $G/H$ fields is rather complicated since the corresponding transformations
are accompanied by some field-dependent gauge transformations and their Lie brackets contain operator central charges. The correct closure of $PSU(2,2|4)$ symmetry,
like that of the $\cN=4$ supersymmetry, is achieved only on shell.
\end{itemize}

As a brief resume, the crucial feature of the spontaneous gauge symmetry breaking in
$\cN=4$ SYM theory is the appearance of {\it massive} multiplets
which correspond to broken directions $G/H$ in the gauge group
$G\,$, while the degrees of freedom corresponding to $H$ remain
massless. At low energies, we can observe only these massless
fields, with the dynamics described by some {\it low-energy
effective action}. In quantum field theory, in order to obtain this
low-energy effective action, one has to integrate out the massive
fields in the functional integral which defines the full effective action.
In the present paper we do not engage with technical details of
this functional integration, but rather discuss the
general structure of the resulting expression for the low-energy
effective action of $\cN=4$ SYM theory. Needless to say,
this low-energy effective action describes $\cN=4$ SYM {\it in the
Coulomb branch}. In the present paper we denote it by $\Gamma$.

\subsection{Low-energy effective action: Derivative expansion}

The computation of low-energy effective action in quantum field
theory is, in general, a complicated problem which is usually
approached by perturbative methods, assuming the series
expansion of the effective action with respect to some small
parameters like the Planck length or coupling constants.
The derivative expansion of the effective action can also be
considered as one of the perturbative methods, which relies upon the
common observation that the fields with long wavelengths at low energies dominate
 over the fields with short wavelengths. It is frequently a good approximation to discard the fields with
short wavelengths which are represented in the effective
action by  terms with higher number of space-time derivatives,
as compared to the terms with lower number of derivatives.
The latter terms involve the fields with longer wavelengths.

To illustrate these ideas, let us consider the effective action for one
scalar field $\phi$. The derivative expansion of the effective
action can be schematically represented as
\be
\Gamma = \sum_{n=0}^\infty \Gamma_{2n}\,,
\ee
where $\Gamma_{2n}$ is a functional which involves just $2n$
space-time derivatives of $\phi$. In particular, $\Gamma_0$
contains no derivatives of $\phi$ and so corresponds to the
(effective) potential for the scalar field $\Gamma_0=-\int d^4x
\,V(\phi)$. The functional $\Gamma_2$ has two space-time derivatives of the
scalar field and corresponds to a finite (or infinite) renormalization
of the wavefunction, if the latter receives perturbative quantum
corrections. The next term is $\Gamma_4$ which involves four
derivatives of the scalar and represents the leading non-trivial quantum
correction to the effective action. The remaining terms, starting with
$\Gamma_6\,$, must be considered as the higher-order corrections to the
low-energy approximation.

The derivative expansion of the effective action straightforwardly
applies to $\cN=4$ SYM theory. We will count the derivative degree
of different terms in the effective action just with respect to the
scalar fields. This means that, after turning off the vector and
spinor fields, the term $\Gamma_{2n}$ in the effective action contains
as the remainder exactly $2n$ space-time derivatives of scalars $\varphi^{IJ}$. It is
important to note that the omitted terms with vector and spinor
fields can be {\it uniquely} restored from the terms with scalar
fields only. Indeed, it is obvious that $\cN=4$ supersymmetry does not mix
those terms in the effective action which contain different numbers of derivatives.

It is well known that in $\cN=4$ SYM theory there are no quantum corrections
to the classical scalar potential (\ref{scalar-potential}), i.e.
$\Gamma_0=0$. Since the effective action in $\cN=4$ SYM theory is UV finite \cite{GRS,V,D},\footnote{The proof of the non-renormalization theorem in the $\cN=2$ harmonic superspace was given in \cite{BKO,BPS15}.} no
wave-function renormalization is needed and so $\Gamma_2 = S_{\rm free}\,$,
where $S_{\rm free}=S|_{g=0}$ is that part of the $\cN=4$ SYM action (\ref{S}) which contains
the kinetic terms of the $\cN=4$ multiplet. The first non-trivial quantum correction
in the effective action starts with $\Gamma_4$,  which will be the basic object of study in the present paper.
The higher-order terms, starting with $\Gamma_6$, will fall beyond our consideration.

To summarize, in the present paper we will study the
low-energy effective action of $\cN=4$ SYM theory in the Coulomb
branch. More precisely, we will be interested only in that part of
this low-energy effective action, which contains, in its component
field expansion, no more than four space-time derivatives of
scalar fields (together with other appropriate terms which involve vector and spinor fields
and are needed for completing the scalar field terms to the invariants of  $\cN=4$ supersymmetry).

\subsection[Wess-Zumino vs.\ $F^4/X^4$ term in the low-energy effective action]
{Wess-Zumino vs.\ $F^4/X^4$ term in the low-energy effective \break action}

In this section we will consider the gauge group $G=SU(2)$
spontaneously broken down to $H=U(1)$. In this case the low-energy
effective action is dominated by one massless $\cN=4$ vector multiplet
which consists of six scalar fields $X_A$, four spinors
$\psi^I_\alpha$ and one abelian vector field $A_m$ with the field strength
$F_{mn}=\partial_m A_n - \partial_n A_m$.

The leading four-derivative
quantum correction to the $\cN=4$ SYM low-energy effective
action is known to contain, among its components, the so-called
$F^4/X^4$ term \cite{DS,S16}
\be
\label{F4X4}
\frac{1}{(8\pi)^2}\int d^4x \frac{1}{(X_A X_A)^2}\Big[F_{m n}F^{n k}F_{k l}F^{l m}-\frac14(F_{p q}F^{p q})^2\Big].
\ee
It was argued that this part of the effective action is {\it one-loop exact} \cite{DS,BKO} and does not receive
non-perturbative corrections \cite{Dorey}. This $F^4/X^4$ term appears as one
of the terms in the component field expansion of the so-called
non-holomorphic effective potential of the $\cN=2$ superfield
strength $W$ and its conjugate $\bar W$ \cite{Henningson}
\be
{\cal H}(W,\bar W)=\frac1{(4\pi)^2} \ln \frac{W}\Lambda \ln\frac{\bar W}{\Lambda}\,.
\label{non-hol}
\ee
Here, $\Lambda$ is some parameter of dimension one the dependence on which
completely disappears after passing to the component form of the effective
action. The details of the construction of the $\cN=4$ SYM low-energy
effective action in $\cN=2$ superspace will be discussed in
sect.\ \ref{SectN2}. It is important to mention that the
non-holomorphic effective potential (\ref{non-hol}) was derived
perturbatively in \cite{deWit,Lindstrom}, using the $\cN=1$
superfield methods and, later, in \cite{non-hol2} and
\cite{BFM,non-hol3} with the use of $\cN=2$ projective and harmonic
superspace techniques, respectively.

Another interesting term in the $\cN=4$ SYM low-energy effective
action is the so-called Wess-Zumino term which involves the scalar
fields only \cite{TZ,Intriligator}:
\be
\label{WZterm}
-\frac{1}{60\pi^2}\int d^5x\,\varepsilon^{M N K L P}\varepsilon^{A B C D E F} \,
\frac{1}{|X|^6}
X_A \partial_M X_B \partial_N X_C \partial_K X_D \partial_L X_E \partial_P X_F \,,
\ee
where $|X|^2=X_A X_A$. Here it is presented in the form of the
integral over a five-dimensional space-time, but it can always be
rewritten as a functional in the conventional four-dimensional
Minkowski space, since the integrand in (\ref{WZterm}) is a closed five-form.
We will show in sect.\ \ref{SectWZ} that there are various
four-dimensional representations of the same Wess-Zumino term (\ref{WZterm}).
They prove to be good starting points for construction of the superfield low-energy
effective actions in various harmonic superspaces. Here it is
important to note that the coefficient $-\frac{1}{60\pi^2}$ in front of this action
 is {\it exact} and, for topological reasons, can only be a multiple of
an integer  (see, e.g., \cite{Patani,Braaten}).

It will be demonstrated in sect.\ \ref{SectWZ} that the
four-dimensional form of the Wess-Zumino term (\ref{WZterm}) contains
four space-time derivatives of scalar fields. Thus it is one of
the terms in the four-derivative part of the full low-energy $\cN=4$
SYM effective action $\Gamma_4$. Recall that the term (\ref{F4X4})
also belongs to $\Gamma_4\,$, since each Maxwell field
strength in it involves one space-time derivative. Thus, these
two terms should be related to each other by the
abelian version of the $\cN=4$ supersymmetry transformations (\ref{N4susy}).

In practice, to check this suggestion, i.e. to prove that (\ref{F4X4}) and
(\ref{WZterm}) are indeed related to each other by the abelian version of
the $\cN=4$ supersymmetry (\ref{N4susy}), is a rather difficult task  since, apart from (\ref{F4X4}) and (\ref{WZterm}),
$\Gamma_4$ contains a lot of other terms depending on the bosonic $X_A$, $A_m$ and the fermionic $\psi^I_\alpha, \,\bar\psi_{\dot\alpha I}$ fields of the $\cN=4$ vector
multiplet. Recovering all these terms in the effective action is
an extremely involved  routine, unless one uses the superspace
techniques. One of the aims of the present  paper is to demonstrate that
the solution to this problem indeed becomes trivial  in the appropriate superfield approaches based on extended
superspaces. We will show that the two terms (\ref{F4X4}) and
(\ref{WZterm}) originate from the same $\cN=4$ superfield
expressions, for which reason the coefficients in front of them prove
to be firmly related.

This property has an important consequence: The whole four-derivative part
$\Gamma_4$ of the low-energy effective action in the $\cN=4$ SYM
action can be found {\it without performing any perturbative
computation}. All what we need to know is that this part contains the
Wess-Zumino term (\ref{WZterm}) the form of which is unique and,
moreover, the coefficient in front of it is fixed by topological reasons. Then,
all other component terms in $\Gamma_4$ can be found by applying the $\cN=4$ supersymmetry transformations.
Just in this sense, the four-derivative part of the $\cN=4$ SYM effective action is
{\it exact}.

\subsection{Low-energy effective action: Why harmonic superspace?}

Finding the totally $\cN=4$ supersymmetric completion of the terms
(\ref{F4X4}) and (\ref{WZterm}) is a non-trivial problem which has
never been solved in the standard component field formulation of
$\cN=4$ SYM theory. It is natural to expect that
the superfield approaches can be useful for solving this problem, since they display
the manifest supersymmetry. In principle, it is possible
to use different superspaces with $1\leq {\cal N}\leq 4$
supersymmetries. Each of them has some specific useful features which we
will discuss in this section.

The simplest and the most developed  approach is based on the standard $\cN=1$
superspace, which is described in details, e.g.,\ in the books
\cite{BK-book,GGRS}. In terms of $\cN=1$ superfields, the
$\cN=4$ gauge multiplet is represented by a triplet of chiral
superfields $\Phi^I$, $I=1,2,3,\,$ and a real gauge superfield $V$
with the chiral superfield strength $W_\alpha$. The general $\cN=1$
superspace action (including various pieces of the effective action)
has the following form
\be
S= \int d^4x d^4\theta\,{\cal L} + \int d^4x d^2\theta\,{\cal
L}_{\rm c} + \int d^4x d^2\bar \theta\, \bar{\cal L}_{\rm c}\,.
\ee
Here, the Lagrangian $\cal L$ is given on the full $\cN=1$ superspace, while ${\cal L}_{\rm
c}$ and $\bar{\cal L}_{\rm c}$ are, respectively,  the chiral superspace Lagrangian and its complex conjugate.
The superfield action can
be rewritten in the component form, using the identities
\be
\int d^4 x d^4\theta\,{\cal L} = \frac1{16}\int d^4 x \, D^2 \bar
D^2 {\cal L}|_{\theta=0}\,,
\qquad
\int d^4 x d^2\theta\,{\cal L}_{\rm c} = \frac1{4}\int d^4 x \, D^2 {\cal
L}|_{\theta=0}\,,
\label{2.15}
\ee
where $D^2=D^\alpha D_\alpha$, $\bar D^2 = \bar D_{\dot\alpha}\bar
D^{\dot\alpha}\,$, and  $D_\alpha$, $\bar D_{\dot\alpha}$ are
covariant spinor derivatives which obey the anticommutation
relations
\be
\{D_\alpha , \bar D_{\dot\alpha} \} = -2i
\sigma^m_{\alpha\dot\alpha}\partial_m\,.
\label{2.16}
\ee
The relations (\ref{2.15}) and (\ref{2.16}) imply that the full
superspace integration measure ensures two space-time derivatives in the
component field action.

When using the $\cN=1$
superspace to describe the four-derivative part  $\Gamma_4$ of the
effective action, one has to deal with a superfield
Lagrangian $\cal L$ which depends on three chiral superfields $\Phi^I$
and $\cN=1$ superfield strength $W_\alpha$ (and their conjugates).
One of the terms in $\Gamma_4$ has the form
\be
\int d^4x d^4\theta\frac{W^\alpha W_\alpha \bar W_{\dot\alpha}\bar
W^{\dot\alpha}}{(\Phi^I \bar \Phi_I)^2}\propto
\int d^4x \frac{1}{(X_A X_A)^2}\Big[F_{m n}F^{n k}F_{k l}F^{l m}-\frac14(F_{p q}F^{p q})^2\Big].
\label{2.17}
\ee
The terms with pure (anti)chiral superfields, which complete (\ref{2.17}) by $\cN=4$ supersymmetry,
involve four covariant spinor derivatives $D_\alpha$ and $\bar D_{\dot\alpha}$ that generate, after passing to the component fields,
two more space-time derivatives besides the two already  brought by the full superspace integration measure.
There is plenty of such terms, and it appears difficult to find the fully $\cN=4$ supersymmetric completion of (\ref{2.17}).
Thus this problem does not seem to be simpler than the previously discussed
purely component construction in the standard Minkowski space.
Note that the solution of this problem in the $\cN=1$ superspace
has never been presented in the fully $\cN=4$ supersymmetric and $SU(4)$ invariant form.

Let us now consider the $\cN=2$ superspace with Grassmann coordinates
$\theta_i^\alpha$ and $\bar\theta^i_{\dot\alpha}$, $i=1,2$. The
superspace integration measure in the full $\cN=2$ superspace effectively contains
 eight covariant spinor derivatives,
\be
\int d^4x d^8 \theta \,{\cal L}\propto \int d^4x (D^1)^2 (\bar D^1)^2 (D^2)^2
 (\bar D^2)^2 {\cal L} |_{\theta=0}\,,
\ee
which gives rise to four space-time derivatives in the
component field Lagrangian owing to the anticommutation relations
\be
\{D^i_\alpha , \bar D_{j\dot\alpha}  \} =
-2i\delta^i_j\sigma^m_{\alpha\dot\alpha} \partial_m\,.
\ee
Thus the $\cN=2$ superspace is most appropriate  for the description of the
four-derivative part of the effective action $\Gamma_4\,$, because the corresponding superfield Lagrangian $\cal L$ must be a function of
just $\cN=2$ superfields {\it without any derivatives} on them. This enormously simplifies
the problem of construction of the low-energy
effective action $\Gamma_4$ in $\cN=4$ SYM theory. The fully
$\cN=4$ supersymmetric expression for $\Gamma_4$ in the
$\cN=2$ superspace was presented in \cite{BuIv}. We will review
the details of this action in sect.\ \ref{SectN2}.

When $\cN=4$ SYM theory is formulated in the $\cN=2$
superspace, $\cN=2$ supersymmetry is realized manifestly and
off the mass shell, while the extra (hidden) $\cN=2$ supersymmetry is realized by
transformations which mix different $\cN=2$ superfields and possess the correct closure only on the mass shell. It is
important to note that the off-shell realizations of matter
hypermultiplets and gauge multiplets in the $\cN=2$ superspace
require special techniques such as the harmonic superspace
\cite{GIOS-84,GIKOS,GIOS1} or the projective superspace
\cite{KLR,LR-projective,LR-projective2}. These two approaches
provide elegant and natural descriptions of field theories with
extended supersymmetry. In fact, they have
much in common and are related to each other
\cite{Kuzenko-projective}. Nevertheless, as regards the quantum calculations, the harmonic superspace approach is
much more elaborated (see, e.g., \cite{GIOS-book}). Just for this reason we prefer
to employ it while studying the low-energy effective action in $\cN=4$ SYM theory.
As we will show in subsequent sections, there are in fact a few $\cN=4$ harmonic superspaces which provide very simple
and nice expressions for $\Gamma_4$.

It is known that the $\cN=3$ and $\cN=4$ SYM models are equivalent
on the mass shell \cite{GIOS-book}. This is also true for their low-energy effective actions. The amazing
feature of $\cN=3$ SYM theory is that it admits an off-shell $\cN=3$
superfield formulation \cite{GIKOS-N3a,GIKOS-N3}. This formulation
is based on $\cN=3$ harmonic superspace with $SU(3)$ harmonic
variables. Thus, it is natural to fulfill  the study of the $\cN=3$ SYM low-energy effective
action, employing the techniques  of the  $\cN=3$ harmonic superspace. The expression for $\Gamma_4$ in the $\cN=3$ harmonic superspace was found in \cite{BISZ}.
This construction will be reviewed in sect. \ref{SectN3}.

\section{Various forms of the Wess-Zumino term for scalar fields}
\label{SectWZ}

The Wess-Zumino term for scalar fields in the $\cN=4$ SYM action
(\ref{WZterm}) is represented by the five-dimensional
integral of the exact five-form with explicit
$SO(6)$ symmetry. Using the Stokes theorem this expression can
always be represented in the form of four-dimensional
integral which is implicitly invariant under $SO(6)$. As we will
show, there are several four-dimensional representations
of this term which differ in the manifestly realized subgroups of the full
R-symmetry group $SO(6)$. All these forms naturally appear in different
superfield formulations of the low-energy $\cN=4$ SYM effective
action.

We will start with a $d$-dimensional generalization of (\ref{WZterm}) and further present the
results for the particular $d=4$ case. The material of this section is essentially based on the papers \cite{BelSam1,BelSam2,BISZ}.

\subsection{$SO(d+2)$-invariant Wess-Zumino term}

Let us consider $d+2$ scalar fields $X_A$, $A=1,\ldots, d+2\,,$ in the
$d+1$-dimensional Minkowski space. For $X_A X_A\ne0$ we can introduce
the normalized scalars $Y_A$
\be
Y_A=\frac{X_A}{|X|}\,, \quad |X|=\sqrt{X_A X_A}\,.
\ee
Since
\be
Y_A Y_A=1\,,
\ee
these normalized scalars parametrize the sphere
$S^{d+1}=SO(d+2)/SO(d+1)$. The volume form on this sphere reads
\bea
\omega_{d+1} &=& \frac{\varepsilon^{A_1\dots A_{d+2}}}{(d+1)!}Y_{A_1}dY_{A_2}\wedge dY_{A_3}\wedge\dots\wedge dY_{A_{d+2}} \nn\\
&=& d^{d+1}x\frac{\varepsilon^{A_1\dots A_{d+2}}}{(d+1)!}\,
\varepsilon^{M_1\dots M_{d+1}}Y_{A_1}\partial_{M_1}Y_{A_2}\dots\partial_{M_{d+1}}Y_{A_{d+2}} \,.\label{Volform}
\eea
In terms of this form the $d+1$ dimensional generalization of
(\ref{WZterm}) is given by
\be
\label{dWZ}
S_{\rm WZ}^{(d)}=-N\frac{(d/2)!}{\pi^{d/2}}\int_{\Omega_Y} \omega_{d+1} \ .
\ee
Here $\Omega_Y$ is a hemisphere in $S^{d+1}$ whose boundary, $\partial\Omega_Y$,
is the image of the $d$-dimensional space-time, viewed as a large $S^d$,
under the map $Y_A(x)$ \cite{Witten:1983tw,Gaume}.
For any integer $N$, choosing another hemisphere
shifts $S_{\rm WZ}^{(d)}$ by $2\pi$ $\times$ an integer.

Let us now split the index $A$ into $a=1,\dots,n$ and $a'=n+1,\dots n+m$, where we defined $m=d+2-n$.
With the normalization $\varepsilon^{1\dots(n+m)}=\varepsilon^{1\dots n}\varepsilon^{n+1\dots n+m}$,
we can rewrite (\ref{Volform}) in the more unfolded form
\be
\omega_{d+1}=\frac{1}{m}\omega_{n-1}\wedge d\omega'_{m-1}
+(-)^n\frac{1}{n}d\omega_{n-1}\wedge\omega'_{m-1} \,,
\ee
where
\bea
\omega_{n-1} &=& \frac{\varepsilon^{a_1\dots a_n}}{(n-1)!}Y_{a_1} dY_{a_2}\wedge\dots\wedge dY_{a_n}\,,
\nn\\
\omega'_{m-1} &=& \frac{\varepsilon^{a'_1\dots a'_m}}{(m-1)!}Y_{a'_1} dY_{a'_2}\wedge\dots\wedge dY_{a'_m} \, .
\eea
Introducing $y=Y_a Y_a=1-Y_{a'}Y_{a'}$, we find the following useful identities
\be
\label{dyom}
dy\wedge\omega_{n-1}=\frac{2}{n}y d\omega_{n-1}\,, \qquad
dy\wedge\omega'_{m-1}=-\frac{2}{m}(1-y) d\omega'_{m-1} \, ,
\ee
where we used the identity $dY_a\wedge dY_{a_2}\wedge \cdots \wedge dY_{a_n} = \frac{1}{n!} \varepsilon_{a a_2\ldots a_n}
\varepsilon^{b b_2\ldots b_n}dY_b\wedge dY_{b_2}\wedge \cdots \wedge dY_{b_n}$. Also, in various manipulations with forms
the cyclic identity $f^a \varepsilon^{a_1 a_2\ldots a_n}  + (-)^n f^{a_n}\varepsilon^{a a_1\ldots a_{n-1}} + \ldots = 0$ is useful.
Expressing $d\omega_{n-1}$ and $d\omega'_{m-1}$ from (\ref{dyom}) and substituting these expressions into (\ref{Volform}), we obtain the convenient
representation for the volume form
\be
\label{dyomom}
\omega_{d+1}=(-)^{n}\frac{dy\wedge\omega_{n-1}\wedge\omega'_{m-1}}{2y(1-y)} \ .
\ee
Next, we take the ansatz~\footnote{
The volume form $\omega_{d+1}$ is closed, but not exact.
This is consistent with (\ref{omf}) only if $f(y)$ is singular
at some value of $y$ in the interval $0\leq y\leq 1$.}
\be
\label{omf}
\omega_{d+1}=d\Big(f(y)\omega_{n-1}\wedge\omega'_{m-1}\Big) \, ,
\ee
and also bring it to the form (\ref{dyomom}), using the identities (\ref{dyom}).
We then immediately find that $f(y)$ must satisfy the following differential equation
\be
\label{feq}
\frac{d}{dy}f(y)+\frac{1}{2}\left(\frac{n}{y}-\frac{m}{1-y}\right)f(y)=\frac{(-1)^{n}}{2y(1-y)} \, .
\ee
Its general solution is given by~\footnote{
$B(n,m)=\Gamma(n)\Gamma(m)/\Gamma(n+m)$ is the Euler beta function, and
$B_y(n,m)=\int_0^y dt\, t^{n-1}(1-t)^{m-1}$ is the incomplete beta function
satisfying $B_1(n,m)=B(n,m)$.}
\be
\label{fnm}
f(y)=\frac{(-1)^n}{2y^{n/2}(1-y)^{m/2}}
\left\{B_y\left(\frac{n}{2},\frac{m}{2}\right)-C\, B\left(\frac{n}{2},\frac{m}{2}\right)\right\} \, ,
\ee
where $C$ is a constant of integration. The solution is regular at
$y=0$ if $C=0$ and regular at $y=1$ if $C=1$.
Choosing $f(y)$ that is non-singular in $\Omega_Y$ and using
Stokes' theorem, we obtain the $d$-dimensional form of the
Wess-Zumino term with manifest $SO(n)\times SO(m)$ invariance,
\be
\label{WZnm}
S_{\rm WZ}^{(d)}=-N\frac{(d/2)!}{\pi^{d/2}}
\frac{\varepsilon^{a_1\dots a_n}}{(n-1)!}
\frac{\varepsilon^{a'_1\dots a'_m}}{(m-1)!}
\int_{\partial\Omega_Y}\!\! f(Y_a Y_a)
Y_{a_1}dY_{a_2}\dots dY_{a_n}
Y_{a'_1}dY_{a'_2}\dots dY_{a'_m}
\ee
(recall that $d=n+m-2$\,). The residual transformations from $SO(d+2)$ vary the integrand
in this expression into an exact $d$-form, which is consistent with the
fact that $S_{\rm WZ}^{(d)}$ is $SO(d+2)$ invariant. The proof is based on the use of (\ref{feq})
and the cyclic identity mentioned earlier.

\subsection{$SO(6)$ Wess-Zumino term with manifest $SO(5)$}

Now we consider the case $d=4$ which corresponds to the
four-dimensional Minkowski space. In this case the Wess-Zumino
term (\ref{dWZ}) has manifest $SO(6)$ symmetry
\be
\label{WZso6}
S_{\rm WZ}^{(4)}=-\frac{N}{60\pi^2}\int_{\Omega_Y} \varepsilon^{A B C D E F}
Y_A dY_B\wedge dY_C\wedge dY_D\wedge dY_E\wedge dY_F \,.
\ee
This expression is reduced to (\ref{WZterm}) for $N=1$. Using (\ref{WZnm})
with $n=5$ and $m=1$, we then obtain the four-dimensional form of
this Wess-Zumino term with manifest $SO(5)$ invariance,
\bea
\label{WZso5}
S_{\rm WZ}^{(4)} &=& \frac{N}{60\pi^2}\int_{\partial\Omega_Y}\varepsilon^{abcde}\;\frac{g(z)}{Y_6^5}
Y_a dY_b\wedge dY_c\wedge dY_d\wedge dY_e \nn\\
&=& \frac{N}{60\pi^2}\int d^4x\,\varepsilon^{mnpq}
 \varepsilon^{abcde}\;\frac{g(z)}{X_6^5}X_a\partial_m X_b\partial_n X_c\partial_p
  X_d\partial_q X_e \, ,
\eea
where $m=0,1,2,3$ is the four-dimensional space-time index,
$a=1,2,3,4,5$ is the $SO(5)$ index, and we defined $g(z)=-5(1-y)^3 f(y)$ with
\be
z^2=\frac{y}{1-y}=\frac{Y_a Y_a}{Y_6^2}=\frac{X_a X_a}{X_6^2} \, .
\ee
This function satisfies the equation
\be
z\frac{d}{dz}g(z)+5g(z)=\frac{5}{(1+z^2)^3} \,.\label{zequat}
\ee
The solution of (\ref{zequat}), such that it is regular at $z=0$, with $g(0)=1$,
is given by the expression
\be
\label{gsum}
g(z)=\frac{5}{8z^5}\left[3\arctan{z}-\frac{z(3+5z^2)}{(1+z^2)^2}\right]
=\frac{5}{2}\sum_{n=0}^\infty\frac{(n+2)(n+1)}{2n+5}(-z^2)^n \, .
\ee

\subsection{$SO(6)$ Wess-Zumino term with manifest $SO(4)\times SO(2)$}
\label{Sect3.3}
When $n=4$ and $m=2$, the solution to (\ref{feq}) that is regular at $y=0$ is simply
\be
f(y)=\frac{1}{4(1-y)} \, .
\ee
The form of the Wess-Zumino term (\ref{WZso6}) with manifest $SO(4)\times SO(2)$ invariance is then
\bea
S_{\rm WZ}^{(4)} &=& -\frac{N}{12\pi^2}\int_{\partial\Omega_Y}\varepsilon^{abcd}\varepsilon^{a' b'}
Y_a dY_b\wedge dY_c\wedge dY_d\wedge \frac{Y_{a'} dY_{b'}}{Y_{c'} Y_{c'}} \nn\\
&=& -\frac{N}{12\pi^2}\int d^4x\, \varepsilon^{m n p q}
 \varepsilon^{a b c d}\varepsilon^{a' b'}
\frac{X_a\partial_m X_b\partial_n X_c\partial_p X_d}{(X_e X_e+X_{d'} X_{d'})^2}
\frac{X_{a'}\partial_q X_{b'}}{X_{c'} X_{c'}} \,,\label{S4WZ}
\eea
where now $a=1,2,3,4$ is the $SO(4)$ index, $a'=5,6$ is the $SO(2)$ index,
and $1-y=Y_{c'}Y_{c'}\,$.
Making the polar decomposition
\bea
X_6+i X_5=X e^{i\alpha} \ ,
\eea
we can rewrite (\ref{S4WZ}) as
\be
\label{WZso4so2}
S_{\rm WZ}^{(4)}=\frac{N}{12\pi^2}\int d^4x\,\varepsilon^{m n p q}\varepsilon^{abcd}
\frac{X_a\partial_m X_b\partial_n X_c\partial_p X_d}{(X_e X_e+X^2)^2}
\partial_q\alpha \,.
\ee
In this form of $S_{\rm WZ}^{(4)}$, the $SO(2)$ group acts as constant shifts of $\alpha$.

\subsection{$SO(6)$ Wess-Zumino term with manifest $SO(3)\times SO(3)$}

Using (\ref{WZnm}) with $n=3$ and $m=3$, we obtain the form of the Wess-Zumino
term (\ref{WZso6}) with manifest $SO(3)\times SO(3)$ invariance,
\be
\label{WZ-biharm}
S_{\rm WZ}^{(4)} = -\frac{N}{2\pi^2}\int_{\partial\Omega_Y}\varepsilon^{abc}
 \varepsilon^{a' b' c'}f(y)
(Y_a dY_b\wedge dY_c)\wedge (Y_{a'}\, dY_{b'}\wedge dY_{c'}) \,,
\ee
where $y=Y_a Y_a=1-Y_{a'}Y_{a'}$ and the function $f(y)$ is given by (\ref{fnm}).

Let us introduce the function
\be
g(z)= -8 f(y)\,,
\ee
where
\be
z^2 = \frac{y}{1-y} = \frac{Y^a Y^a}{Y^{a'}Y^{a'}}\,.
\ee
As a corollary of eq.\ (\ref{feq}), this function obeys
\be
z \frac{d}{dz}g(z) + 3 \frac{1-z^2}{1+z^2} g(z) =8\,.
\ee
The solution of this equation which is regular at $z=0$, with $g(0)=\frac83$, is given by
\be
g(z)=\frac{z^4-1}{z^2}+\frac{(z^2+1)^3}{z^3}\arctan z\,.
\label{2}
\ee
This function defines the Wess-Zumino term (\ref{WZ-biharm}) in the form
\be
\label{WZ-biharm+}
S_{\rm WZ}^{(4)} = \frac{N}{16\pi^2}\int_{\partial\Omega_Y}\varepsilon^{abc}
 \varepsilon^{a' b' c'}g(z)
(Y_a dY_b\wedge dY_c)\wedge (Y_{a'}\, dY_{b'}\wedge dY_{c'}) \,.
\ee

Note that the group $SO(3)\times SO(3)$ is locally isomorphic to $SU(2)\times SU(2)$. Therefore, as we will see in sect.\ \ref{SectWZ-biharm},
the Wess-Zumino term in the form (\ref{WZ-biharm}) appears as a component in the $\cN=4$ SYM low-energy effective action
in the bi-harmonic $\cN=4$ superspace.

\subsection{Wess-Zumino term and $SU(3)$ symmetry}
The Lie group $SO(6)\simeq SU(4)$ has the following maximal subgroups:\footnote{By definition, the subgroup $H$ of a group $G$
is called maximal if there is no other proper subgroup of $G$ that contains $H$. Note that this definition does not assume
that the maximal subgroup is unique, unless additional conditions are imposed.}
$SO(5)$, $SO(4)\times SO(2)$, $SO(3)\times SO(3)$ and $SU(3)\times U(1)$. In the previous sections we considered
three different forms of the Wess-Zumino term which correspond to the first three subgroups: $SO(5)$, $SO(4)\times SO(2)$ and
$SO(3)\times SO(3)$. It remains to consider the last possibility related to $SU(3)\times U(1)$. As we will show here,
in contrast to the former cases this symmetry group does not admit a manifest realization in the four-dimensional
form of the Wess-Zumino term.

We start with the $SO(6)$ covariant Wess-Zumino term (\ref{WZterm}) and rewrite it in the form with the explicit $SU(3)$ symmetry.
To this end, using six real scalars $Y^A$, we construct three complex $SU(3)$ triplet scalars $f^i$, $i=1,2,3$, as
\bea
&&
f^1=Y^1+iY^2\,,\quad
f^2=Y^3+iY^4\,,\quad
f^3=Y^5+iY^6\,,\nn\\&&
\bar f_1=Y^1-iY^2\,,\quad
\bar f_2=Y^3-iY^4\,,\quad
\bar f_3=Y^5-iY^6\,.
\label{ff_}
\eea
Like $Y^A$, the scalars $f^i$ take values on the five-sphere with the unit radius
\be
f^i \bar f_i =1\,.
\label{norm}
\ee
In terms of these complex scalars the Wess-Zumino action (\ref{WZterm}) exhibits manifest $SU(3)$ symmetry:
\bea
S_{\rm WZ}&=&\frac i{48\pi^2}\varepsilon^{MNKLP}
\varepsilon_{ijk}\varepsilon^{i'j'k'}
\int_{\cal M} d^5x[-(f^i\partial_M f^j \partial_N f^k)
\partial_K(\bar f_l \partial_L \bar f_m \partial_P \bar f_n)
\nn\\&&
+\,
\partial_K(f^i\partial_M f^j \partial_N f^k)
(\bar f_{i'} \partial_L \bar f_{j'} \partial_P \bar f_{k'})
]\,.
\label{WZ2}
\eea

Let us introduce the following 2-forms
\be
\omega_2=\varepsilon_{ijk}f^i df^j\wedge df^k\,,\qquad
\bar\omega_2=\varepsilon^{ijk}\bar f_i d\bar f_j \wedge d\bar
f_k\,.
\label{omega2}
\ee
In terms of these forms the action (\ref{WZ2}) acquires the concise form
\be
S_{\rm WZ}=\frac i{48\pi^2}\int_{\cal M} (d\omega_2\wedge \bar\omega_2
-\omega_2\wedge d\bar\omega_2)\,.
\label{WZ3}
\ee
It is easy to check that this action is real.

The equation (\ref{norm}) has the obvious corollary
\be
d f^i \bar f_i + f^i d\bar f_i=0\,.
\ee
As a consequence, the differential forms (\ref{omega2}) obey the important constraint
\be
\omega_2\wedge d\bar\omega_2=-d\omega_2\wedge \bar\omega_2\,,
\ee
or
\be
d(\omega_2\wedge\bar\omega_2)=0\,.
\label{10}
\ee
Using this relation, the action (\ref{WZ3}) cab be cast in the form
\be
S_{\rm WZ}=\frac i{24\pi^2}\int_{\cal M} d\omega_2\wedge\bar \omega_2\,.
\label{WZ4}
\ee

Let us define some complex constant triplet $c^i$ with the non-vanishing norm,
$c^i \bar c_i\ne0$. With the help of this triplet we can construct the scalar objects
\be
y=f^i \bar c_i\,,\qquad
\bar y=\bar f_i c^i\,,
\ee
which obey the identities
\be
dy\wedge\omega_2=\frac y3d\omega_2\,,
\qquad
d\bar y\wedge\bar \omega_2=\frac{\bar y}3d\bar\omega_2\,.
\ee
Owing to these identities, the action (\ref{WZ3}) admits the form
\be
S_{\rm WZ}=\frac{i}{8\pi^2}\int_{\cal M} \frac 1y dy\wedge
\omega_2\wedge\bar\omega_2
=\frac{i}{8\pi^2}\int_{\cal M} d\ln y \wedge
\omega_2\wedge\bar\omega_2\,.
\ee
Equivalently, it can be rewritten in the self-conjugated form
\be
S_{\rm WZ}=\frac{i}{16\pi^2}\int_{\cal M} d\ln\frac{ y}{\bar y} \wedge
\omega_2\wedge\bar\omega_2\,.
\label{S-omega}
\ee

The identity (\ref{10}) allows us to apply the Stokes theorem to rewrite the
action (\ref{S-omega}) as an integral over the boundary of $\cal M$ \be
S_{\rm WZ}=\frac{i}{16\pi^2}\int_{\cal M}
d[\ln\frac{ y}{\bar y}\, \omega_2\wedge\bar\omega_2]
=\frac{i}{16\pi^2}\int_{\partial\cal M}
\ln\frac{ y}{\bar y}\,
\omega_2\wedge\bar\omega_2+\chi_4\,.
\label{WZ5}
\ee
Here, $\chi_4$ is an arbitrary closed  4-form, $d\chi_4=0\,$.
For simplicity in what follows we choose this form to be vanishing,  $\chi_4=0\,$.
The boundary $\partial{\cal M}$ can be identified with the four-dimensional Minkowski space.

Let us express the action (\ref{WZ5}) in terms of the scalars (\ref{ff_})
\be
S_{\rm WZ}=\frac{i}{16\pi^2}
\varepsilon^{mnpq}\varepsilon_{ijk}\varepsilon^{i'j'k'}
\int d^4x\, \ln\frac{f^l \bar c_l}{\bar f_{l'} c^{l'}}
(f^i \partial_m f^j \partial_n f^k)
(\bar f_{i'} \partial_p \bar f_{j'} \partial_q \bar f_{k'})\,.\label{3.42}
\ee
Recall that the scalars $f^i$ have unit norm [eq. (\ref{norm})]. They are expressed through the unconstraint scalars $\varphi^i$ as
\be
f^i=\varphi^i/\sqrt{\varphi^l\bar\varphi_l}\,,\qquad
\bar f_i=\bar\varphi_i/\sqrt{\varphi^l\bar\varphi_l}\,.
\label{3.39}
\ee
Being written through $\varphi^i$ and $\bar\varphi_i$, the Wess-Zumino action (\ref{3.42}) reads
\be
S_{\rm WZ}=\frac{i}{16\pi^2}
\varepsilon^{mnpq}\varepsilon_{ijk}\varepsilon^{i'j'k'}
\int d^4x\, \ln\frac{\varphi^l \bar c_l}{\bar \varphi_{l'} c^{l'}}
\frac{
(\varphi^i \partial_m \varphi^j \partial_n \varphi^k)
(\bar \varphi_{i'} \partial_p \bar \varphi_{j'} \partial_q \bar \varphi_{k'})}{
(\varphi^i\bar\varphi_i)^3}\,.
\label{SWZ0}
\ee

It is important to note that the constants $c^i$ break the
manifest $SU(3)$ symmetry. Nevertheless, it is possible to show that under the $SU(3)$ transformations of the scalars
the Lagrangian in (\ref{SWZ0}) is shifted by  a total space-time derivative, so that the action enjoys
a non-manifest  $SU(3)$ invariance (and in fact $SO(6)$
invariance as well, since we started from the covariant action (\ref{WZterm})). This is a specific feature of the subgroup
$SU(3)$ of $SU(4)$ as compared to the other maximal subgroups $SO(5)$, $SO(4)\times SO(2)$ and $SO(3)\times SO(3)$.

\subsection{The origin of the Wess-Zumino term}

One can wonder why the case of the group $SU(3)\times U(1)$ is so different from the cases of other maximal subgroups of $SO(6)$
considered in this section.
To answer this question, we have to recall the origin of the Wess-Zumino terms in the low-energy effective actions.

The appearance of Wess-Zumino terms in low-energy quantum effective actions is related to chiral anomalies of the global (``flavor'')  symmetries
\cite{Wess:1971yu,Witten:1983tw}.
In a four-dimensional gauge theory, with the gauge group $G_\text{g}$ and the global symmetry group $G_\text{gl}$, the anomaly
with respect to  $G_\text{gl}$ can be generated in a `global-gauge-gauge' or a `global-global-global' triangle diagram. In the former case,
the global symmetry is broken at the quantum level: The Noether current of $G_\text{gl}$ is not conserved and the quantum effective action
has a non-zero variation under $G_\text{gl}$. However, if only the `global-global-global' diagram is anomalous, $G_\text{gl}$
is \emph{not} broken at the quantum level: The $G_\text{gl}$ current is conserved and the effective action is invariant.
Yet, the anomaly manifests itself in the presence of the Wess-Zumino term in the quantum effective action, and the necessity of such a term can be understood on the basis
of the 't~Hooft anomaly-matching condition \cite{'tHooft:1979bh,Weinberg:1996kr}.

It is pertinent to recall what the 't~Hooft anomaly-matching argument is. Consider a model which involves chiral fermions
interacting with the gauge fields corresponding to a gauge symmetry $G_{\rm g}$ spontaneously broken down to $H_{\rm g}\subset G_{\rm g}$
by means of the Higgs mechanism. Assume
that there is a quantum anomaly of this gauge symmetry. If we integrate out, in the functional integral, some number of fields (including chiral fermions)
which have become massive due to the Higgs mechanism, we obtain an effective theory for the remaining light fields.
One may think that the contribution to the anomaly in the effective theory changes due to a fewer number of the remaining chiral fermions.
However, the anomaly is known to be \emph{exact} and so should have the same strength in the effective theory,
when part of chiral fields has been integrated out. It cannot depend on any scalar field vacuum values which trigger spontaneous breaking of gauge symmetry
and/or masses of the heavy fields and so must preserve its form in any branch of the theory.
Respectively,  the missing contribution to the anomaly in the effective theory is accounted for
just by the Wess-Zumino terms for Goldstone bosons which appear in the process of spontaneous gauge symmetry breaking, and this is the essence of
the 't~Hooft anomaly-matching condition. If the chiral fermions belong to the adjoint
representation of the anomalous gauge group, like the gauge fields, the coefficients in front of the directly calculated anomalies in the original and effective theories
are ${\rm dim}G$ and ${\rm dim} H$, respectively (up to the same overall numerical coefficient). Then the coefficient in the Wess-Zumino term should
be proportional to $({\rm dim}G_{\rm g} - {\rm dim} H_{\rm g})$. This coefficient coincides with the number
of chiral fermions which acquired mass due to the Higgs mechanism and do not show up in the effective theory. The $G_{\rm g}$ variation of such a Wess-Zumino term
makes the precisely same contribution to the anomalous current as the missed fermions \cite{Witten:1983tw,Manoh}.

To summarize, the quantum effective action of the light fields
in the theories with the heavy fields integrated out should necessarily involve the Wess-Zumino term with a fixed coefficient,
and it can be directly found by the explicit quantum calculations (see, e.g., \cite{TZ}). The real virtue of the 't~Hooft anomaly-matching argument
is that in fact {\it there is no need} to make such calculations in order to uncover this Wess-Zumino term.

It is important to realize that the 't~Hooft anomaly-matching argument can be also successfully
applied to find the Wess-Zumino term in the effective theory, when the {\it global} symmetries are anomalous, rather
than the local gauge symmetry. Indeed, if we have some global symmetry with the group $G_\text{gl}$
we can formally make it local by introducing external gauge fields which couple to the corresponding Noether currents.
Then, if $G_\text{gl}$ is potentially anomalous, i.e. there are chiral fermions in the theory, after the gauging just mentioned there
will explicitly appear the anomaly proportional to the number of  these chiral fermions. If  $G_\text{gl}$ is spontaneously broken,
the above arguments are applicable and we find out the Wess-Zumino term in the effective theory,
such that it remains non-vanishing even after switching off the background gauge field and coming back
to the original case with $G_\text{gl}$ acting as the global symmetry. Thus it should be present in the effective action of the corresponding
light fields {\it prior to} any gauging. The coefficient in front
of such Wess-Zumino term should be proportional to the number of chiral fermions which are missing in the effective theory.

This is precisely what happens in $\cN=4$ SYM theory which has the global $SU(4)$ R-symmetry with anomalous
`global-global-global' diagram \cite{Sokatchev98}. With respect to this R-symmetry, $\cN=4$ SYM is a chiral theory, because the left and right
gauginos $\psi^I_{\alpha}$ and $\bar\psi_{\dot\alpha I}$ belong to the representations ${\bf 4}$ and $\underline{\bf 4}$ which
are not equivalent to each other.\footnote{This has to be contrasted with the gauge group, with respect to which both gauginos belong to
the same adjoint representation and so cannot produce any anomaly.}
When the gauge group $G_\text{g}$ is spontaneously broken down to a subgroup $H_\text{g}$, and
the $({\rm dim}\,G_\text{g}- {\rm dim} H_\text{g})$ massive gauginos
are integrated out, the Wess-Zumino term \cite{TZ} appears in the effective action, with the coefficient proportional
to $({\rm dim}\,G_\text{g}- {\rm dim} H_\text{g})\,$, so that the 't~Hooft anomaly matching condition is satisfied \cite{Weinberg:1996kr,Intriligator}.
Since the scalar fields which receive the vacuum expectation values are in the \emph{adjoint} of $G_\text{g}$,
the unbroken group $H_\text{g}$ necessarily includes an $U(1)$ subgroup, and, as a result, the theory ``sits'' on the Coulomb branch.

At this point it is important to note that, though the $\cN=4$ SYM theory in flat Minkowski space is finite and free of anomalies,
this ceases to be true when it couples to $\cN=4$ conformal supergravity \cite{FT1,FT2}. In the latter case there is one-loop quantum anomaly
of the local superconformal symmetry $PSU(2,2|4)$ which contains $SU(4)_R$ as a subgroup. The $\cN=4$ conformal supergravity multiplet involves
vector fields which couple to the $SU(4)_R$ Noether currents of $\cN=4$ SYM theory. These vector fields give the origin of the Wess-Zumino
term in the $\cN=4$ SYM effective action, according to the 't~Hooft anomaly-matching argument. The Wess-Zumino term survives upon
switching off the supergravity fields and plays an important role in securing the rigid $\cN=4$ supersymmetry (and conformal supersymmetry) of
the $\cN=4$ SYM effective action in the flat Minkowski space.

As we have shown in this section, in order to write the Wess-Zumino term (\ref{WZterm}) as a four-dimensional integral one is forced to
sacrifice part of the manifest $SO(6)$ R-symmetry. The
't Hooft anomaly-matching argument \cite{'tHooft:1979bh,Weinberg:1996kr}
tells us that all anomalous R-symmetry generators must transform the
four-dimensional Wess-Zumino term into a total divergence, and therefore
\emph{anomalous R-symmetry subgroups cannot be made manifest.} On the other hand, with respect to the non-anomalous subgroups of $SO(6)$
(for which left and right fermions are transformed by the same representation) the density of the Wess-Zumino term should reveal a {\it manifest} invariance.

Recall that the spinor fields of the $\cN=4$ SYM supermultiplet carry the representation ${\bf 4} + {\bf \underline{4}}$ of $SU(4)$.
This representation splits into the following representations of the four maximal subgroups of $SO(6)\simeq SU(4)$ (we write this splitting only for the {\bf 4} part):
\be
\label{msubs}
\ba[b]{rclcrcl}
&& SU(3)\times U(1), &\quad& {\bf 4} &=& {\bf 3}_{+1}+{\bf 1}_{-3} \\
SO(5) &\simeq& USp(4), &\quad& {\bf 4} &=& {\bf 4} \\
SO(4)\times SO(2) &\simeq& SU(2)\times SU(2)\times U(1), &\quad&
{\bf 4} &=& ({\bf 2},{\bf 1})_{+1}+({\bf 1},{\bf 2})_{-1} \\
SO(3)\times SO(3) &\simeq& SU(2)\times SU(2), &\quad& {\bf 4} &=& ({\bf 2},{\bf 2})\,.
\ea
\ee
The first subgroup is anomalous, whereas the other three are non-anomalous. The anomaly is absent for the $USp(4)$
and $SU(2)\times SU(2)$ subgroups because the multiplets {\bf 4} of $USp(4)$ and {\bf 2} of $SU(2)$ are equivalent to the conjugated ones. The potential
$U(1)$ anomaly for the $SU(2)\times SU(2)\times U(1)$ subgroup cancels due to the symmetric $U(1)$ charge assignments of
${\bf 4}= ({\bf 2},{\bf 1})_{+1}+({\bf 1},{\bf 2})_{-1}$. Thus only symmetries under these non-anomalous subgroups can be made manifest
in the four-dimensional representation  of the Wess-Zumino term. The $SU(3)$ group, being anomalous,  cannot be made manifest. This is exactly what we see
in the action (\ref{SWZ0}), which involves the constant triplet  $c^i$ which explicitly breaks the manifest $SU(3)$ symmetry.

In the next sections we will show that the
Wess-Zumino terms with $SO(5)$ and $SO(3)\times SO(3)$ manifest symmetry
naturally appear from formulations of the $\cN=4$ SYM effective
action in the $\cN=4$ harmonic superspaces with $USp(4)$ and
$SU(2)\times SU(2)$ harmonic variables. The $SO(4)\times SO(2)$ form of the Wess-Zumino term is inherent to the $\cN=2$ harmonic superspace
formulation of $\cN=4$ SYM theory. The Wess-Zumino term in the form (\ref{SWZ0}) originates
from the $\cN=3$ SYM low-energy effective action in the $\cN=3$ harmonic superspace. It is worth pointing out in advance that all these
Wess-Zumino terms are generated by the superfield expressions for $\cN=4$ SYM effective action which are almost uniquely, up to an overall constant,
determined by the requirements of ${\cal N}=4$ supersymmetry and/or superconformal $PSU(2,2|4)$ symmetry, without any need in the explicit
perturbative calculations. The overall coefficient is further fixed by the purely topological reasoning, since it multiplies
the component Wess-Zumino term.

\section[Low-energy effective action in $\cN=2$ harmonic  superspace]{Low-energy effective action in $\cN=2$ harmonic \break superspace}
\label{SectN2}
In this section we construct the low-energy effective action in $\cN=4$ SYM theory in terms of superfields given on the $\cN=2$ harmonic superspace.
The exposition in this section is essentially based on the results of the paper \cite{BuIv}.
To make the consideration more pedagogical we start with a brief review of the basic concepts
of the $\cN=2$ harmonic superspace which was originally introduced in \cite{GIKOS}. The detailed description of
the principles of the harmonic superspace is given in the book \cite{GIOS-book}.

\subsection{Brief review of $\cN=2$ harmonic superspace}

The $\cN$-extended Minkowski superspace is parametrized by the coordinates
\be
z^M=(x^m,\theta_i^\alpha, \bar\theta^{\dot\alpha i})\,,
\label{zM}
\ee
where $x^m$, $m=0,1,2,3$, are the Minkowski space coordinates, while
$\theta_i^\alpha$ and their conjugate $\bar\theta^{\dot\alpha i}
=\overline{\theta^\alpha_i}$, $i=1,2,\ldots,\cN$, $\alpha,\dot\alpha=1,2$, are the anticommuting Grassmann coordinates. In this superspace,
$\cN$-extended Poincar\'e  supersymmetry is realized by the following infinitesimal coordinate transformations
\be
\delta \theta_i^\alpha = \epsilon_i^\alpha\,,\qquad
\delta \bar\theta^{\dot\alpha i} = \bar\epsilon^{\dot\alpha i} \,, \qquad
\delta x^m = i(\epsilon^i \sigma^m \bar\theta_i - \theta^i \sigma^m
\bar\epsilon_i)\,.
\label{N2susy}
\ee
The generators of these transformations as differential operators on the superspace can be chosen in the form
\bea
&& Q^i_\alpha = i\frac\partial{\partial\theta_i^\alpha}
 + \bar\theta^{\dot\alpha i} \sigma^m_{\alpha\dot\alpha} \partial_m\,,\qquad
\bar Q_{\dot\alpha i} = -i\frac\partial{\partial\bar\theta^{\dot\alpha i} }
-\theta_i^\alpha \sigma^m_{\alpha\dot\alpha}\partial_m\,, \nonumber \\
&& \{ Q^i_\alpha , Q^j_\beta \} = \{ \bar Q_{\dot\alpha i} , \bar Q_{\dot\beta j} \}
=0\,,\qquad
\{ Q^i_\alpha , \bar Q_{\dot\alpha j} \} = -2i \delta^i_j
\sigma^m_{\alpha\dot\alpha} \partial_m\,.\label{QN}
\eea
The corresponding covariant spinor derivatives which anticommute with the supercharges are defined as
\bea
&& D^i_\alpha = \frac\partial{\partial\theta_i^\alpha} + i \bar\theta^{\dot\alpha i}
 \sigma^m_{\alpha\dot\alpha}\partial_m\,,\qquad
\bar D_{\dot\alpha i} = -\frac\partial{\partial\bar\theta^{\dot\alpha i}}
-i \theta_i^\alpha \sigma^m_{\alpha\dot\alpha}\partial_m\,,
\label{D-explicit}\\
&& \{ D^i_\alpha , D^j_\beta \} = \{ \bar D_{\dot\alpha i} , \bar D_{\dot\beta j} \}
=0\,,\qquad
\{ D^i_\alpha , \bar D_{\dot\alpha j} \} = -2i \delta^i_j
\sigma^m_{\alpha\dot\alpha} \partial_m\,.
\label{D-algebra}
\eea

The above formulas are valid for any $\cN$. In the rest of this section we will consider the particular case $\cN=2$, with the indices $i,j =1, 2$
corresponding to the automorphism $SU(2)$ group.

By definition, the harmonic superspace, besides the familiar coordinates (\ref{zM}), contains additional bosonic coordinates $u^\pm_i$ which
parametrize the $SU(2)$ group manifold. These extra bosonic coordinates (harmonics) can be viewed as the unitary
matrices which obey the following defining property
\be
u^{+i} u^-_j - u^{-i} u^+_j = \delta^i_j\,.
\ee
The rule of complex conjugation for them is
\be
\overline{u^+_i} = u^{-i}\,.
\ee
The harmonics carry the indices $\pm$ which denote their $U(1)$ charges. We allow the superfields to be functions on the $SU(2)$ group,
$\Phi=\Phi(z,u)$. In what follows we will consider only those superfields which are represented by the harmonic series with the definite $U(1)$ charges
\be
\Phi^{(q)}(z,u) = \sum_{n=0}^\infty \varphi^{(i_1\ldots i_{n+q}j_1\ldots j_n)}(z)
u^+_{i_1} \ldots u^+_{i_{n+q}} u^-_{j_1} \ldots u^-_{j_n}\,.
\label{harmonic-series}
\ee
The coefficients of this harmonic expansion, $\varphi^{(i_1\ldots i_{n+q}j_1\ldots j_n)}(z)$,
are the conventional $\cN=2$ superfields which carry the external $SU(2)$ spin $s$,
such that $2s = |2n+q|$. This means that the superfields $\Phi^{(q)}(z,u)$ are functions on the two-sphere $S^2=SU(2)/U(1)$
rather than on the full $SU(2)$.
The series (\ref{harmonic-series}) is nothing else than the expansion over spherical harmonics on $S^2$.

One can define three independent covariant derivatives,
\be
\partial^{++} = u^{+i}\frac\partial{\partial u^{-i}}\,,\quad
\partial^{--} = u^{-i}\frac\partial{\partial u^{+i}}\,,\quad
\partial^0 = u^{+i}\frac\partial{\partial u^{+i}}
- u^{-i}\frac\partial{\partial u^{-i}}\,,
\label{harm-d}
\ee
which obey the commutation relations of the Lie algebra $su(2)$
\be
[\partial^{++},\partial^{--}]=\partial^0\,,\quad
[\partial^0, \partial^{++}] = 2 \partial^{++}\,,\quad
[\partial^0, \partial^{--}] = - 2 \partial^{--}\,.
\label{commutators-harmonic}
\ee
It is easy to see that the derivative $\partial^0$ counts the $U(1)$ charge of superfields
\be
\partial^0 \Phi^{(q)} = q \Phi^{(q)}\,.
\ee

Using the harmonic variables, we can define the $U(1)$ projections of the Grassmann variables and covariant spinor derivatives
\bea
\label{theta+}
&&\theta^\pm_\alpha = u^\pm_i \theta^i_\alpha\,,\qquad
\bar\theta^\pm_{\dot\alpha} = u^\pm_i \bar\theta^i_{\dot\alpha}\,,\\
&&D^\pm_\alpha = u^\pm_i D^i_\alpha\,,\qquad
 \bar D^\pm_{\dot\alpha} = u^\pm_i \bar D^i_{\dot\alpha}\,.
\label{D+}
\eea
Projecting the anticommutation relations (\ref{D-algebra}) for $\cN=2$ on the harmonics, we observe that
the derivatives $D^+_\alpha $ and $\bar D^+_{\dot\alpha}$ form the mutually anticommuting set
\be
\{ D^+_\alpha ,  D^+_{\beta} \} =
\{ \bar  D^+_{\dot\alpha} ,  \bar D^+_{\dot\beta} \}
=\{ D^+_\alpha, \bar D^+_{\dot\beta} \} =0\,,
\ee
while the non-trivial anticommutators are
\be
 \{ D^-_\alpha , \bar D^+_{\dot\alpha} \}
 =-\{ D^+_\alpha , \bar D^-_{\dot\alpha} \}
 = 2i\sigma^m_{\alpha\dot\alpha}
 \partial_m\,.
\ee
These anticommutation relations are completely equivalent to the $\cN=2\,$ case of the algebra (\ref{D-algebra}).

The rules of (complex) conjugation in the harmonic superspace deserve some comments. First of all,
it should be noted that the standard complex conjugation is not suitable since it maps the superfield
of the charge $q$ into the superfield of the charge $-q\,,$
\be
\overline{\Phi^{(q)}(z,u)} = \Phi^{(-q)}(z,u)\,.
\ee
Thus it seems impossible to define a real superfield in the harmonic superspace, unless $q\ne0$. It turns out, however, that
in the harmonic superspace there exists a generalized conjugation ``$\,\widetilde{\phantom{m}}\,$'' which does not change the harmonic $U(1)$ charge
 and allows to define the appropriate reality conditions. By definition \cite{GIKOS}, its action on the harmonic-independent superfields
 coincides with the conventional complex conjugation
\be
\widetilde{\varphi^{i_1\ldots i_n}(z)}
=\bar\varphi_{i_1\ldots i_n}(z)\,,
\label{4.17}
\ee
while its action on the harmonics is postulated to be
\be
\widetilde{u^\pm_i} = u^{\pm i}\,,\qquad
\widetilde{u^{\pm i}} = - u^\pm_i\,.
\label{tilde-u}
\ee
Using these rules, it is easy to see that the generalized conjugation acts on the Grassmann variables (\ref{theta+}) as
\be
\widetilde{\theta^\pm_\alpha} = \bar\theta^\pm_{\dot\alpha}\,,
\qquad
\widetilde{\bar\theta^\pm_{\dot\alpha}} = -\theta^\pm_\alpha\,.
\label{tilde-theta+}
\ee
The properties (\ref{tilde-u}) and (\ref{tilde-theta+}) show that the operation $\,\widetilde{~}\,$ is rather
a pseudo-conjugation, since it squares to $-1$ on the objects with odd charge $q$:
\be
\widetilde{\widetilde{\Phi^{(q)}(z,u)}} = (-1)^q \Phi^{(q)}(z,u)
\ee
(the same is true for the $\,\widetilde{~}\,$ conjugation of the harmonic variables and the harmonic projections of the Grassmann coordinates).
Hence, for the superfields
with the even $U(1)$ charge $q=2n$ it becomes possible to impose the reality condition
\be
{\widetilde{\Phi^{(2n)}(z,u)}} = \Phi^{(2n)}(z,u)\,.
\ee

The basic advantage of dealing with the $\cN=2$ superspace extended by the harmonic variables is that it contains invariant subspaces
with the fewer number of Grassmann coordinates, which are different from the standard {\it chiral} subspaces and are closed
under the generalized $\,\widetilde{~}\,$-conjugation. One of such subspaces, which is usually referred to as the {\it analytic subspace},
is spanned by the coordinates
\be
\zeta_A = (x^m_A, \theta^+_\alpha, \bar\theta^+_{\dot\alpha},u^\pm_i)\,,\qquad
x^m_A = x^m - 2i \theta^{(i} \sigma^m \bar\theta^{j)}
u^+_i u^-_j\,.
\label{analyt-coord}
\ee
Indeed, $x^m_A$ are real under the $\,\widetilde{~}\,$ conjugation, $\widetilde{x^m_A}=x^m_A$, and the set of Grassmann
variables $(\theta^+_\alpha, \bar\theta^+_{\dot\alpha})$ is also closed under this conjugation, as follows
from (\ref{tilde-theta+}). The $\cN=2$ supersymmetry
is realized on the coordinates (\ref{analyt-coord}) by the transformations
\bea
\delta x^m_A &=& -2i (\epsilon^i\sigma^m \bar\theta^+
+\theta^+\sigma^m \bar\epsilon^i)u^-_i\,,
 \nn\\
\delta \theta^+_\alpha &=& u^+_i \epsilon^i_\alpha\,,\qquad
 \delta\bar\theta^+_{\dot\alpha} =
  u^+_i \bar\epsilon^i_{\dot\alpha}\,,\nn\\
\delta u^\pm_i &=&0\,,
\eea
which leave the set (\ref{analyt-coord}) intact. The covariant spinor derivatives (\ref{D+}) in the analytic basis
$(\zeta_A, \theta^-_\alpha, \bar\theta^-_{\dot\alpha})$  have the following form
\bea
&&D^+_\alpha = \frac\partial{\partial \theta^{-\alpha}}\,,\qquad
 \bar D^+_{\dot\alpha} = \frac\partial{\partial \bar\theta^{-\dot\alpha}}\,,
 \label{D+analyt}\\
&&D^-_\alpha = -\frac\partial{\partial\theta^{+\alpha}}
 + 2i \bar\theta^{-\dot\alpha} \sigma^m_{\alpha\dot\alpha} \frac\partial{  \partial x^m_A}\,,\qquad
 \bar D^-_{\dot\alpha} = -\frac\partial{\partial\theta^{+\dot\alpha}}
  -2i \theta^{-\alpha} \sigma^m_{\alpha\dot\alpha} \frac\partial{
  \partial x^m_A}\,.
\eea

Some harmonic superfield $\Phi_A$ is said to be {\it analytic} if it is annihilated by the covariant
spinor derivatives $D^+_\alpha$ and $\bar D^+_{\dot\alpha}\,$,
\be
D^+_\alpha \Phi_A = \bar D^+_{\dot\alpha} \Phi_A =0\,.
\label{analyt-constr}
\ee
Since these derivatives are short in the analytic coordinates, see (\ref{D+analyt}), the analyticity constraints (\ref{analyt-constr})
are just the Grassmann Cauchy-Riemann conditions \cite{GrassmAnal} which imply that the superfield $\Phi_A$
is independent of $\theta^-_\alpha$ and $\bar\theta^-_{\dot\alpha}$ in the analytic basis:
\be
\Phi_A = \Phi_A(x^m_A, \theta^+_\alpha, \bar\theta^+_{\dot\alpha},u^\pm_i)\,.
\ee

For completeness, in this subsection we also give the analytic basis form of the harmonic derivatives (\ref{harm-d}):
\begin{subequations}
\bea
D^{++}&=& \partial^{++} -2i \theta^+\sigma^m \bar\theta^+ \frac\partial{\partial x^m_A}
 +\theta^{+\alpha} \frac\partial{\partial\theta^{-\alpha}}
 +\bar\theta^{+\dot\alpha} \frac\partial{\partial\bar\theta^{-\dot\alpha}}\,,
 \label{D++}\\
D^{--}&=& \partial^{--} -2i \theta^-\sigma^m \bar\theta^-
 \frac\partial{\partial x^m_A}
 +\theta^{-\alpha} \frac\partial{\partial\theta^{+\alpha}}
  +\bar\theta^{-\dot\alpha} \frac\partial{\partial\bar\theta^{+\dot\alpha}}
  \,,\label{D--} \\
D^0 &=& \partial^0 + \theta^{+\alpha} \frac\partial{\partial\theta^{+\alpha}}
 -\theta^{-\alpha}\frac\partial{\partial\theta^{-\alpha}}
 +\bar\theta^{+\dot\alpha} \frac\partial{\partial\bar\theta^{+\dot\alpha}}
 -\bar\theta^{-\dot\alpha} \frac\partial{\partial\bar\theta^{-\dot\alpha}}
 \,.
\eea
\end{subequations}
The commutation relations between these derivatives form
of course the same algebra as (\ref{commutators-harmonic}):
\be
[D^{++},D^{--}]=D^0\,,\quad
[D^0, D^{++}] = 2 D^{++}\,,\quad
[D^0, D^{--}] = - 2 D^{--}\,.
\label{commutators-harmonic1}
\ee

\subsection{Classical action of $\cN=4$ SYM in $\cN=2$ harmonic superspace}

The $\cN=4$ vector multiplet consists of the hypermultiplet
($\cN=2$ matter multiplet) and the $\cN=2$ vector multiplet. In this section we give an overview of these multiplets in the $\cN=2$
harmonic superspace and then present the $\cN=4$ SYM classical action in terms of these superfields.

\subsubsection{$q$-hypermultiplet}
The Fayet-Sohnius hypermultiplet \cite{FS} in harmonic superspace is described by a charged superfield $q^+$ and its conjugate $\tilde q^+$
subject to the analyticity constraints
\be
D^+_\alpha q^+ = \bar D^+_{\dot\alpha} q^+ =0\,.
\label{q-analyt}
\ee
Their free classical action reads \cite{GIKOS}
\be
S_q^{\rm free} = -\int d\zeta^{-4} du\, \tilde q^+ D^{++} q^+\,.
\label{free-q}
\ee
Here $D^{++}$ is the harmonic derivative in the analytic basis given by (\ref{D++})  and the integration measure on the analytic superspace
is defined in such a way that the following properties hold
\begin{subequations}
\label{int-analyt-rule}
\bea
&&\int d\zeta^{-4} (\theta^+)^2 (\bar\theta^+)^2 f(x) =
 \int d^4x \, f(x)\,,\\&&
\int du \,1 = 1\,,\qquad
\int du \,u^+_{(i_1} \ldots u^+_{i_m} u^-_{j_1}
 \ldots u^-_{j_n)} =0\qquad(m+n>0)\,.
\label{int-du}
\eea
\end{subequations}
Note that the analytic measure $d\zeta^{-4}$ is charged, so any Lagrangian given on the analytic superspace should carry
the harmonic  $U(1)$ charge $+4$.
The rule of integration over the harmonic variables (\ref{int-du}) implies that the integral of any monomial of harmonics
in a non-singlet irreducible representation
of $SU(2)$ vanishes.

The classical action (\ref{free-q}) yields the equation of motion for the superfield $q^+$
\be
D^{++} q^+ = 0\,.
\ee
It is possible to show that in the central basis with coordinates $(z^M,u)$ this equation has the simple solution
\be
q^+(z,u) = u^+_i q^i(z)\,,
\ee
that is $q^+$ is linear in harmonics. The analyticity constraints
(\ref{q-analyt}) acquire the form of the following constraints on  $q^i(z)$ \cite{FS}
\be
D^{(i}_\alpha q^{j)}=0\,,\qquad
\bar D^{(i}_{\dot\alpha} q^{j)}=0\,.
\ee
It is known that these constraints eliminate all auxiliary fields in $q^i$ and put the physical scalar and spinor fields on the mass shell.

In some cases it is convenient to combine the superfield $q^+$ and its conjugate $\tilde q^+$ into a doublet $q^+_a$
\be
q^+_a = (q^+ , -\tilde q^+)\,,\qquad
\widetilde{q^+_a} = q^{+a} = \left(
\begin{array}{c}
\tilde q^+ \\ q^+
\end{array}
 \right)\,.
\label{q-doublet}
\ee
In terms of these superfields the classical action (\ref{free-q}) reads
\be
S_q^{\rm free} = \frac12\int d\zeta^{-4} du \, q^+_a D^{++} q^{+a}\,.
\ee
This action is manifestly invariant under the so-called Pauli-G\"ursey $SU(2)$ symmetry
which transforms $q^+_a$ as a doublet.

\subsubsection{$\cN=2$ SYM theory in harmonic superspace}
Let us consider now the vector gauge multiplet in the $\cN=2$ superspace. The geometric approach to the gauge theory in the
$\cN=2$ superspace is based on extending the $\cN=2$ superspace derivatives $D_M= (\partial_m, D^i_\alpha, \bar D_{\dot\alpha i})$
by the gauge superfield connections
\be
D_M \longrightarrow {\cal D}_M = D_M + i A_M\,,
\label{A-connections}
\ee
and imposing the following constraints \cite{Sohnius78}
\begin{subequations}
\label{D-algebra-gauged}
\bea
\{ {\cal D}^i_\alpha , {\cal D}^j_\beta  \} &=& -2i \varepsilon^{ij}
 \varepsilon_{\alpha\beta} \bar W\,,\label{bar-W-N2}\\
\{ \bar{\cal D}_{\dot\alpha i}, \bar{\cal D}_{j\dot\beta} \} &=&
 -2i \varepsilon_{ij} \varepsilon_{\dot\alpha\dot\beta} W\,,\label{W-N2}\\
\{ {\cal D}^i_\alpha , \bar{\cal D}_{\dot\alpha j} \} &=& -2i \delta^i_j
{\cal D}_{\alpha\dot\alpha}\,.
\eea
\end{subequations}
Here $W$ and $\bar W$ are the superfield strengths which obey the Bianchi identities
\begin{subequations}
\bea
\label{chiral-W}
&\bar {\cal D}_{\dot\alpha i} W =0 \,,\qquad
{\cal D}^i_\alpha \bar W =0\,,&\\
\label{Bianchi-id}
&{\cal D}^{\alpha i} {\cal D}^j_\alpha W
 = \bar{\cal D}^i_{\dot\alpha} \bar {\cal D}^{\dot\alpha j} \bar W\,.&
\eea
\end{subequations}
The equations (\ref{chiral-W}) show that the superfield $W$
is chiral and $\bar W$ is antichiral. Therefore, the $\cN=2$ SYM action is given as an integral over the chiral or antichiral subspaces of the $\cN=2$ superspace
\be
S^{\cN=2}_{\rm SYM} = \frac14 \tr \int d^4x d^4\theta\, W^2
=\frac14 \tr \int d^4x d^4\bar\theta\, \bar W^2\,.
\label{N2-SYM-action}
\ee
Here we assume that the integrals over the Grassmann coordinates are normalized so that the following properties are valid
\be
\int d^4\theta \, \theta^4 =1\,,\quad
\int d^4\bar \theta\, \bar\theta^4 =1\,,\quad
\int d^8\theta \, \theta^4 \bar \theta^4 =1\,,
\label{int-chiral-rule}
\ee
where
\be
\theta^4 = (\theta^+)^2 (\theta^-)^2\,,\qquad
\bar\theta^4 = (\bar\theta^+)^2 (\bar\theta^-)^2\,.
\ee

The gauge connections introduced in (\ref{A-connections}) and their superfield strengths appearing in (\ref{bar-W-N2}) and (\ref{W-N2})
are defined up to the gauge transformations
\be
A'_M = -i e^{i\tau} ({\cal D}_M e^{-i\tau})\,,\quad
W' = e^{i\tau} W e^{-i\tau}\,,\quad
\bar W' = e^{i\tau} \bar W e^{-i\tau}\,, \label{tau0}
\ee
where $\tau=\tau(z)$ is a real ${\cal N}=2$ superfield gauge parameter. The action (\ref{N2-SYM-action})
is obviously invariant under these transformations.
The $\cN=2$ gauge theory introduced through the gauge connections defined in the standard ${\cal N}=2$ superspace as above is usually
referred to as the $\tau$-frame gauge theory.

The $\cN=2$ SYM Lagrangian (\ref{N2-SYM-action}) is expressed in terms of the {\it constrained} chiral (antichiral)
superfield strengths $W$ or $\bar W$.
For many application it is necessary to have an expression for the Lagrangian in terms of {\it unconstrained} gauge prepotentials
of these superfield strengths. The harmonic superspace approach naturally provides such a formulation, as is explained below.

The algebra of covariant spinor derivatives (\ref{D-algebra-gauged}) entails the corollaries
\be
\{ {\cal D}^+_\alpha , {\cal D}^+_\beta \} =
\{ \bar{\cal D}^+_{\dot\alpha} , \bar{\cal D}^+_{\dot\beta} \}
=\{ {\cal D}^+_\alpha , \bar{\cal D}^+_{\dot\beta} \} =0\,,
\label{int-cond}
\ee
where
\be
{\cal D}^\pm_\alpha = u^\pm_i {\cal D}^i_{\alpha}\,,\qquad
\bar{\cal D}^\pm_{\dot\alpha} = u^\pm_i \bar{\cal D}^i_{\dot\alpha}\,.
\label{def-D+}
\ee
The relations (\ref{int-cond}) are just the integrability conditions for the existence of the covariantly analytic superfields:
\be
{\cal D}^+_\alpha \Phi(z,u) =0\,,\qquad
\bar{\cal D}^+_{\dot\alpha} \Phi(z,u) =0\,.
\label{cov-analyt}
\ee

The solution to these constraints can be found with the help of the so-called bridge superfield $b=b(z,u)$.
The integrability conditions (\ref{int-cond})
imply the following  representation for the $+$ projections of the gauge-covariant spinor derivatives
\be
{\cal D}^+_\alpha = e^{-i b}D^+_\alpha e^{ib}\,,\qquad
\bar{\cal D}^+_{\dot\alpha} = e^{-i b}\bar D^+_{\dot\alpha} e^{ib}\,.
\label{calDD}
\ee
Without loss of generality the bridge superfield can be chosen real, $\tilde b(z,u)=b(z,u)\,$.
As follows from (\ref{calDD}), this superfield is defined modulo gauge transformations,
\be
e^{ib'} = e^{i\lambda}e^{ib}e^{-i\tau}\,,
\label{gauge-tr-bridge}
\ee
where $\tau=\tau(z)$ is an arbitrary real harmonic-independent superfield parameter (it coincides with that appearing in (\ref{tau0})),
while $\lambda=\lambda(z,u)$ is an arbitrary real analytic superfield,
$\tilde \lambda = \lambda$,
$D^+_\alpha\lambda = \bar D^+_{\dot\alpha}\lambda =0$. Now, the general solution to (\ref{cov-analyt}) in the analytic basis is given by
\be
\Phi(z,u)= e^{-ib} \Phi_A(z,u)\,,
\ee
where $\Phi_A(z,u)$ is the analytic superfield (\ref{analyt-constr}).
Thus, with the help of the bridge superfield we can bring all the differential operators and the superfields
into the so-called $\lambda$-frame, which, being combined with the choice of the analytic coordinate basis, yields what  is called
``$\lambda$-representation''. In the $\lambda$-representation, the covariantly analytic superfields become manifestly analytic
and the covariant spinor derivatives $D^+_\alpha$ and $\bar D^+_{\dot\alpha}$ acquire the ``short'' form
without gauge connections. At the same time, the harmonic derivatives (\ref{D++}) and (\ref{D--}) acquire non-trivial gauge connections
\be
{\cal D}^{++} = D^{++} + i V^{++} = e^{ib} D^{++} e^{-ib}\,,\qquad
{\cal D}^{--} = D^{--} + i V^{--} = e^{ib} D^{--} e^{-ib}\,.
\label{D--covariant}
\ee
Since the bridge superfield is real with respect to the $\,\widetilde{~}\,$ conjugation, these new gauge connections are also real
\be
\widetilde{V^{++}} = V^{++}\,,\qquad
\widetilde{V^{--}} = V^{--}\,.
\ee
Moreover, the superfield $V^{++}$ is analytic
\be
D^+_\alpha V^{++} = \bar D^+_{\dot\alpha} V^{++} =0
\ee
as a consequence of the commutation relations $[D^+_\alpha,{\cal D}^{++}]
=[\bar D^{++}_{\dot\alpha},{\cal D}^{++}]=0$.

It is important to point out that the superfields $V^{++}$ and $V^{--}$ introduced in (\ref{D--covariant}) are not independent.
They are related to each other by the ``harmonic flatness condition''
\be
D^{++} V^{--} - D^{--} V^{++} +i[V^{++},V^{--}] =0\,,
\label{V++V--}
\ee
which is a corollary of one of the
commutation relations of the algebra (\ref{commutators-harmonic1}) rewritten in the $\lambda$-frame, $[{\cal D}^{++},{\cal D}^{--}]= D^0$.
It was demonstrated in \cite{Zupnik86,Zupnik87} that the equation (\ref{V++V--}) can be uniquely solved for $V^{--}$ in terms of $V^{++}$
as the series
\be
V^{--}(z,u) = \sum_{n=1}^\infty
\int du_1 \ldots du_n \frac{(-i)^n V^{++}(z,u_1)\ldots V^{++}(z,u_n)}{
(u^+ u^+_1)(u^+_1 u^+_2)\ldots (u^+_n u^+)}\,.
\label{V--V++}
\ee
This expression involves the harmonic distributions introduced in \cite{GIOS1} and described in detail in \cite{GIOS-book}.

The superfields $V^{++}$ and $V^{--}$ are defined by (\ref{D--covariant})
up to the gauge transformations
\be
V^{\pm\pm}{}' =-ie^{i\lambda}D^{\pm\pm}e^{-i\lambda}+e^{i\lambda}V^{\pm\pm}e^{-i\lambda}\,,
\label{V-gauge}
\ee
which follow from (\ref{gauge-tr-bridge}). Since the superfield $V^{++}$ is analytic and otherwise unconstrained, while $V^{--}$
is expressed through
$V^{++}$, just $V^{++}$ is the fundamental gauge prepotential of $\cN=2$ SYM theory.
The superfield strengths $W$, $\bar W$ and the classical action (\ref{N2-SYM-action}) can be expressed through this prepotential.

Since the covariant spinor derivatives in the $\tau$-frame (\ref{def-D+}) are linear in harmonics, the following simple commutation relations
hold in this frame
\be
[D^{--}, {\cal D}^+_\alpha] = {\cal D}^-_\alpha\,,\qquad
[D^{--}, \bar{\cal D}^+_{\dot\alpha}] = \bar{\cal D}^-_{\dot\alpha}\,.
\ee
Let us rewrite these commutators in the $\lambda$-frame using the rules (\ref{calDD}) and (\ref{D--covariant}),
\be
[({\cal D}^{--})_\lambda, ({\cal D}^+_\alpha)_\lambda] = ({\cal D}^-_\alpha)_\lambda\,,\qquad
[({\cal D}^{--})_\lambda, (\bar{\cal D}^+_{\dot\alpha})_\lambda] = (\bar{\cal D}^-_{\dot\alpha})_\lambda\,,
\label{commutators-lambda}
\ee
and take into account the fact that in the $\lambda$-frame the covariant spinor derivatives
${\cal D}^+_\alpha$ and $\bar{\cal D}^+_{\dot\alpha}$ are short,
$({\cal D}^+_\alpha)_\lambda = D^+_\alpha$ and $(\bar{\cal D}^+_{\dot\alpha})_\lambda=\bar D^+_{\dot\alpha}$.
Then, the commutation relations (\ref{commutators-lambda}) amount to the following expressions for the spinor connections
\be
(V^-_\alpha)_\lambda = -D^+_\alpha V^{--}\,,\qquad
(\bar V^-_{\dot\alpha})_\lambda = -\bar D^+_{\dot\alpha} V^{--}\,.
\label{V-}
\ee

Contracting the anticommutators (\ref{bar-W-N2}) and (\ref{W-N2})
with the harmonics $u^+_i\,,  u^-_j\,$, we find the expressions for the superfield strengths,
\be
W = -\frac i4 \{ \bar{\cal D}^+_{\dot\alpha} ,\bar{\cal D}^{-\dot\alpha}  \}  \,,\qquad
\bar W = -\frac i4 \{ {\cal D}^{+\alpha} , {\cal D}^-_\alpha\}\,.
\ee
Using the expressions (\ref{V-}), we represent these superfield strengths in terms of the non-analytic harmonic gauge connection $V^{--}$
\be
W_\lambda= -\frac14 \bar D^+_{\dot\alpha} \bar D^{+\dot\alpha} V^{--}\,,\qquad
\bar W_\lambda = -\frac14 D^{+\alpha} D^+_\alpha V^{--}\,.
\label{Wlambda}
\ee
Owing to (\ref{V--V++}), the superfield strengths are functions of the analytic gauge prepotential $V^{++}$.
This makes it possible to express the $\cN=2$ SYM classical action (\ref{N2-SYM-action}) via $V^{++}$ \cite{Zupnik87}
\be
S_{\rm SYM}^{\cN=2} = \frac12 \sum_{n=2}^\infty
\frac{(-i)^n}{n} \tr\int d^{12}z du_1\ldots du_n
\frac{V^{++}(z,u_1)\ldots V^{++}(z,u_n)}{(u^+_1u^+_2)\ldots
(u^+_n u^+_1)}\,.
\label{S-SYM-harm}
\ee
The derivation of this action from (\ref{N2-SYM-action}) requires some algebra, the details of which can be found, e.g.,\
in \cite{GIOS-book}. As was demonstrated in \cite{GIOS2}, the $\cN=2$ SYM classical action in the form (\ref{S-SYM-harm})
is most suitable for quantization and studying quantum aspects of $\cN=2$ gauge theories in superspace.

Using the unconstrained analytic prepotential $V^{++}$, it is rather trivial to promote the free hypermultiplet $q^+$ action (\ref{free-q})
to the gauge invariant one; this is accomplished just through the replacement $D^{++} \rightarrow {\cal D}^{++}$:
\be
S_q =-\int d\zeta^{-4} du\, \tilde q^+ {\cal D}^{++} q^+
= -\int d\zeta^{-4} du\, \tilde q^+( D^{++} + i V^{++}) q^+\,.
\label{SqV}
\ee
Here we assume that the $q$-hypermultiplet transforms in some representation of the gauge group
\be
q^+{}' = e^{i\lambda}q^+\,,\qquad
\tilde q^+{}' = \tilde q^+ e^{-i\lambda}\,, \label{Gaugeq}
\ee
and $V^{++}$ takes values in the matrix algebra of the generators of this representation.
The classical action is invariant under the gauge transformations (\ref{Gaugeq}) supplemented by
the corresponding variation (\ref{V-gauge}) of the gauge superfield $V^{++}$.

If the $q$-hypermultiplet transforms in the adjoint representation of the gauge group, the action (\ref{SqV})
possesses the Pauli-G\"ursey $SU(2)$ symmetry.
Using the notations (\ref{q-doublet}), it can be rewritten as
\be
S_q = \frac12 \tr\int d\zeta^{-4} du \, q^+_a{\cal D}^{++} q^{+a}\,,
\label{Sq-ad}
\ee
where the covariant harmonic derivative acts on the hypermultiplet according to the rule
\be
{\cal D}^{++} q^{+a} = D^{++} q^{+a} + i [V^{++}, q^{+a}] \,.
\ee

\subsubsection{$\cN=4$ SYM classical action}
In the $\cN=2$ harmonic superspace, the $\cN=4$ vector gauge multiplet is represented   by the $\cN=2$ gauge multiplet $V^{++}$
and the hypermultiplet $q^+$. Both these multiplets should belong to the same adjoint representation of the gauge group.
The $\cN=4$ SYM action is given by the sum of the actions (\ref{S-SYM-harm}) and (\ref{Sq-ad}) for these multiplets,
\begin{subequations}
\label{S-N4}
\bea
S_{\rm SYM}^{\cN=4} &=& S_{\rm SYM}^{\cN=2} + S_q\,,\\
S_{\rm SYM}^{\cN=2} &=& \frac12 \sum_{n=2}^\infty
\frac{(-i)^n}{n} \tr\int d^{12}z du_1\ldots du_n
\frac{V^{++}(z,u_1)\ldots V^{++}(z,u_n)}{(u^+_1u^+_2)\ldots
(u^+_n u^+_1)}\,,\\
S_q &=&\frac12 \tr\int d\zeta^{-4} du \,
q^+_a ( D^{++} q^{+a} + i [V^{++}, q^{+a}])\,.
\eea
\end{subequations}
The total  action is invariant under the following hidden $\cN=2$ supersymmetry transformations
\begin{subequations}
\label{hidden-suzy-full}
\bea
\delta V^{++} &=&(\epsilon^{\alpha a}\theta^+_\alpha +
 \bar\epsilon^a_{\dot\alpha}\bar\theta^{+\dot\alpha}) q^+_a\,,\\
\delta q^+_a &=& -\frac1{32}(D^+)^2 (\bar D^+)^2  [
(\epsilon^{\alpha}_a\theta^-_\alpha +
 \bar\epsilon_{\dot\alpha a}\bar\theta^{-\dot\alpha})V^{--}]
 \nn\\&=&\frac18 (D^+)^2[ (\epsilon^{\alpha}_a\theta^-_\alpha +
 \bar\epsilon_{\dot\alpha a}\bar\theta^{-\dot\alpha}) W_\lambda ]
+ \frac18 (\bar D^+)^2[ (\epsilon^{\alpha}_a\theta^-_\alpha +
 \bar\epsilon_{\dot\alpha a}\bar\theta^{-\dot\alpha}) \bar W_\lambda ]
\nn\\&&-\frac18 (\epsilon^{\alpha}_a\theta^-_\alpha +
 \bar\epsilon_{\dot\alpha a}\bar\theta^{-\dot\alpha}) (D^+)^2 W_\lambda\,,
\eea
\end{subequations}
where $\epsilon^{\alpha a}$ and $\bar\epsilon^a_{\dot\alpha}$ are new anticommuting parameters and
$W_\lambda\,$, $\bar W_\lambda$ are defined in (\ref{Wlambda}).
It is possible to show that the algebra of these transformations
is closed modulo terms proportional to the classical equations of motion. Therefore, in this formulation only $\cN=2$ supersymmetry is closed off shell.

In conclusion of this section we present the harmonic superspace formulation of the abelian $\cN=4$ SYM theory. In this case
the action (\ref{S-N4}) acquires the simple form
\be
S^{\cN=4}  = \frac18  \int d^4x d^4\theta\, W^2
+\frac18  \int d^4x d^4\bar\theta\, \bar W^2
+\frac12 \int d\zeta^{-4} du \,
q^+_a  D^{++} q^{+a}\,.
\label{S-N4-ab}
\ee
Recall that the hypermultiplet obeys the {\it off-shell} analyticity constraint
\be
D^+_\alpha q^+_a = \bar D^+_{\dot\alpha} q^+_a =0\,,
\label{q-analyt1}
\ee
while the $\cN=2$ gauge superfield strengths $W$ and $\bar W$ are chiral and anti-chiral
\begin{subequations}
\be
\bar D^\pm_{\dot\alpha} W =0\,,\qquad
D^\pm_\alpha \bar W =0\,,
\label{W-anti-chiral}
\ee
and also obey the Bianchi identity
\be
(D^\pm)^2 W = (\bar D^\pm)^2 \bar W\,.
\label{W-bianchi}
\ee
\end{subequations}
The relations (\ref{W-anti-chiral}) and (\ref{W-bianchi}) follow from
(\ref{chiral-W}) and (\ref{Bianchi-id}), respectively.
The equations of motion for these superfields implied by the action (\ref{S-N4-ab}) read
\begin{subequations}
\label{on-shell-constr}
\bea
&&D^{++} q^+_a = 0\,, \label{hyper-eom}\\
&&
(D^\pm)^2 W =0\,,\quad
(\bar D^\pm)^2 \bar W=0\,.
\eea
\end{subequations}
They are obtained by varying (\ref{S-N4-ab}) with respect to the analytic unconstrained prepotential $V^{++}\,$.
In what follows, the equations (\ref{on-shell-constr}) will be referred to as the {\it on-shell} constraints.

Note that the hypermultiplet equation of motion (\ref{hyper-eom}) in the central basis implies
that $q^+_a$ is linear in harmonics, $q^+_a = u^+_i q^i_a$.
Thus, we can define the superfield
\be
q^-_a = D^{--} q^+_a = u^-_i q^i_a\,,
\ee
which obeys
\be
D^{--} q^-_a = 0\,,\qquad
D^-_\alpha q^-_a = \bar D^-_{\dot\alpha} q^-_a =0
\label{anlyt-q-}
\ee
as a consequence of (\ref{hyper-eom}) and (\ref{q-analyt1}). In the analytic basis, $q^-_a$ is defined in the same way, but with the appropriate
analytic-basis covariant derivatives.

When the superfields $W$, $\bar W$ and $q^+_a$ obey both off- and on-shell constraints (\ref{q-analyt1})--(\ref{anlyt-q-}),
the transformations of hidden $\cN=2$ supersymmetry (\ref{hidden-suzy-full}) are simplified to
\begin{subequations}
\label{hidden-susy-simplified}
\bea
&&\delta W = \frac12 \bar\epsilon^{\dot\alpha a }
 \bar D^-_{\dot\alpha} q^+_a\,,\qquad
 \delta \bar W = \frac12 \epsilon^{\alpha a }
  D^-_\alpha q ^+_a\,,\\
&& \delta q^+_a = \frac14 (\epsilon^\alpha_a D^+_\alpha W
 + \bar\epsilon^{\dot\alpha}_a \bar D^+_{\dot\alpha} \bar W )
 \,,\qquad
\delta q^-_a = \frac14 (\epsilon^\alpha_a D^-_\alpha W
 + \bar\epsilon^{\dot\alpha}_a \bar D^-_{\dot\alpha} \bar W )\,.
\eea
\end{subequations}
This form of hidden supersymmetry is useful for checking the invariance of the action functionals up to terms vanishing on the equations of motion.
We will employ these transformations in the next subsection for constructing the $\cN=4$ SYM low-energy effective action in the $\cN=2$ harmonic
superspace.

\subsection{Derivation of the effective action}

Our goal is to find the four-derivative part of the
$\cN=4$ SYM low-energy effective action $\Gamma$.
In the component formulation, this action should include both the term $F^4/X^4$ (\ref{F4X4}) and the Wess-Zumino term (\ref{WZterm}), as well as all their $\cN=4$ supersymmetric completions.

Recall that the $F^4/X^4$ term in the $\cN=2$ superspace is described by the non-holomorphic potential (\ref{non-hol}) \cite{DS,S16}:
\be
\int d^{12} z \, {\cal H}(W,\bar W)\,,\qquad
{\cal H}(W,\bar W) = c \,\ln \frac W\Lambda \ln\frac{\bar W}\Lambda\,,
\label{non-hol1}
\ee
where $\Lambda$ is an arbitrary scale.
The value of the constant $c$ was calculated in \cite{deWit,Lindstrom,non-hol2,non-hol3} (see also the review \cite{EChaYa}). In particular,
for the case of the gauge group $SU(2)$ spontaneously broken down to $U(1)$ the value of this coefficient is
\be
c=\frac1{(4\pi)^2}\,.
\label{c}
\ee

The $\cN=4$ SYM low-energy effective action should be an $\cN=4$ supersymmetric completion
of the $\cN=2$ non-holomorphic potential (\ref{non-hol1}):
\begin{subequations}
\bea
\Gamma &=& \int d^{12}z du \,{\cal L}_{\rm eff} (W,\bar W, q^\pm_a)\,,
\label{Gamma-full}
\\
{\cal L}_{\rm eff} (W,\bar W, q^\pm_a) &=&
{\cal H}(W,\bar W) + {\cal L}(W,\bar W,q^\pm_a)\,.
\label{H-L}
\eea
\end{subequations}
The part of the effective Lagrangian ${\cal L}(W,\bar W,q^\pm_a)$ should be fixed from the requirement that the effective action $\Gamma$
is invariant under $\cN=4$ supersymmetry. Since we are interested in the on-shell low-energy effective action, it will be sufficient
to impose the condition that $\Gamma$ is invariant under the hidden $\cN=2$ supersymmetry transformations in the on-shell form
(\ref{hidden-susy-simplified}).

To begin with, we compute the variation of the $\cN=2$ non-holomorphic effective action under the $\cN=2$ supersymmetry transformations
(\ref{hidden-susy-simplified})
\be
\delta \int d^{12}z du\, {\cal H}(W,\bar W) = \frac c2
\int d^{12} z du\, \frac{q^{+a}}{\bar W W}
 (\epsilon^\alpha_a D^-_\alpha W + \bar\epsilon^{\dot\alpha}_a
  \bar D^-_{\dot\alpha} \bar W )\,.
 \label{delta-H}
\ee
The Lagrangian ${\cal L}(W, \bar W , q^\pm_a)$ must be determined from the condition that its variation cancels (\ref{delta-H}).
We introduce the quantity
\be
{\cal L}_1 = - c \frac{q^{+a} q^-_a}{\bar W W}
\ee
and observe that it transforms according to the rule
\be
\delta \frac{q^{+a} q^-_a}{\bar W W} =
\frac{q^{+a}}{2\bar W W}
 (\epsilon^\alpha_a D^-_\alpha W + \bar\epsilon^{\dot\alpha}_a
  \bar D^-_{\dot\alpha} \bar W )
   + (q^{+a}q^-_a)\delta\left(\frac1{\bar W W} \right)
   +D^{--} \left( \frac{\delta q^{+a} q^+_a}{\bar W W} \right).
\label{delta-1}
\ee
Then, in the expression
\be
{\cal L}^{(1)}_{\rm eff,1} = {\cal H}(W,\bar W) + {\cal L}_1
=c\ln\frac W\Lambda \ln\frac{\bar W}{\Lambda}
-c \frac{q^{+a} q^-_a}{\bar W W}
\label{L-1}
\ee
the variation of the non-holomorphic potential (\ref{delta-H}) is canceled by the variation of ${\cal L}_1$,
but the contributions from the second term in
(\ref{delta-1}) remain non-canceled.

The variation of (\ref{L-1}) can be brought to the form
\bea
\delta {\cal L}_{\rm eff,1} &=&\frac c2 \int d^{12}z du
 \frac{q^{+b} q^-_b }{(\bar W W)^2}(\bar W \bar\epsilon^{\dot\alpha a}
  \bar D^-_{\dot\alpha} q^+_a + W \epsilon^{\alpha a}D^-_\alpha q^+_a)
\nn\\
&=&-\frac c3 \int d^{12}z du
\frac{q^{+b} q^-_b }{(\bar W W)^2} q^{+a}(
\bar\epsilon^{\dot\alpha}_a \bar D^-_{\dot\alpha} \bar W
+\epsilon^\alpha_a D^-_\alpha W)\,,
\label{delta-L-1}
\label{4.81}
\eea
where we have integrated by parts and used the equations (\ref{q-analyt1})--(\ref{anlyt-q-}), as well as cyclic identities for the $SU(2)$ doublet indices.
Now let us consider the quantity
\be
{\cal L}_{\rm eff,2}  = {\cal L}_{\rm eff,1}
+\frac c3 \left(\frac{q^{+a} q^-_a}{\bar WW } \right)^3
\equiv {\cal L}_{\rm eff,1} + {\cal L}_2\,,
\ee
where ${\cal L}_{\rm eff,1}$ is given by (\ref{L-1}).
The coefficient in the new term ${\cal L}_2$ has been fixed so that
the variation of the numerator of this term cancels (\ref{delta-L-1}). The rest of the full variation of ${\cal L}_2$ once again survives,
and in order to cancel it, one is led to add the term
\be
{\cal L}_3 = -\frac{2c}{9} \left(
\frac{q^{+a}q^-_a}{\bar WW}
\right)^3
\ee
to ${\cal L}_1+{\cal L}_2$, and so on.

The above consideration suggests that the hypermultiplet-dependent part of the effective Lagrangian (\ref{H-L}) has the form of the power series
\be
{\cal L}= \sum_{n=1}^\infty {\cal L}_n
 = c\sum_{n=1}^\infty c_n \left(
 \frac{q^{+a}q^-_a}{\bar WW}
 \right)^n ,
\label{L-series}
\ee
where $c_n$ are some coefficients. We have already found that
$c_1 = -1$, $c_2= \frac13$, $c_3=-\frac29$. Now we are prepared to determine the form of the generic coefficient $c_n$.

Consider two adjacent terms in the series (\ref{L-series})
\be
c_{n-1} \left(
 \frac{q^{+a}q^-_a}{\bar WW}
 \right)^{n-1}
+c_n \left(
 \frac{q^{+a}q^-_a}{\bar WW}
 \right)^n
 \label{4.85}
\ee
and assume that the variation of the numerator of the first term has already been used to cancel the remaining part of
the variation of preceding term under the full superspace integral. Then we rewrite the rest of the full variation of the first
term using the same manipulations as in (\ref{4.81}) and require that this part should be canceled by the variation
of the numerator of the second term in (\ref{4.85}). This gives rise to the following recursive relation between the coefficients $c_{n-1}$ and $c_n$:
\be
c_n = -2 \frac{(n-1)^2}{n(n+1)} c_{n-1}\,.
\ee
Taking into account that $c_1 = -1$, we find the value of the generic coefficient
\be
c_n = \frac{(-2)^n}{n^2(n+1)}\,.
\ee

As a result, we find the full hypermultiplet completion of the non-holomorphic potential in the form
\bea
{\cal L}(W,\bar W,q^\pm_a)
\equiv {\cal L}(Z)&=&
c\sum_{n=1}^\infty
\frac1{n^2(n+1)} Z^n
\nn\\&=&c\left[
(Z-1)\frac{\ln(1-Z)}{Z}
+{\rm Li}_2(Z)-1
\right],
\label{L-solution}
\eea
where
\be
Z= -2\frac{q^{+a}q^-_a}{\bar WW}\,.
\label{X}
\ee
Here ${\rm Li}_2(Z)$ is the Euler dilogarithm which is represented by the power series expansion ${\rm Li}_2(Z) = \sum_{n=1}^\infty \frac1{n^2} Z^n\,$.

It is worth to note that the expression (\ref{X}) is harmonic-independent for the on-shell hypermultiplets
which are linear in harmonics, $q^\pm_a =u^\pm_i q^i_a$. Indeed, (\ref{X}) can be identically rewritten as
\be
Z= -\frac{q^{ia} q_{ia}}{ \bar WW}\,.
\ee
As a consequence, the effective Lagrangian (\ref{L-solution}) is harmonic-independent and one can omit the integration
over the harmonics in (\ref{Gamma-full}). Taking this into account, we rewrite the final answer for the four-derivative
part of the $\cN=4$ SYM low-energy effective action in the $\cN=2$ superspace as
\be
\Gamma = \int d^{12}z \left[
c\ln \frac W\Lambda \ln\frac{\bar W}{\Lambda}
+{\cal L}\left(-\frac{q^{ia} q_{ia}}{ \bar WW}\right)
\right],\qquad
{\cal L}(Z) = c\sum_{n=1}^\infty
\frac{Z^n}{n^2(n+1)}
\,.
\label{Gamma-N2-final}
\ee
The $\cN=4$ SYM low-energy effective action in this form was first obtained in the paper \cite{BuIv}, using
the procedure described in this section. In the subsequent papers \cite{BIP,BBP,BP2005}, the expression
(\ref{Gamma-N2-final}) was reproduced by direct calculations within the quantum perturbative theory in $\cN=2$ harmonic superspace.

It should be noted that the low-energy effective action (\ref{Gamma-N2-final}) is scale invariant.
It is possible to show that it respects also the $SU(2,2|2)$ superconformal symmetry realized
on the superfields $W$, $\bar W$ and $q^\pm_a$. The on-shell closure of this symmetry with the hidden ${\cal N}=2$ supersymmetry is
just the superconformal $PSU(2,2|4)$ symmetry. To avoid a possible confusion, we would like also to point out that the expression
(\ref{Gamma-N2-final}) with $Z$ (\ref{X}) as the argument in ${\cal L}$ (and with an integral over harmonics restored)
is an off-shell invariant of the manifest ${\cal N}=2$ supersymmetry.  The on-shell conditions need to be imposed only
when we prove the hidden second on-shell ${\cal N}=2$ supersymmetry of this ${\cal N}=2$ superfield expression.


\subsection{Component structure}

The abelian $\cN=2$ on-shell vector multiplet consists of one complex scalar $\phi$, $SU(2)$ doublet of spinors $\lambda^i_\alpha$
and a gauge vector $A_m$ with the Maxwell field strength $F_{mn}=\partial_m A_n - \partial_n A_m$. The on-shell hypermultiplet contains
$SU(2)$ doublet of complex scalars $f^i$ and two spinors $\psi_\alpha$, $\bar\chi_{\dot\alpha}$. We adopt
the following two essential simplifications, while considering the component structure of the effective action:
(i) we discard all spinor and auxiliary fields and (ii) we assume that the bosonic fields obey free classical equations of motion.
 Though these constraints are very strong, they suffice to determine the bosonic core of the low-energy effective action
 which is non-vanishing on the mass shell. Taking these constraints into account, we find the component structure of the superfields
 $W$, $\bar W$ and $q^+$, $\tilde q^+$ in the form
\bea
\label{W-comp}
W&=&i\sqrt2\phi-2\sqrt2\theta^-\sigma^m\bar\theta^+\partial_m \phi
-\theta^+_\alpha \theta^-_\beta
\sigma^{m\alpha}{}_{\dot\alpha}\sigma^{n\beta\dot\alpha}F_{mn}\,,
\nn\\
\bar{W}&=&-i\sqrt2\bar\phi+2\sqrt2\theta^+\sigma^m\bar\theta^-\partial_m \bar\phi
-\bar\theta^-_{\dot\beta}\bar\theta^+_{\dot\alpha}
\sigma^{m\dot\alpha}{}_{\alpha}\sigma^{n\alpha\dot\beta}F_{mn}\,,
\eea
and
\bea
\label{q-comp}
q^{+} &=& f^{i}u^+_i+2i\theta^+\sigma^m\bar\theta^+\partial_m f^{i}u^-_i \,,\nn\\[5pt]
\tilde q^{+} &=& -\bar f^{i}u^+_i-2i\theta^+\sigma^m\bar\theta^+\partial_m \bar f^{i}u^-_i \,.
\eea
The component fields in these expressions were normalized in agreement with the notations of \cite{GIOS-book}.

\subsubsection{$F^4/X^4$ term}
To derive the $F^4/X^4$ term in the $\cN=4$ SYM effective action, it is sufficient to consider a constant Maxwell field strength $F_{mn}$
 and discard all derivatives of the scalars. Then, we substitute (\ref{W-comp}) and (\ref{q-comp}) into (\ref{Gamma-N2-final}) and
 integrate over all Grassmann coordinates according to the rules (\ref{int-chiral-rule})
\bea
\Gamma_{F^4/X^4} &=& \frac{c}{4}\int d^4x
\frac{F_{m n}F^{n k}F_{k l}F^{l m}-\frac14(F_{p q}F^{p q})}{\phi^2\bar\phi^2}\sum_{n=0}^\infty
(n+1)\left(\frac{-f^i\bar f_i}{\phi\bar\phi}\right)^n \nn\\
&=& \frac{c}{4}\int d^4x
\frac{F_{m n}F^{n k}F_{k l}F^{l m}-\frac14(F_{p q}F^{p q})}{(\phi\bar\phi+f^i\bar f_i)^2} \,.
\label{Gamma-F4}
\eea
Here we used the identity for $\sigma$-matrices
\be
\tr\tilde\sigma^m\sigma^n\tilde\sigma^p
\sigma^q=-2i\varepsilon^{mnpq}
+2(\eta^{mn}\eta^{pq}+\eta^{np}\eta^{mq}-\eta^{mp}\eta^{nq})
\,,\qquad
\varepsilon^{0123}=1\,.
\label{trace-4sigma}
\ee
Now it remains to express the complex scalars $f^i$ and $\phi$ via the six real scalars $X_A$, $A=1,\ldots,6$,
\be
\label{fphiX}
f^1=X_1+i X_2\,, \quad f^2=X_3+i X_4\,, \quad \phi=X_6+i X_5
\,.
\ee
Then, with $c$ given in (\ref{c}), the considered part of the low-energy effective action  takes exactly the form
of the $F^4/X^4$ term (\ref{F4X4})
\be
\label{F4X4_}
\Gamma_{F^4/X^4}=\frac{1}{(8\pi)^2}\int d^4x \frac{1}{(X_A X_A)^2}\Big[F_{m n}F^{n k}F_{k l}F^{l m}-\frac14(F_{p q}F^{p q})^2\Big].
\ee

\subsubsection{Wess-Zumino term}
In order to single out the Wess-Zumino term in the component structure of the low-energy effective action (\ref{Gamma-N2-final}),
it is sufficient to consider another approximation: We discard the Maxwell field $F_{mn}$,
but keep the space-time derivatives of the scalars.

First of all, we point out that the non-holomorphic potential $\ln \frac W\Lambda
\ln\frac{\bar W}{\Lambda}$ cannot make a contribution to the Wess-Zumino term because it involves only
two out of six scalar fields. Thus we have to consider only that part of the effective action (\ref{Gamma-N2-final}) which
is described by the function ${\cal L}\,$,
\be
\Gamma_{\rm WZ} = \int d^4x d^8\theta \,{\cal L}(W,\bar W,q^\pm_a)\,.
\label{100}
\ee
Here we assume that the superfields contain only scalar fields in their component field expansion.

For deriving the Wess-Zumino term we will use the rule of integration over the Grassmann variables which
is equivalent to (\ref{int-chiral-rule})
\be
\int d^8\theta\, {\cal L}=\bar{D}^4 D^4 {\cal L}|_{\theta=0}\,, \quad
\bar{D}^4 D^4=\frac{1}{2^8}
\bar D^+_{\dot\alpha}\bar D^{+\dot\alpha}
\bar D^-_{\dot\beta}\bar D^{-\dot\beta}
D^{+\alpha}D^+_\alpha D^{-\beta}D^-_\beta \,.
\label{B18}
\ee
Thus we have to hit the function $\cal L$ by eight covariant spinor derivatives. While doing so, we should take into account that
for the superfields $W$, $\bar W$ and $q^{\pm}_a$ obeying the on-shell constraints (\ref{q-analyt1})--(\ref{anlyt-q-})
a lot of identities can be derived, e.g.,
\bea
&&(D^-)^2 q^+_a=(\bar D^-)^2 q^+_a=0\,,\quad
(D^+)^2 q^-_a=(\bar D^+)^2 q^-_a=0\,,\nn\\&&
(D^+)^2{W}=(D^-)^2{W}=D^{+\alpha}D^-_\alpha {W}=0\,,\nn\\&&
(\bar D^+)^2\bar{W}=(\bar D^-)^2\bar {W}=\bar D^{+\dot\alpha}\bar D^-_{\dot\alpha}
\bar {W}=0\,,
\label{note2}
\eea
and
\bea
2i\partial_{\alpha\dot\alpha}q^+_a&=&\bar
D^+_{\dot\alpha}D^-_\alpha q^+_a=-D^+_\alpha \bar
D^-_{\dot\alpha}q^+_a=D^+_\alpha\bar D^+_{\dot\alpha}q^-_a
=-\bar D^+_{\dot\alpha}D^+_\alpha q^-_a\,,\nn\\
2i\partial_{\alpha\dot\alpha}q^-_a&=&
D^-_\alpha \bar D^+_{\dot\alpha}q^-_a=-\bar
D^-_{\dot\alpha}D^+_\alpha q^-_a
=\bar D^-_{\dot\alpha}D^-_\alpha q^+_a=-D^-_\alpha\bar
D^-_{\dot\alpha}q^+_a\,,\nn\\
2i\partial_{\alpha\dot\alpha}{W}&=&-\bar D^-_{\dot\alpha}D^+_\alpha{W}
=\bar D^+_{\dot\alpha} D^-_\alpha{W}\,.
\label{l128}
\eea
Using these identities, we find
\bea
\bar{D}^4 D^4 {\cal L}(W,\bar{W},q_a^{\pm})
&=&-\frac{\partial^4{{\cal L}}}{\partial q^+_a \partial q^+_b \partial q^-_c \partial q^-_d}
\partial^{\alpha\dot\beta} q^-_d \partial_{\dot\alpha\alpha} q^+_c
\partial^{\beta\dot\alpha}q^+_b
\partial_{\beta\dot\beta}q^-_a
\nn\\&&
-\frac{\partial^4{{\cal L}}}{\partial{W}\partial q^+_a\partial q^+_b\partial q^-_c}
\partial^{\alpha\dot\beta}{W}\partial_{\alpha\dot\alpha}
q^+_c \partial^{\beta\dot\alpha}q^+_b
\partial_{\beta\dot\beta}q^-_a
\nn\\&&
-\frac{\partial^4{{\cal L}}}{\partial{W}\partial q^+_a\partial q^-_c \partial q^-_d}
\partial^{\alpha\dot\beta}q^-_d \partial_{\alpha\dot\alpha}q^+_c
\partial^{\beta\dot\alpha}{W}
\partial_{\beta\dot\beta}q^-_a+\ldots
\,. ~~~~~~~
\label{l130}
\eea
Here, we have explicitly written  only terms with cyclic contraction of the spinor indices
of the space-time derivatives, since only such expressions can produce, by the identity (\ref{trace-4sigma}), the antisymmetric $\varepsilon$-tensor.
Now we set to zero the Grassmann variables in (\ref{l130}) and obtain the following representation for (\ref{100})
\bea
\Gamma_{\rm WZ} &=& 2i \,\varepsilon^{mnpq}\int d^4x du\bigg[
\frac{\partial^4{{\cal L}}(z)}{\partial f^+_a \partial f^+_b \partial f^-_c \partial f^-_d}
\partial_m f^-_d \partial_n f^+_c
\partial_p f^+_b
\partial_q f^-_a
\nn\\
&&\hspace{-50pt}
+\frac{\partial^4{{\cal L}}(z)}{\partial\phi\partial f^+_a\partial f^+_b\partial f^-_c}
\partial_m \phi
\partial_n f^+_c \partial_p f^+_b
\partial_q f^-_a
+\frac{\partial^4{{\cal L}}(z)}{\partial\phi\partial f^+_a\partial f^-_c \partial f^-_d}
\partial_m f^-_d \partial_n f^+_c
\partial_p \phi
\partial_q f^-_a\bigg], \quad
\label{l135}
\eea
where
\be
z = Z|_{\theta=0}=-\frac{f^{+a}f^-_a}{\bar \phi\phi}
=-\frac{f^i\bar f_i}{\bar \phi\phi}\,,
\ee
and
\be
{\cal L}(z) = c\sum_{n=1}^\infty \frac{z^n}{n^2(n+1)}\,.
\label{Lz}
\ee

The expression (\ref{l135}) is not manifestly real. However, its imaginary part can be shown to be a total $x$-derivative and so
vanishes under the space-time integral. Applying the integration by parts, the remaining real part can be represented in the form:
\bea
\Gamma_{\rm WZ} &=& i\,\varepsilon^{mnpq}\int d^4x\, \left(\frac{\partial_m\phi}{\phi}-\frac{\partial_m\bar\phi}{\bar\phi}\right)
\bigg\{
\partial_q f^i_a \partial_n f_i^c \partial_p f^j_c f_j^a
\frac{2{{\cal L}}^{(2)}+z{{\cal L}}^{(3)}}{(\phi\bar\phi)^2}\nn\\
&&\hspace{-35pt}
-\left(\frac1{12}\partial_n f^i_c f^c_k \partial_q f^k_a f^a_j
 \partial_p f^j_b f^b_i
 +\frac18 f^{ak}f_{ak}\partial_n f^i_c \partial_q f_j^c
  \partial_p f^j_b f_i^b\right)\frac{3{{\cal L}}^{(3)}+z{{
  \cal L}}^{(4)}}{(\phi\bar\phi)^3}
 \bigg\}\,.~~~~~~~~
 \label{l142}
\eea
Here we have also expressed the partial derivatives of
${\cal L}$ in terms of usual derivatives ${\cal L}^{(n)}={d^n {\cal L}(z)}/{d z^n}$. With $f^{i}_a=(f^i,\bar f^i)$ and $f_i^a=(-\bar f_i,f_i)$,
we then obtain
\be
\Gamma_{\rm WZ}=i \,\varepsilon^{mnpq}\int d^4x\,
\Big[6{{\cal L}}^{(2)}+6z{{\cal L}}^{(3)}+z^2{{\cal L}}^{(4)}\Big]
\frac{\partial_n f^i\partial_p \bar f_i(
  \partial_q f^j \bar f_j -\partial_q \bar f_j f^j)}{(\phi\bar\phi)^2}\,
\partial_m\ln\frac{\phi}{\bar\phi}\,.
\label{l144}
\ee
Using (\ref{fphiX}) and performing the polar decomposition of $\phi$,
\bea
\phi=X_6+i X_5= X e^{i\alpha} \,,
\eea
we find
\be
\Gamma_{\rm WZ}=-\frac{4 }3\varepsilon^{mnpq}\varepsilon^{a'b'c'd'}
\int d^4x\, \Big[6{\cal L}^{(2)}+6z{\cal L}^{(3)}+z^2{\cal L}^{(4)}\Big]
\frac{X_{a'} \partial_n X_{b'} \partial_p X_{c'} \partial_q X_{d'}}{X^4}\partial_m\alpha
\,,
\label{l150}
\ee
where $a',b'=1,2,3,4$ are $SO(4)$ indices
and $\varepsilon^{1234}=1$.
Finally, we observe that the function (\ref{Lz}) obeys the equation
\be
6{{\cal L}}^{(2)}(z)+6z{{\cal L}}^{(3)}(z)+z^2{{\cal L}}^{(4)}(z)=\frac c{(z-1)^2}
=c\frac{X^4}{(X_{e'} X_{e'}+X^2)^2}\,.
\label{difur}
\ee
After substituting this into the expression (\ref{l150}), the latter becomes
\be
\label{WZfromN2_}
\Gamma_{\rm WZ}=\frac43 c\,\varepsilon^{mnpq}\varepsilon^{a'b'c'd'}
\int d^4x\,
\frac{X_{a'} \partial_m X_{b'} \partial_n X_{c'} \partial_p X_{d'}
}{(X_{e'} X_{e'}+X^2)^2}\partial_q \alpha\,.
\ee
With $c$ defined in (\ref{c}), it perfectly  matches the expression (\ref{WZso4so2}).

The Wess-Zumino term (\ref{WZfromN2_}) in the component field formulation of the $\cN=4$ SYM low-energy effective action (\ref{Gamma-N2-final})
was found for the first time in \cite{BelSam1}, although attempts to derive this term were undertaken in the preceding papers \cite{Argyres,BP2006}.

As we have shown in sect.\ \ref{Sect3.3}, the Wess-Zumino term in the form (\ref{WZfromN2_})
has a manifest symmetry under the group $SO(4)\times SO(2)$
which, in the considered setting,  is locally isomorphic to $SU(2)_R\times SU(2)_{PG}\times U(1)$. Here, the group $SU(2)_R$ corresponds
to the R-symmetry  of the $\cN=2$ superspace,
while  $SU(2)_{PG}$ is the Pauli-G\"ursey group which acts on the index $a$ of the hypermultiplet $q^+_a$ in (\ref{S-N4-ab}).
The last $U(1)$ factor
is the phase rotation of the $\cN=2$ superfield strengths $W$ and $\bar W$ in (\ref{S-N4-ab}). Thus it is absolutely natural
that the Wess-Zumino term in the $\cN=4$ SYM low-energy effective action appears in the $\cN=2$ harmonic superspace approach just
in the form (\ref{WZfromN2_}) with manifest $SO(4)\times SO(2)$ symmetry.

\section[Low-energy effective action in $\cN=3$ harmonic
superspace]{Low-energy effective action in $\cN=3$ harmonic \break
superspace}
\label{SectN3}

Classical action of $\cN=3$ SYM theory in harmonic superspace was constructed in the pioneering papers \cite{GIKOS-N3a,GIKOS-N3}.
On the mass shell, this theory is known to be equivalent to $\cN=4$ SYM \cite{GIOS-book}. Since no $\cN=4$ off-shell superfield description for
$\cN=4$ SYM theory is known so far, the $\cN=3$ harmonic superspace provides the maximal number of manifest supersymmetries. As a consequence,
it appears very efficient at quantum level. For instance, the quantum finiteness of $\cN=3$ SYM theory can be easily proved just by analyzing
the dimension of the propagator for gauge superfield in the $\cN=3$ harmonic superspace \cite{Delduc}. What is more important for the present
consideration, $\cN=3$ supersymmetry, combined with the requirement of scale invariance, prove to be so strong that these symmetries fix uniquely,
up to an overall coefficient,
the leading part of the $\cN=3$ SYM low-energy effective action \cite{BISZ}. In the present section, we explicitly construct this effective action,
reviewing the results of \cite{BISZ}.

To make our consideration more pedagogical, we start by explaining basics of the $\cN=3$ harmonic superspace and gauge theory in it.
The detailed exposition of $\cN=3$ SYM theory is given in the book \cite{GIOS-book}.

\subsection{$\cN=3$ harmonic superspace setup}

The standard $\cN=3$ superspace is parametrized by the coordinates (\ref{zM}), where the indices $i,j=1,2,3$ correspond
now to the $SU(3)$ R-symmetry group.
The covariant spinor derivatives $D^i_\alpha$ and $\bar D_{i\dot\alpha}$ in this superspace have the same form as in (\ref{D-explicit})
and obey the anti-commutation relations (\ref{D-algebra}). We extend this superspace by the harmonic variables $u^I_i=(u^1_i,u^2_i,u^3_i)$
and their conjugates, $\bar u_I^i=(\bar u_1^i, \bar u_2^i,\bar u_3^i)$, which obey the following defining properties
\be
u^I_i \bar u^i_J=\delta^I_J\,,\quad
u^I_i \bar u^j_I=\delta_i^j\,,\quad
\varepsilon^{ijk}u^1_i u^2_j u^3_k=1\,.
\label{N3harmonics}
\ee
These properties show that the harmonics $u^I_i$, $\bar{u}^j_J$ form the $SU(3)$ matrices in the fundamental and co-fundamental
representations.

The eight independent harmonic derivatives on $SU(3)$ are defined as the differential operators
\be
\partial^I_J = u^I_i \frac\partial{\partial u^J_i} -
 \bar u_J^i \frac\partial{\partial \bar u_I^i}\,,
\label{u-partial}
\ee
which can be interpreted as the generators of the right $SU(3)$ shifts of $(u^I_i, \bar u^j_J)$.\footnote{The generators of the left shifts are
$\partial^i_j = \bar u^i_J \frac\partial{\partial \bar u_J^j} -
 u^I_j \frac\partial{\partial u^I_i}$ and they produce the standard $SU(3)$ rotations of the triplet indices $i, j$ of the harmonic variables\,.}
Correspondingly, they are subject to the commutation relations of the $SU(3)$ algebra
\be
[\partial^I_J,\partial^K_L] = \delta^K_J \partial^I_L -
\delta^I_L \partial^K_J\,.
\label{partial-algebra}
\ee
The more convenient notation for the covariant derivatives is as follows
\begin{subequations}
\label{harm-deriv}
\bea
D^I_J &=& \partial^I_J \qquad \mbox{for} \quad I\ne J\,,
\label{harm-deriv-a}\\
S_1 &=& \partial^1_1 - \partial^2_2 \,,\qquad
S_2 = \partial^2_2 - \partial^3_3\,.
\eea
\end{subequations}
The operators $S_1$ and $S_2$ are two independent mutually commuting $U(1)$ charge operators. In this notation,
the non-zero commutation relations in (\ref{partial-algebra})
are rewritten as
\begin{subequations}
\label{5.5}
\begin{align}
&[D^1_2, D^2_3] = D^1_3\,,&&
[D^1_3,D^3_2] = D^1_2\,,&&
[D^2_1,D^1_3] = D^2_3\,,\\
&{}[S_1,D^1_3] = D^1_3\,,&&
[S_1,D^1_2] = 2D^1_2\,,&&
[S_1,D^2_3] = -D^2_3\,,\\
&{}[S_2,D^1_3]=D^1_3\,,&&
[S_2,D^1_2]=-D^1_2\,,&&
[S_2,D^2_3]= 2D^2_3\,,\\
&{}[D^1_2,D^2_1] = S_1\,,&&
[D^2_3,D^3_2]=S_2\,,&&
[D^1_3,D^3_1]=S_1+S_2\,.
\label{5.5d}
\end{align}
\end{subequations}

By analogy with the ${\cal N}=2$ harmonic superspace, in the $\cN=3$ harmonic superspace we will consider only those superfields
which possess definite $U(1)$ charges $(q_1,q_2)$ with respect to the operators $S_1$ and $S_2$:
\be
S_1 \Phi^{(q_1,q_2)}(z,u) = q_1 \Phi^{(q_1,q_2)}(z,u)\,,\qquad
S_2 \Phi^{(q_1,q_2)}(z,u) = q_2 \Phi^{(q_1,q_2)}(z,u)\,.
\ee
These equations effectively restrict the harmonic dependence
of the fields originally defined  on the full $SU(3)$ group manifold to the coset
$SU(3)/[U(1)\times U(1)]$. We will assume that the superfields are smooth function on this coset, such that they can always be represented
by power series expansions over the harmonic variables.

The defining constraints (\ref{N3harmonics}) can be viewed as the orthogonality and completeness relations for the harmonic variables.
They allow one to form the harmonic projections of any objects with $SU(3)$ indices just by contracting the latter with
the complementary  $SU(3)$ indices of the harmonics.
For instance, for the Grassmann coordinates and covariant spinor derivatives we have
\begin{align}
\theta_i^\alpha &\longrightarrow \theta_I^\alpha = \theta_i^\alpha \bar u_I^i\,,
&
\bar\theta^{i\dot\alpha}&\longrightarrow \bar\theta^{I\dot\alpha} =
 \bar\theta^{i\dot\alpha} u^I_i\,,\\
D^i_\alpha & \longrightarrow D^I_\alpha = D^i_\alpha u^I_i\,,&
\bar D_{i\dot\alpha} & \longrightarrow \bar D_{I\dot\alpha} =
\bar D_{i\dot\alpha} \bar u_I^i\,.
\label{D-I}
\end{align}
The covariant spinor derivatives (\ref{D-I}) obey the following anti-commutation relations
\be
\{ D^I_\alpha ,\bar D_{J\dot\alpha} \} = -2i\delta^I_J \sigma^m_{\alpha\dot\alpha}
\partial_m\,,\quad
\{ D^I_\alpha,D^J_\beta \} = \{ \bar D_{I\dot\alpha}, \bar D_{J\dot\beta} \} =0\,.
\label{D-I-N3}
\ee

The full $\cN=3$ harmonic superspace with the coordinates
$(x^m,\theta_I^\alpha,\bar\theta^{I\dot\alpha},u)$ contains the {\it analytic subspace} parametrized by the coordinates
\be
\{\zeta_A, u \}=\{x_A^{m},\theta_2^\alpha,\theta_3^\alpha,
\bar\theta^{1\dot\alpha},\bar\theta^{2\dot\alpha},u \}\,,\qquad
x_A^{m}=x^m-i\theta_1\sigma^m\bar\theta^1+i\theta_3\sigma^m\bar\theta^3\,.
\label{anal-coord}
\ee
It is straightforward to show that this subspace is closed under $\cN=3$ supersymmetry, by analogy with the $\cN=2$ analytic
subspace (\ref{analyt-coord}).

The basis $\{\zeta_A, u, \theta^\alpha_1, \bar\theta^{3 \dot\alpha}\}$ of the full ${\cal N}=3$ harmonic superspace
is called {\it analytic basis}.
The covariant spinor derivatives $D^I_\alpha$ and $\bar D_{I\dot\alpha}$ in this basis acquire the form
\begin{align}
D^1_\alpha&=\frac\partial{\partial\theta_1^\alpha}\,,&
\bar D_{1\dot\alpha}&=-\frac\partial{\partial\bar\theta^{1\dot\alpha}}-
 2i\theta_1^\alpha \sigma^m_{\alpha\dot\alpha}\frac\partial{\partial x_A^{m}}\,,\nn\\
D^2_\alpha&=\frac\partial{\partial\theta_2^\alpha}+
 i\bar\theta^{2\dot\alpha}\sigma^m_{\alpha\dot\alpha}\frac\partial{\partial x_A^{m}}\,,
 &
\bar D_{2\dot\alpha}&=-\frac\partial{\partial\bar\theta^{2\dot\alpha}}-
 i\theta_2^\alpha \sigma^m_{\alpha\dot\alpha} \frac\partial{\partial x_A^{m}}\,,\nn\\
D^3_\alpha&=\frac\partial{\partial\theta_3^\alpha}+
 2i\bar\theta^{3\dot\alpha}\sigma^m_{\alpha\dot\alpha}\frac\partial{\partial x_A^{m}}\,,
 &
\bar D_{3\dot\alpha}&=-\frac\partial{\partial\bar\theta^{3\dot\alpha}}\,.
\label{cov-spinor-N3}
\end{align}
We observe that the anticommuting derivatives
$D^1_\alpha$ and $\bar D_{3\dot\alpha}$ become short. Hence, the analytic superfields (i.e.\ those living on
the analytic superspace (\ref{anal-coord}))
can be covariantly defined by the Grassmann Cauchy-Riemann conditions
\be
D^1_\alpha \Phi_A(z, u) =\bar  D_{3\dot\alpha} \Phi_A(z, u) = 0\quad
 \Rightarrow \quad\Phi_A(z, u) = \hat\Phi_A(\zeta_A, u)\,.
\label{analyt-N3}
\ee

The harmonic derivatives $D^1_2$, $D^2_3$ and $D^1_3$ in the analytic basis have the form
\bea
D^1_2&=&\partial^1_2+i\theta_2^\alpha\bar\theta^{1\dot\alpha}
 \sigma^m_{\alpha\dot\alpha}
 \frac\partial{\partial x_A^{m}}+\bar\theta^{1\dot\alpha}
  \frac\partial{\partial\bar\theta^{2\dot\alpha}}-
  \theta_2^\alpha\frac\partial{\partial\theta_1^\alpha}\,,\nn\\
D^2_3&=&\partial^2_3+i\theta_3^\alpha\bar\theta^{2\dot\alpha}
 \sigma^m_{\alpha\dot\alpha}
 \frac\partial{\partial x_A^{m}}+
  \bar\theta^{2\dot\alpha}\frac\partial{\partial\bar\theta^{3\dot\alpha}}-
  \theta_3^\alpha\frac\partial{\partial\theta_2^\alpha}\,,\nn\\
D^1_3&=&\partial^1_3+2i\theta_3^\alpha\bar\theta^{1\dot\alpha}
 \sigma^m_{\alpha\dot\alpha}
 \frac\partial{\partial x_A^{m}}+
 \bar\theta^{1\dot\alpha}\dfrac\partial{\partial\bar\theta^{3\dot\alpha}}-
 \theta_3^\alpha\frac\partial{\partial\theta_1^\alpha}\,.
\label{harm-deriv-AB}
\eea
One can check that they commute with the covariant spinor derivatives
$D^1_\alpha$ and $\bar D_{3\dot\alpha}$
\be
[D^1_2,D^1_\alpha] = [D^2_3,D^1_\alpha] = [D^1_3,D^1_\alpha]=0\,,\qquad
[D^1_2,\bar D_{3\dot\alpha}] = [D^2_3, \bar D_{3\dot\alpha}]
 = [D^1_3,\bar D_{3\dot\alpha}]=0\,,
\label{5.16}
\ee
and, hence, preserve the Grassmann harmonic analyticity. The other three harmonic derivatives,
\bea
D^2_1&=&\partial^2_1-i\theta_1^\alpha\bar\theta^{2\dot\alpha}\sigma^m_{\alpha\dot\alpha} \frac\partial{\partial x_A^{m}}
 +\bar\theta^{2\dot\alpha}
  \frac\partial{\partial\bar\theta^{1\dot\alpha}}-
  \theta_1^\alpha\frac\partial{\partial\theta_2^\alpha}\,,\nn\\
D^3_2&=&\partial^3_2-i\theta_2^\alpha\bar\theta^{3\dot\alpha}\sigma^m_{\alpha\dot\alpha} \frac\partial{\partial x_A^{m}}+
  \bar\theta^{3\dot\alpha}\frac\partial{\partial\bar\theta^{2\dot\alpha}}-
  \theta_2^\alpha\frac\partial{\partial\theta_3^\alpha}\,,\nn\\
D^3_1&=&\partial^3_1-2i\theta_1^\alpha\bar\theta^{3\dot\alpha}\sigma^m_{\alpha\dot\alpha} \frac\partial{\partial x_A^{m}}+
 \bar\theta^{3\dot\alpha}\frac\partial{\partial\bar\theta^{1\dot\alpha}}-
 \theta_1^\alpha\frac\partial{\partial\theta_3^\alpha},
\label{harm-deriv-AB_}
\eea
do not possess this property.

Like in the $\cN=2$ harmonic superspace, the conventional complex conjugation is not useful as it does not preserve the analyticity.
Therefore, it is customary to use the generalized complex conjugation denoted by $\,\widetilde{\phantom{m}}\,$ and defined by the following properties:
On the harmonic-independent objects it coincides with the usual complex conjugation, see eq.\ (\ref{4.17}), while on the harmonic variables
it acts according to the rules \footnote{Here we use the convention for the $\,\widetilde{\phantom{m}}\,$-conjugation adopted in \cite{N3-BI,BISZ} which
is somewhat different from the convention used in \cite{GIOS-book}.}
\be
u^1_i \stackrel{\sim}{\longleftrightarrow} \bar u_3^i \,,\quad
u^2_i \stackrel{\sim}{\longleftrightarrow} -\bar u_2^i \,,\quad
u^3_i \stackrel{\sim}{\longleftrightarrow} \bar u_1^i \,.
\ee
Using these rules, one can find the conjugation properties of the Grassmann variables,
\be
\theta_1^\alpha \stackrel{\sim}{\longleftrightarrow} \bar\theta^{3\dot\alpha}\,,\quad
\theta_2^\alpha \stackrel{\sim}{\longleftrightarrow} -\bar\theta^{2\dot\alpha}\,,\quad
\theta_3^\alpha \stackrel{\sim}{\longleftrightarrow} \bar\theta^{1\dot\alpha}\,,
\ee
as well as of the harmonic covariant derivatives (\ref{harm-deriv-AB}),
\be
\widetilde{D^1_3 f} = - D^1_3 \tilde f\,,\qquad
\widetilde{D^1_2 f} = D^2_3 \tilde f\,,
\label{5.20}
\ee
where $f$ is an arbitrary function depending on the superspace coordinates $(x^m,\theta_i^\alpha,\bar\theta^{i\dot\alpha})$ and harmonics $u$.

It is easy to see that the analytic subspace with the coordinates (\ref{anal-coord}) is closed under the $\,\widetilde{\phantom{m}}\,$-conjugation,
but not under the conventional complex conjugation.

\subsection{Gauge theory in $\cN=3$ harmonic superspace}

In this section we shortly review the superspace description of $\cN=3$ SYM theory.

The constraints of this theory  in the conventional $\cN=3$ superspace were introduced in \cite{Sohnius78},
while their harmonic superspace version was discussed in the book \cite{GIOS-book} (see also \cite{AFSZ}).
Here we limit our attention only to the {\it abelian} case,
which is sufficient for constructing the low-energy effective action in the Coulomb branch.

In the standard geometric approach, the gauge theory is introduced through adding gauge connections to the superspace derivatives,
as in eq.\ (\ref{A-connections}).
In the $\cN=3$ case, the analogs of the constraints (\ref{D-algebra-gauged}) read
\begin{subequations}
\label{N3constraints}
\bea
\{ {\cal D}^i_\alpha , {\cal D}^j_\beta \} &=& -2i\,\varepsilon_{\alpha\beta} \bar W^{ij}\,,\label{N3constraints-a}\\
\{ \bar{\cal D}_{i\dot\alpha} , \bar {\cal D}_{j\dot\beta} \} &=&2i\,
\varepsilon_{\dot\alpha\dot\beta} W_{ij}\,,\label{N3constraints-b}\\
\{ {\cal D}^i_\alpha , \bar{\cal D}_{j\dot\alpha}  \}&=&
-2i \delta^i_j {\cal D}_{\alpha\dot\alpha}\,,
\eea
\end{subequations}
where $W_{ij}=-W_{ji}$ and its conjugate $\bar W^{ij}=\overline{W_{ij}}$ are the superfield strengths for the $\cN=3$ gauge vector multiplet.
The constraints (\ref{N3constraints}) imply the following Bianchi identities for these superfield strengths \cite{Sohnius78}
\begin{subequations}
\label{Wconstr}
\bea
&&D^i_\alpha W_{jl}=\frac12(\delta^i_j D^k_\alpha W_{kl}-\delta^i_l D^k_\alpha W_{kj})
\,,\\
&&\bar D_{i\dot\alpha}W_{jk}+\bar D_{j\dot\alpha}W_{ik}=0\,.
\eea
\end{subequations}
It is known that these constraints kill all unphysical (auxiliary) components in the superfield strengths,
simultaneously yielding the free equations of motion for the physical components of the $\cN=3$ vector multiplet.

Let us introduce the harmonic projections of the superfield strengths
\begin{align}
\bar W^{12}&=u^1_i u^2_j\bar  W^{ij}\,,&
\bar W^{23}&=u^2_i u^3_j\bar  W^{ij}\,,&
\bar W^{13}&=u^1_i u^3_j\bar  W^{ij}\,,
\nn\\
W_{12}&=\bar u_1^i \bar u_2^j W_{ij}\,,&
W_{23}&=\bar u_2^i \bar u_3^j W_{ij}\,,&
W_{13}&=\bar u_1^i \bar u_3^j W_{ij}\,.
\label{W-projections}
\end{align}
For these superfields one can deduce many off- and on-shell constraints which follow from (\ref{Wconstr}). Here we will need
only the independent constraints
for the superfield strengths $\bar W^{12}$ and $W_{23}$. They can be grouped into the three sets:
\begin{itemize}
\item[(i)] Grassmann shortness constraints which originate from the harmonic projections of (\ref{Wconstr}):
\bea
&&
D^1_\alpha \bar W^{12}=D^2_\alpha \bar W^{12}=\bar D_{3\dot\alpha}
\bar W^{12}=0\,,\nn\\&&
D^1_\alpha W_{23}=\bar D_{2\dot\alpha}W_{23}=\bar
D_{3\dot\alpha}W_{23}=0\,;
\label{anal}
\eea
\item[(ii)] Grassmann linearity constraints which are also corollaries of (\ref{Wconstr}):
\bea
&&(D^3)^2 \bar W^{12}=(\bar D_1)^2 \bar W^{12}=(\bar D_2)^2
\bar W^{12}=(\bar D_1 \bar D_2)\bar W^{12}=0\,,\nn\\&&
(D^2)^2W_{23}=(D^3)^2 W_{23}=(D^2 D^3)W_{23}=(\bar D_1)^2
W_{23}=0\,;
\label{linearity}
\eea
\item[(iii)] Harmonic shortness constraints which are direct consequences of the definitions (\ref{W-projections}) and
the form of the harmonic derivatives (\ref{harm-deriv-a}):
\bea
&&
D^2_1\bar W^{12} =
D^1_2 \bar W^{12}=D^2_3 \bar W^{12}=D^1_3 \bar W^{12}=0\,,\nn\\&&
D^1_2 W_{23}=D^2_3 W_{23}=D^1_3 W_{23}=D^3_2 W_{23}=0\,.
\label{harm-short}
\eea
\end{itemize}
The general solution of the equations (\ref{anal})--(\ref{harm-short}) is given by the following
$\theta$-expansions of $\bar W^{12}$ and $W_{23}$ written in the analytic basis
\bea
\label{W-comp-N3}
W_{23}&=&\varphi^1
+i\theta_2^\alpha\bar\theta^{2\dot\alpha}\partial_{\alpha\dot\alpha}\varphi^1
-2i\theta_2^\alpha\bar\theta^{1\dot\alpha}\partial_{\alpha\dot\alpha}\varphi^2
-2i\theta_3^\alpha\bar\theta^{1\dot\alpha}\partial_{\alpha\dot\alpha}\varphi^3
\nn\\&&
+\,4i\theta_2^\alpha\theta_3^\beta F_{\alpha\beta}
+\bar\theta^{1\dot\alpha}\bar\lambda_{\dot\alpha}
+\theta_2^\alpha\lambda_{3\alpha}-\theta_3^\alpha\lambda_{2\alpha}
\nn\\&&
+\,i\theta_2^\alpha\bar\theta^{2\dot\alpha}\bar\theta^{1\dot\beta}
 \partial_{\alpha\dot\alpha}\bar\lambda_{\dot\beta}
+i\theta_2^\beta\theta_3^\alpha \bar\theta^{2\dot\alpha}
 \partial_{\alpha\dot\alpha}\lambda_{2\beta}
+2i\theta_2^\beta\theta_3^\alpha\bar\theta^{1\dot\alpha}
 \partial_{\alpha\dot\alpha}\lambda_{1\beta}
 \nn\\&&
+\,2\theta_2^\alpha\theta_3^\beta\bar\theta^{1\dot\alpha}\bar\theta^{2\dot\beta}
\partial_{\alpha\dot\alpha}\partial_{\beta\dot\beta}\varphi^3\,,
\nn\\
\bar W^{12}&=&\bar\varphi_3
-i\theta_2^\alpha\bar\theta^{2\dot\alpha}\partial_{\alpha\dot\alpha}\bar\varphi_3
+2i\theta_3^\alpha\bar\theta^{1\dot\alpha}\partial_{\alpha\dot\alpha}\bar\varphi_1
+2i\theta_3^\alpha\bar\theta^{2\dot\alpha}\partial_{\alpha\dot\alpha}\bar\varphi_2
\nn\\&&
+\,4i\bar\theta^{1\dot\alpha}\bar\theta^{2\dot\beta}\bar F_{\dot\alpha\dot\beta}
+\theta_3^\alpha\lambda_\alpha
-\bar\theta^{2\dot\alpha}\bar\lambda^1_{\dot\alpha}
+\bar\theta^{1\dot\alpha}\bar\lambda^2_{\dot\alpha}
\nn\\&&
+\,i\theta_2^\alpha\theta_3^\beta\bar\theta^{2\dot\alpha}
 \partial_{\alpha\dot\alpha}\lambda_\beta
+i\bar\theta^{1\dot\alpha}\bar\theta^{2\dot\beta}\theta_2^\alpha
 \partial_{\alpha\dot\alpha}\bar\lambda^2_{\dot\beta}
+2i\bar\theta^{1\dot\alpha}\bar\theta^{2\dot\beta}\theta_3^\alpha
 \partial_{\alpha\dot\alpha}\bar\lambda^3_{\dot\beta}
\nn\\&&
+\,2\bar\theta^{1\dot\alpha}\bar\theta^{2\dot\beta}
  \theta_2^\alpha\theta_3^\beta
  \partial_{\alpha\dot\alpha}\partial_{\beta\dot\beta}\bar\varphi_1
  \,.
  \label{W-N3}
\eea
Here
\be
\varphi^I=u^I_i \varphi^i\,,\qquad
\bar\varphi_I=\bar u_I^i \bar\varphi_i\,,
\ee
and $\varphi^i$ is a triplet of physical scalar fields subject to the Klein-Gordon equation $\square\varphi^i=0\,$.
The four spinor fields are accommodated by the $SU(3)$ singlet
$\lambda_\alpha$ and the triplet $\lambda_{I\alpha}=\bar
u_I^i\lambda_{i\alpha}\,$, all satisfying the free equations of motion,
$\partial^{\alpha\dot\alpha}\lambda_\alpha=\partial^{\alpha\dot\alpha}\lambda_{i\alpha}=0$.
The fields $F_{\alpha\beta}=F_{(\alpha\beta)}$ and
$\bar F_{\dot\alpha\dot\beta}=\bar F_{(\dot\alpha\dot\beta)}$ are
spinorial components of the Maxwell field strength $F_{mn}=\partial_m A_n-\partial_n
A_m$, $\partial^m F_{mn}=0$.

Similarly to (\ref{D-I}), the gauge-covariant spinor derivatives have harmonic projections ${\cal D}^I_\alpha = {\cal D}^i_\alpha u^I_i$
and $\bar{\cal D}_{I\dot\alpha} = \bar{\cal D}_{i\dot\alpha} \bar u_I^i$. As follows from (\ref{N3constraints}),
the derivatives ${\cal D}^1_\alpha$ and $\bar{\cal D}_{3\dot\alpha}$ form the set of anticommuting operators
\be
\{ {\cal D}^1_\alpha , {\cal D}^1_\beta \} =0\,,\quad
\{ \bar{\cal D}_{3\dot\alpha} , \bar{\cal D}_{3\dot\beta} \} =0\,,\quad
\{ {\cal D}^1_\alpha , \bar{\cal D}_{3\dot\alpha}\} =0\,.\label{N3IntegrCon}
\ee
These relations are just the integrability conditions for the  existence of the covariantly analytic superfields defined by
\be
{\cal D}^1_\alpha \Phi =0\,,\qquad
\bar{\cal D}_{3\dot\alpha } \Phi =0
\label{cov-analyt-N3}\,.
\ee
The explicit solution to these constraints can be found using the bridge superfield $b=b(z,u)$ which
solves the integrability conditions (\ref{N3IntegrCon}):
\be
{\cal D}^1_\alpha = e^{-i b}D^1_\alpha e^{ib}\,,\qquad
\bar{\cal D}_{3\dot\alpha} = e^{-i b}\bar D_{3\dot\alpha} e^{ib}\,.
\label{calDD-N3}
\ee
Without loss of generality, the bridge superfield  can be chosen real, $\tilde b(z,u)=b(z,u)$.
Like in $\cN=2$ SYM theory, $b(z,u)$ in (\ref{calDD-N3}) is defined modulo the gauge transformations
\be
e^{ib'} = e^{i\lambda}e^{ib}e^{-i\tau}\,,
\label{gauge-tr-bridge-N3}
\ee
where $\tau=\tau(z)$ is an arbitrary real harmonic-independent superfield, while $\lambda=\lambda(z,u)$ is an arbitrary tilde-real
and analytic superfield,
$\tilde \lambda = \lambda$, $D^1_\alpha\lambda = \bar D_{3\dot\alpha}\lambda =0\,$. Using (\ref{calDD-N3}),
the general solution to (\ref{cov-analyt-N3})
can be written as
\be
\Phi(z,u)= e^{-ib} \Phi_A(z,u)\,,
\ee
where $\Phi_A(z,u)$ is the manifestly analytic ${\cal N}=3$ superfield (\ref{analyt-N3}).

Thus, the introduction of the bridge superfield allows one to bring all the differential operators and superfields to the
$\lambda$-representation, in which the covariantly
analytic superfields become manifestly analytic and the covariant spinor derivatives $D^1_\alpha$ and $\bar D_{3\dot\alpha}$
cease to contain the gauge connections.

On the contrary, the harmonic derivatives (\ref{harm-deriv-AB}) and (\ref{harm-deriv-AB_}) acquire  gauge connections in the $\lambda$-frame
\be
{\cal D}^I_J =e^{ib} D^I_J e^{-ib}= D^I_J + i V^I_J \,.
\label{D-harm-covariant}
\ee
As stems from (\ref{5.20}), the superfields $V^I_J$ have the following properties under the $\,\widetilde{\phantom{m}}\,$-conjugation
\be
\widetilde{V^1_3} = -V^1_3\,,\quad
\widetilde{V^3_1} = -V^3_1\,,\quad
\widetilde{V^1_2} = V^2_3\,,\quad
\widetilde{V^2_1} = V^3_2\,.
\ee
The gauge transformations (\ref{gauge-tr-bridge-N3}) imply that these superfields transform as
\be
\delta V^I_J =-D^I_J \lambda\,.
\label{V-gauge-N3}
\ee

The commutation relations (\ref{5.16}) have the gauge covariant counterparts
\be
[D^1_2,{\cal D}^1_\alpha] = [D^2_3,{\cal D}^1_\alpha] = [D^1_3,{\cal D}^1_\alpha]=0\,,\qquad
[D^1_2,\bar{\cal D}_{3\dot\alpha}] = [D^2_3, \bar{\cal D}_{3\dot\alpha}]
 = [D^1_3,\bar{\cal D}_{3\dot\alpha}]=0\,.
\ee
Transferring these constraints to the $\lambda$-frame,  one observes that the
superfields $V^1_3$, $V^1_2$ and $V^2_3$ are analytic
\be
D^1_\alpha(V^1_3, V^1_2, V^2_3)=0\,,\qquad
\bar D_{3\dot\alpha}(V^1_3, V^1_2, V^2_3) =0\,,
\label{V-analyt}
\ee
while the other three gauge connections $V^3_1$, $V^2_1$ and $V^3_2$ are not. The analytic
superfields $V^1_3$, $V^1_2$ and $V^2_3$ are the fundamental prepotentials of ${\cal N}=3$ SYM theory, analogs of the analytic
prepotential $V^{++}$ of ${\cal N}=2$ SYM theory.

The difference from the ${\cal N}=2$ case is nevertheless as follows. The harmonic commutators (\ref{5.5})
can also be rewritten in the $\lambda$-frame.
One of these relations is the equation
\be
[{\cal D}^1_2, {\cal D}^2_3] = {\cal D}^1_3\,,
\ee
which implies that the analytic gauge connection $V^1_3$ is expressed through the other two
analytic connections $V^1_2$ and $V^2_3$
\be
V^1_3 = D^1_2 V^2_3 - D^2_3 V^1_2\,.
\ee
Therefore, in what follows we will consider only the analytic connections $V^1_2$ and $V^2_3$ as the independent basic ones.
Next, the commutators (\ref{5.5d}) in the $\lambda$-frame are
\be
[{\cal D}^1_2,{\cal D}^2_1]= S_1\,,\qquad
[{\cal D}^2_3,{\cal D}^3_2]= S_2\,,
\label{5.40}
\ee
where the operators $S_1$ and $S_2$ do not have gauge connections, since the bridge superfield $b$ is uncharged.
As a consequence of (\ref{5.40}), the non-analytic gauge connections $V^2_1$ and $V^3_2$ are related to the basic analytic
ones $V^1_2$ and $V^2_3$ by the corresponding
harmonic flatness conditions
\be
D^1_2 V^2_1 = D^2_1 V^1_2 \,,\qquad
D^2_3 V^3_2 = D^3_2 V^2_3\,.
\ee
In contrast to the $\cN=2$ case, eq.\ (\ref{V--V++}), the explicit solutions of these equations are not known because
harmonic distributions with the $SU(3)$ harmonics are not well worked out so far. Nevertheless, given that the solution
of these equations exists and is unique,
we can treat the superfields $V^2_1$ and $V^3_2$ as some functions of $V^1_2$ and $V^2_3$
\be
V^2_1 = V^2_1(V^1_2,V^2_3)\,,\qquad
V^3_2 = V^3_2(V^1_2,V^2_3)\,.
\label{5.43}
\ee

Taking harmonic projections of the anticommutation relations (\ref{N3constraints-a}) and (\ref{N3constraints-b}),
we find the expressions for the superfield strengths,
\be
\bar W^{12} = \frac i4 \{ {\cal D}^{1\alpha}, {\cal D}^2_\alpha \}\,,\qquad
W_{23} = \frac i4 \{ \bar{\cal D}_{2\dot\alpha}, \bar{\cal D}_3^{\dot\alpha} \}\,.
\ee
Recall that, in the $\lambda$-frame, the derivatives ${\cal D}^1_\alpha=D^1_\alpha$ and $\bar{\cal D}_{3\dot\alpha}=\bar D_{3\dot\alpha}$
contain no gauge connections, unlike the derivatives ${\cal D}^2_\alpha = D^2_\alpha + iV^2_\alpha$ and
$\bar{\cal D}_{2\dot\alpha} =\bar D_{2\dot\alpha}+i \bar V_{2\dot\alpha}$. Hence, in the $\lambda$-frame we have
\be
\bar W^{12} = -\frac14 D^{1\alpha} V^2_\alpha\,,\qquad
W_{23} = \frac14 \bar D_{3\dot\alpha} \bar V_2^{\dot\alpha}\,.
\label{5.45}
\ee
The spinor gauge connections $V^2_\alpha$ and $\bar V_{2\dot\alpha}$ can be expressed through the non-analytic harmonic gauge connections $V^2_1$ and $V^3_2$
in virtue of the following commutation relations in the $\lambda$-frame
\begin{subequations}
\bea
&&{\cal D}^2_\alpha = -[{\cal D}^1_\alpha , {\cal D}^2_1] \quad\Rightarrow\quad
 V^2_\alpha = -D^1_\alpha V^2_1\,,\\
&&\bar{\cal D}_{2\dot\alpha} = [\bar{\cal D}_{3\dot\alpha},{\cal D}^3_2]
\quad\Rightarrow\quad
\bar V_{2\dot\alpha} = \bar D_{3\dot\alpha} V^3_2\,.
\eea
\end{subequations}
These solutions for $V^2_\alpha$ and $\bar V_{2\dot\alpha}$
allow us to express the superfield strengths (\ref{5.45}) as
\be
\bar W^{12} = \frac14 D^{1\alpha} D^1_\alpha V^2_1\,,\qquad
W_{23} = \frac14 \bar D_{3\dot\alpha} \bar D_3^{\dot\alpha} V^3_2\,.
\label{W-anal}
\ee
In these expressions, the gauge connections  $V^2_1$ and $V^3_2$ are some functions  of the unconstrained
analytic gauge prepotentials $V^1_2$ and $V^2_3\,$,
as is defined by (\ref{5.43}). One can easily check that the superfield strengths (\ref{W-anal}) are invariant under the gauge transformations
(\ref{V-gauge-N3}). Note also that the $\,\widetilde{\phantom{m}}\,$-conjugation maps $\bar W^{12}$ and $W_{23}$ into each other
\be
\bar W^{12} = \widetilde{W_{23}}\,.
\ee

\subsection{Superconformal transformations}

The $\cN=3$ superconformal group $SU(2,2|3)$, besides the $\cN=3$ super Poincar\'e transformations, contains
dilatation (with the parameter $a$), $\gamma_5$-transformation (with the parameter $b$),
conformal boosts (with the parameters $k_{\alpha\dot\alpha}$), S-supersymmetry (with the parameters
$\eta^i_\alpha$, $\bar\eta_{i\dot\beta}$) and $SU(3)$
R-symmetry transformations (with the parameters $\lambda_i^j$, $\overline{\lambda_i^j}=-\lambda_j^i$,
$\lambda_i^i=0$). The realization of this supergroup on the analytic coordinates
(\ref{anal-coord}) was found in \cite{GIO-N3},
\bea
\delta_{\rm sc} x_A^{\alpha\dot\alpha}&=&ax_A^{\alpha\dot\alpha}+
 k_{\beta\dot\beta}x_A^{\alpha\dot\beta}x_A^{\beta\dot\alpha}-
 4k_{\beta\dot\beta}\theta_2^\beta\bar\theta^{2\dot\alpha}
  \theta_2^\alpha\bar\theta^{2\dot\beta}+
  4ix_A^{\alpha\dot\beta}\bar\theta^{1\dot\alpha}\bar
  u_1^i\bar\eta_{i\dot\beta}\nonumber\\
&& +\,2ix_{A-}^{\alpha\dot\beta}\bar\theta^{2\dot\alpha}
  \bar u_2^i \bar\eta_{i\dot\beta}+
  4ix_A^{\beta\dot\alpha}\theta_3^\alpha u^3_i\eta^i_\beta+
  2ix_{A+}^{\beta\dot\alpha}\theta_2^\alpha u^2_i\eta^i_\beta\nonumber\\
&& -\,4i\lambda_i^j\theta_3^\alpha\bar\theta^{1\dot\alpha}u^3_j\bar u_1^i-
  2i\lambda_i^j\theta_2^\alpha\bar\theta^{1\dot\alpha}u^2_j\bar u_1^i-
  2i\lambda_i^j\theta_3^\alpha\bar\theta^{2\dot\alpha}u^3_j\bar
  u_2^i\,,
\nn\\
\delta_{\rm sc}\theta_2^\alpha&=&(a/2+ib)\theta_2^\alpha+
  k_{\beta\dot\beta}x_{A+}^{\alpha\dot\beta}\theta_2^\beta-
  4i(\theta_2^\alpha u^2_i+\theta_3^\alpha u^3_i)\theta_2^\beta\eta^i_\beta
  \nonumber\\
&& +\,x_{A+}^{\alpha\dot\beta}\bar u_2^i\bar\eta_{\dot\beta i}+
   \lambda_i^j(\theta_2^\alpha u^2_j+\theta_3^\alpha u^3_j)\bar
   u_2^i\,,
   \nn\\
\delta_{\rm sc}\theta_3^\alpha&=&(a/2+ib)\theta_3^\alpha+
  k_{\beta\dot\beta}x_{A-}^{\alpha\dot\beta}\theta_3^\beta-
  4i\theta_3^\alpha\theta_3^\beta u^3_i\eta^i_\beta
 +x_{A-}^{\alpha\dot\beta}\bar u_3^i\bar\eta_{\dot\beta i}+
   \lambda_i^j\theta_3^\alpha u^3_j\bar u_3^i\,,
\nn\\
\delta_{\rm sc}\bar\theta^{1\dot\alpha}&=&(a/2-ib)\bar\theta^{1\dot\alpha}+
 k_{\beta\dot\beta}x_{A+}^{\beta\dot\alpha}\bar\theta^{1\dot\beta}+
 4i\bar\theta^{1\dot\beta}\bar\theta^{1\dot\alpha}\bar
  u_1^i\bar\eta_{\dot\beta i}+x_{A+}^{\beta\dot\alpha}u^1_i\eta^i_\beta-
  \lambda_i^j\bar\theta^{1\dot\alpha}\bar u_1^i u^1_j\,,
\nn  \\
\delta_{\rm sc}\bar\theta^{2\dot\alpha}&=&(a/2-ib)\bar\theta^{2\dot\alpha}+
 k_{\beta\dot\beta}x_{A-}^{\beta\dot\alpha}\bar\theta^{2\dot\beta}+
 4i\bar\theta^{2\dot\beta}(\bar\theta^{1\dot\alpha}\bar u_1^i+
 \bar\theta^{2\dot\alpha}\bar u_2^i)\bar\eta_{\dot\beta i}
 \nonumber\\
&&  +\,x_{A-}^{\beta\dot\alpha}u^2_i\eta^i_\beta-
  \lambda_i^j(\bar\theta^{1\dot\alpha}\bar u_1^i +
  \bar\theta^{2\dot\alpha}\bar u_2^i)u^2_j\,,
  \label{dxA}
\end{eqnarray}
where $x_{A\pm}^{\alpha\dot\alpha}=x_{A}^{\alpha\dot\alpha}\pm
2i\theta_2^\alpha\bar\theta^{2\dot\alpha}$.
For preserving the analyticity, the harmonic variables should transform according to the rules
\begin{align}
&\delta_{\rm sc} u^1_i=u^2_i\lambda^1_2+u^3_i\lambda^1_3\,,&&
 \delta_{\rm sc}\bar u_1^i=0\,,\nn\\
&\delta_{\rm sc} u_i^2=u_i^3\lambda^2_3\,,&&
 \delta_{\rm sc}\bar u_2^i=-\bar u^i_1\lambda^1_2\,,\nn\\
&\delta_{\rm sc} u^3_i=0\,,&&
 \delta_{\rm sc}\bar u_3^i=-\bar u_2^i\lambda^2_3-\bar u_1^i\lambda^1_3\,,
\label{du}
\end{align}
where
\be
\lambda^I_J=-4ik_{\beta\dot\beta}\theta_J^\beta\bar\theta^{I\dot\beta}-
 4i(\bar\eta_{\dot\beta i}\bar\theta^{I\dot\beta}\bar u_J^i+
  \theta_J^\beta\eta^i_\beta u^I_i)+u^I_i\bar u_J^j\lambda^i_j\,.
\label{lambda}
\ee

In this paper we will use the so-called passive form of superconformal
transformations of superfields, when the variation is taken at
different points, e.g., $\delta_{\rm sc} W \simeq W'(x')-W(x)$. In this case we have
to take care of transformations of the superspace derivatives and the superspace integration measure.
Nevertheless, this does not lead to extra complications since we will study the part of effective
action which is described by the superfield strengths without derivatives on them. Moreover,
it is possible to show, see, e.g.,\ \cite{GIOS-book}, that the integration measure of the analytic
superspace (\ref{anal-coord}) defined as follows \cite{N3-BI,BISZ},
\be
d\zeta(^{33}_{11})du=\frac1{16^2}d^4x_A du(D^3)^2(D^2)^2(\bar D_1)^2(\bar
D_2)^2\,,
\label{measure}
\ee
is invariant under (\ref{dxA}) and (\ref{du}):
\be
{\rm Ber}\,\frac{\partial(x_A',\theta',u')}{\partial(x_A,\theta,u)}=1\,.
\label{Ber}
\ee

Using the coordinate transformations (\ref{dxA}) and (\ref{du}),
it is straightforward to compute the superconformal variations  of the harmonic derivatives:
\begin{align}
&\delta_{\rm sc} D^1_2=-\lambda^1_2 S_1\,, && \delta_{\rm sc} D^2_1=(\lambda^1_1-\lambda^2_2)D^2_1\,,\nn\\
&\delta_{\rm sc} D^2_3=-\lambda^2_3 S_2\,, && \delta_{\rm sc} D^3_2=(\lambda^2_2-\lambda^3_3)D^3_2\,,\nn\\
&\delta_{\rm sc} D^1_3=\lambda^1_2D^2_3-\lambda^2_3D^1_2-\lambda^1_3(S_1+S_2)\,, &&
\delta_{\rm sc} D^3_1=(\lambda^1_1-\lambda^3_3)D^3_1+\lambda^2_1D^3_2-\lambda^3_2D^2_1\,,\nn\\
&\delta_{\rm sc} D^1_1=\delta_{\rm sc} D^2_2=\delta_{\rm sc} D^3_3=0\,,&&
\delta_{\rm sc} S_1=\delta_{\rm sc} S_2=0\,.
\label{dD}
\end{align}
The gauge-covariant harmonic derivatives (\ref{D-harm-covariant})
must have the same transformation properties (\ref{dD}). Hence, the gauge connections should transform under the superconformal
group according to the rules
\begin{align}
&\delta_{\rm sc} V^1_2=0\,, && \delta_{\rm sc} V^2_1=(\lambda^1_1-\lambda^2_2)V^2_1\,,\nn\\
&\delta_{\rm sc} V^2_3=0\,, && \delta_{\rm sc} V^3_2=(\lambda^2_2-\lambda^3_3)V^3_2\,,\nn\\
&\delta_{\rm sc} V^1_3=\lambda^1_2 V^2_3-\lambda^2_3 V^1_2\,,&&
 \delta_{\rm sc} V^3_1=(\lambda^1_1-\lambda^3_3)V^3_1+
  \lambda^2_1V^3_2-\lambda^3_2V^2_1\,.
\label{dV}
\end{align}

Using (\ref{dxA}) and (\ref{du}) it is also easy to find the
superconformal transformations of the covariant spinor derivatives
$D^1_\alpha$ and $\bar D_{3\dot\alpha}$
\bea
\delta_{\rm sc} D^1_\alpha&=&(-a/2-ib-\lambda^1_1)D^1_\alpha+B_\alpha^\beta D^1_\beta\,,
\nn\\
\delta_{\rm sc}\bar D_{3\dot\alpha}&=&(-a/2+ib+\lambda^3_3)\bar D_{3\dot\alpha}+
 \bar B_{\dot\alpha}^{\dot\beta}\bar D_{3\dot\beta}\,,
\label{delta-D}
\eea
where $\lambda^1_1$ and $\lambda^3_3$ are defined in
(\ref{lambda}) and
\bea
B_\alpha^\beta&=&-k_{\alpha\dot\beta}(x_{A+}^{\beta\dot\beta}+
    4i\theta_1^\beta\bar\theta^{1\dot\beta})-4i\theta_I^\beta
    u^I_j\eta^j_\alpha\,,\nn\\
\bar B_{\dot\alpha}^{\dot\beta}&=&-k_{\beta\dot\alpha}
   (x_{A-}^{\beta\dot\beta}-4i\theta_3^\beta\bar\theta^{3\dot\beta})-
   4i\bar\theta^{I\dot\beta}\bar u_I^j\bar\eta_{\dot\alpha j}\,.
\eea
It is worth pointing out that the spinor derivatives $ D^1_\alpha $ and $\bar D_{3\dot\alpha}$ are not mixed
under the superconformal transformations.

Finally, using the variations of the harmonic gauge connections
(\ref{dV}) and derivatives (\ref{delta-D}), we can find the
superconformal transformations of the superfield strengths
(\ref{W-anal}),
\be
\delta_{\rm sc} W_{23}=A\, W_{23}\,,\qquad
\delta_{\rm sc} \bar W^{12}=\bar A\, \bar W^{12}\,,
\label{dW}
\ee
where
\be
A=-a+2ib+\lambda^2_2+\lambda^3_3+\bar
B^{\dot\alpha}_{\dot\alpha}\,,\qquad
\bar A=-a-2ib-\lambda^1_1-\lambda^2_2+B^\alpha_\alpha\,.
\label{A}
\ee
One can check that the superfields $A$ and $\bar A$ are analytic,
\be
D^1_\alpha A = D^1_\alpha \bar A=0\,,
\qquad
\bar D_{3\dot\alpha}A = \bar D_{3\dot\alpha}\bar A=0\,.
\ee
Hence, the transformations (\ref{dW}) preserve the $\cN=3$ harmonic analyticity.

\subsection{Classical $\cN=3$ SYM action}
Superfield classical off-shell action of $\cN=3$ SYM theory was constructed in \cite{GIKOS-N3a,GIKOS-N3}. For completeness, here
we review this construction, although it will not be used in the next sections, when studying the effective action.
As we will show, the classical action has a very remarkable Chern-Simons form which does not resemble
the superfield classical SYM actions neither in $\cN=1$ nor in $\cN=2$ superspaces. In this section we consider
the general case of {\it non-abelian} gauge theory.

Recall that in the $\tau$-frame the covariant spinor derivatives ${\cal D}^I_\alpha = {\cal D}^i_\alpha u^I_i$
and $\bar{\cal D}_{I\dot\alpha}=\bar{\cal D}_{i\dot\alpha} \bar u_I^i$ possess gauge connections
which are subject to the constraints (\ref{N3constraints}). The harmonic derivatives (\ref{harm-deriv-AB})
and (\ref{harm-deriv-AB_}) are automatically gauge-covariant in the $\tau$-frame and so do not require gauge connections.
It is unclear how to relax the constraints (\ref{N3constraints}) in such a way that they would appear as Euler-Lagrange
equations associated with some superfield action. This becomes possible after passing to the $\lambda$-frame.

In the $\lambda$-frame the covariant spinor derivatives ${\cal D}^1_\alpha$ and $\bar{\cal D}_{3\dot\alpha}$ become short
(they have no gauge connections), but the covariant harmonic derivatives acquire gauge connections (\ref{D-harm-covariant}).
Let us concentrate on the analyticity-preserving derivatives ${\cal D}^1_2$, ${\cal D}^2_3$ and ${\cal D}^1_3$ 
(see (\ref{V-analyt})). As follows from (\ref{5.5}), the mutual commutators of these derivatives read
\be
[{\cal D}^1_3, {\cal D}^1_2] =0\,,\quad
[{\cal D}^2_3, {\cal D}^1_3] =0\,,\quad
[{\cal D}^1_2, {\cal D}^2_3] ={\cal D}^1_3\,.
\label{N3-harm-constraints}
\ee
The basic idea of \cite{GIKOS-N3a,GIKOS-N3} was to treat these equations as {\it constraints} which admit a relaxation
\be
[{\cal D}^1_3, {\cal D}^1_2] =iF^{11}_{32}\,,\quad
[{\cal D}^2_3, {\cal D}^1_3] =iF^{21}_{33}\,,\quad
[{\cal D}^1_2, {\cal D}^2_3] -{\cal D}^1_3 = iF^1_3\,.
\label{N3-harm-constraints1}
\ee
Here $F^{11}_{32}$, $F^{12}_{33}$ and $F^1_3$ are some analytic superfields which can be treated as the field strengths
for the corresponding harmonic superfield connections. In terms of the gauge connections $V^I_J$ these superfield strengths
have the following explicit form
\bea
F^{11}_{32} &=& D^1_3 V^1_2 - D^1_2 V^1_3 + i[V^1_3, V^1_2]\,,\nn\\
F^{21}_{33} &=& D^2_3 V^1_3 - D^1_3 V^2_3 + i[V^2_3, V^1_3]\,,\nn\\
F^1_3 &=& D^1_2 V^2_3 - D^2_3 V^1_2 + i [V^1_2,V^2_3] - V^1_3\,.
\label{N3-harm-strengths-explicit}
\eea

Relaxing the constraints (\ref{N3-harm-constraints}) as in eqs.\ (\ref{N3-harm-constraints1}) amounts to going off shell.
Coming back to the mass shell requires these harmonic superfield strengths to vanish,
\be
F^{11}_{32}=0\,,\quad F^{12}_{33}=0\,,\quad
F^1_3=0\,.
\ee
Remarkably, these constraints can be reproduced as the Euler-Lagrange equations associated with the following
off-shell action \footnote{The overall coefficient in this action is chosen in agreement with
the conventions of \cite{N3-BI}.}
\bea
S^{\cN=3}_{\rm SYM}&=&-\frac1{16}\tr\int d\zeta(^{11}_{33})du\big\{
V^2_3(D^1_3 V^1_2 -D^1_2 V^1_3)
-V^1_2(D^1_3 V^2_3 - D^2_3 V^1_3)\nn\\&&
+V^1_3(D^1_2 V^2_3 - D^2_3 V^1_2)
- (V^1_3)^2
+2iV^1_3[V^1_2, V^2_3]
\big\}\,.
\label{N3-class-action}
\eea
Indeed, the general variation of this action with respect to the unconstrained analytic prepotentials $V^1_2, V^2_3$ and $V^1_3$
reads
\be
\delta S^{\cN=3}_{\rm SYM} =-\frac18\tr
\int d\zeta(^{11}_{33})du\left(
\delta V^1_2 F^{21}_{33}
+\delta V^2_3 F^{11}_{32}
+\delta V^1_3 F^1_3
\right)\,.
\ee

The action (\ref{N3-class-action}) is invariant, modulo a total derivative,  under the non-abelian
 generalization of the gauge transformation (\ref{N3-class-action}),
\be
\delta_\lambda V^I_J =-{\cal D}^I_J \lambda = - D^I_J \lambda
- i [V^I_J,\lambda]\,,
\ee
where $\lambda$ is a real and analytic superfield parameter taking values
in the Lie algebra of the gauge group. Indeed, the gauge variation of (\ref{N3-class-action}),
\be
\delta_\lambda S^{\cN=3}_{\rm SYM} =-\frac18\tr
\int d\zeta(^{11}_{33})du\,\lambda \left(
{\cal D}^1_2 F^{21}_{33}
+{\cal D}^2_3 F^{11}_{32}
+{\cal D}^1_3 F^1_3
\right),
\ee
vanishes owing to the off-shell Bianchi identity for the strengths (\ref{N3-harm-strengths-explicit})
\be
{\cal D}^1_2 F^{21}_{33}
+{\cal D}^2_3 F^{11}_{32}
+{\cal D}^1_3 F^1_3 =0\,.
\ee

The  action (\ref{N3-class-action}) also respects full $SU(2,2|3)$ superconformal symmetry.
To check this, one has to take into account that the analytic measure is superconformally invariant,
see (\ref{Ber}), while the harmonic derivatives and prepotentials  transform according to
the rules (\ref{dD}) and (\ref{dV}), respectively.

The action (\ref{N3-class-action}) has the very specific form as compared
to the $\cN=2$ SYM action (\ref{S-SYM-harm}). The latter is non-polynomial in the gauge prepotential
(in the non-abelian case) while the above $\cN=3$ SYM action has only cubic interaction vertex.
Surprisingly, the superfield Lagrangian of $\cN=3$ SYM theory is of the first order in harmonic derivatives.
The form of this Lagrangian resembles the Chern-Simons Lagrangians, though the action (\ref{N3-class-action})
describes the full-fledged $\cN=3$ super Yang-Mills theory. In fact, as was pointed out in \cite{SV},
the $\cN=3$ superfield Lagrangian does acquire the literal Chern-Simons form for the properly defined
one-form of gauge connection.

In components, the off-shell $\cN=3$ gauge multiplet contains
an infinite tower of auxiliary fields \cite{GIKOS-N3a,GIKOS-N3} (along with an infinite number of gauge degrees of freedom
most of which, however, are brought away in WZ gauge). It is possible to show that,
once all auxiliary fields have been  eliminated from the action, one is left with
the multiplet of physical fields which coincides with the $\cN=4$ gauge multiplet on the mass shell.
The classical action for the physical fields has exactly the form (\ref{S}). Thus,
classically, the $\cN=3$ and $\cN=4$ gauge theories are equivalent on the mass shell.

\subsection{Superconformal effective action}
The aim of this section is to construct the $\cN=3$ superspace prototype of the effective action (\ref{Gamma-N2-final}).
Before solving this problem, let us briefly discuss a closely related issue concerning the $\cN=3$ supersymmetric
generalization of the Born-Infeld theory constructed for the first time in \cite{N3-BI}.

The Lagrangian of the Born-Infeld theory is a non-polynomial function of the abelian field strength $F_{mn}$. Being expanded
in a power series in $F_{mn}$, it starts with the standard Maxwell $F^2$ term, while the next term is $F^4 \equiv F^2 \bar F^2$,
where $F^2 = F^{\alpha\beta}F_{\alpha\beta}$, $\bar F^2 = \bar F^{\dot\alpha\dot\beta}\bar F_{\dot\alpha\dot\beta}$ and $F_{\alpha\beta}$,
$\bar F_{\dot\alpha\dot\beta}$ are the spinorial components of $F_{mn}$. The $\cN=3$ supersymmetric generalization
of this $F^4$ term is given by \cite{N3-BI}
\be
S_4=\frac1{32}\int  d\zeta(^{33}_{11})du\,
\frac{(\bar W^{12} W_{23})^2}{(\bar\Lambda\Lambda)^2}\,,
\label{S4}
\ee
where $\Lambda$ is a coupling constant of dimension one in mass units, which is
introduced to ensure the correct dimension of the integrand.
The analytic measure defined as in (\ref{measure}) is dimensionless,
$[d\zeta(^{33}_{11})du]=0\,$, and $[\bar W^{12}]=[W_{23}]=1\,$.
With this analytic measure, it is
straightforward to check that, together with other component terms, the action (\ref{S4}) yields the standard $F^4$ term,
\be
S_4=\frac12\int d^4x\,\frac{F^2\bar
F^2}{(\bar\Lambda\Lambda)^2}+\ldots\,.
\label{F4}
\ee

Consider now the superconformal variation of the action (\ref{S4})
\be
\delta_{\rm sc} S_{4}=\frac1{16}\int  d\zeta(^{33}_{11})du(A+\bar A)
\frac{(\bar W^{12} W_{23})^2}{(\bar\Lambda\Lambda)^2}\,,
\label{dS4}
\ee
where we made use of the variations of the superfield strengths
(\ref{dW}) and the property of invariance of the analytic measure (\ref{Ber}).
Here $A$ and $\bar A$ are the superfield parameters of superconformal transformations (\ref{A}) collecting
the constant parameters of the superconformal transformations
(\ref{dxA}) and (\ref{du}). We see that the action (\ref{S4}) is not superconformal, since its variation (\ref{dS4})
is non-vanishing. In the present section we will construct a superconformal generalization of (\ref{S4})
and will show that it contains the terms (\ref{F4X4}) and (\ref{WZterm}) in its component-field expansion.

\subsubsection{Scale and $\gamma_5$ invariant $F^4/X^4$ term}

We will denote the superconformal generalization of (\ref{S4}) by $\Gamma$ to stress that it is a part of the $\cN=3$ SYM
low-energy effective action. The action $\Gamma$
should meet the following criteria:
\begin{enumerate}
\item It should be a local functional defined on the analytic superspace and constructed out of the superfield strengths
$\bar W^{12}$ and $W_{23}$ without derivatives on them,
\be
\Gamma=\int d\zeta(^{33}_{11}) du\,{\cal H}^{11}_{33}(\bar W^{12},W_{23})\,.
\label{3.5}
\ee
The analytic Lagrangian density ${\cal H}^{11}_{33}$ is an arbitrary function of its
arguments, such that its external harmonic $U(1)$ charges cancel those of the analytic integration measure.
This is the most general form of the superspace action yielding terms with four-derivatives in components,  since
the analytic measure (\ref{measure}) contains just eight spinor derivatives which can produce four space-time ones
on the component fields.
\item
The action $\Gamma$ should be invariant under the superconformal
transformations (\ref{dW}),
\be
\delta_{\rm sc} \Gamma=0\,.
\ee
As a weaker requirement, in this subsection we will employ only the scale- and
$\gamma_5$-transformations out of the full $SU(2,2|3)$
superconformal group. We will show that this is sufficient to uniquely specify the structure of the action. The check of the full
superconformal symmetry will be performed in the next subsection.
\item
In the component-field expansion the action $\Gamma$ should
reproduce the scale- and $SU(3)$-invariant
$F^4/X^4$ term (\ref{F4}),
\be
\int d^4x\frac{F^2\bar F^2}{(\varphi^i\bar\varphi_i)^2}\,.
\ee
\item
We are interested in the low-energy effective action for massless
fields, with massive ones being integrated out. The massive fields appear
in the Coulomb branch, when the gauge symmetry is broken down
spontaneously. For instance, the $SU(2)$ gauge symmetry is broken down to $U(1)\,$,
when the scalar field corresponding to the Cartan subalgebra of
$su(2)$ acquire non-trivial vevs,
\be
c^i=\langle \varphi^i \rangle\ne0\,,\qquad
\bar c_i=\langle \bar \varphi_i \rangle \ne0\,.
\label{c-N3}
\ee
However, the effective action should be independent of any particular choice
of these constants,
\be
\Gamma({c'}^i,\bar c'_j)=\Gamma(c^i,\bar c_j)\,, \qquad c^i\bar c_i \ne0\,,
\ee
because such a dependence would break superconformal invariance of the action.
\item
Finally, we simplify the problem by considering only those parts of the action (\ref{3.5}), which do not vanish on the mass shell,
i.e., we will assume that the superfield strengths obey the constraints (\ref{anal})--(\ref{linearity}).
We will neglect all terms in the action $\Gamma$ which vanish when these constraints are imposed.
As a consequence, one is free to add to $\Gamma\,$, or to subtract from it, the following expressions which vanish on the mass shell,
\bea
\int d\zeta(^{33}_{11})\, \bar W^{12}{\cal F}(W_{23})&\propto&
\int d^4x (D^3)^2 (D^2)^2 (\bar D_1)^2[{\cal F}(W_{23})  (\bar D_2)^2 \bar
W^{12}] \simeq 0\,,\nn\\
\int d\zeta(^{33}_{11})\, W_{23}{\cal F}( \bar W^{12})&\propto&
\int d^4x (D^3)^2 (\bar D_2)^2 (\bar D_1)^2[{\cal F}( \bar W^{12})
(D^2)^2 W_{23}] \simeq 0\,.\nn\\
\label{prop}
\eea
Here ${\cal F}(W)$ is an arbitrary function of its argument.
We will frequently employ this property, when deriving the action.
\end{enumerate}

Now we turn to constructing the action $\Gamma$ that meets the requirements and properties listed above.

As the first step, we introduce the shifted scalar fields, $\phi^i$ and
$\bar\phi_i$,
\be
\varphi^i=c^i+\phi^i\,,\quad
\bar\varphi_i=\bar c_i+\bar \phi_i\,,\qquad
\langle \phi^i \rangle=\langle \bar\phi_i \rangle=0\,.
\label{vev-shift}
\ee
Next, we define the harmonic projections of these vev
constants
\be
c^1=u^1_i c^i\,,\quad c^2=u^2_i c^i\,\quad
c^3=u^3_i c^i\,,\qquad
\bar c_1 =\bar u_1^i \bar c_i\,,\quad
\bar c_2 =\bar u_2^i \bar c_i\,,\quad
\bar c_3 = \bar u_3^i \bar c_i\,.
\label{cccc}
\ee
Using these objects, we introduce the shifted superfield
strengths, $\bar\omega^{12}$ and $\omega_{23}$,
\be
\bar W^{12}=\bar c_3+\bar\omega^{12}\,,\qquad
W_{23}=c^1+\omega_{23}\,.
\label{shift}
\ee
Under the scale and $\gamma_5$ transformations these shifted
superfields transform inhomogeneously,
\be
\delta_{\rm sc} \bar\omega^{12}=\bar A\bar c_3 +\bar A\bar\omega^{12}\,,\qquad
\delta_{\rm sc} \omega_{23}=Ac^1+A\omega_{23}\,,
\label{scale}
\ee
where $A=-a+2ib$. The case of generic $A$ and $\bar A$ defined in (\ref{A}) will be considered in the next
subsection.

We point out that on shell, when the relations (\ref{prop}) are valid, the non-superconformal action (\ref{S4}) can be
rewritten in terms of $\bar \omega^{12}$ and $\omega_{23}$ as
\be
S_4=\frac1{32}\int  d\zeta(^{33}_{11})du\,
\frac{(\bar \omega^{12} \omega_{23})^2}{(c^i\bar c_i)^2}\,.
\label{S4_}
\ee
Here we substituted $(c^i\bar c_i)^2$ in the denominator instead of
$(\bar\Lambda\Lambda)^2$, because no other dimensionful constants besides the vevs $c^i$ can be
present in the superconformal case.

We seek for a superconformal generalization of the action (\ref{S4_}) in the form
\be
\Gamma=\frac\alpha{8}\int  d\zeta(^{33}_{11})du\frac{(\bar \omega^{12}
\omega_{23})^2}{(c^i\bar c_i)^2}
H\left(\frac{\bar \omega^{12}c^3}{c^i \bar c_i},
\frac{\omega_{23}\bar c_1}{c^i\bar c_i}\right),
\label{G}
\ee
where $H(x,y)$ is some function to be determined and $\alpha$ is a
dimensionless coupling constat.
The arguments $\frac{\bar \omega^{12}c^3}{c^i \bar
c_i}$ and $\frac{\omega_{23}\bar c_1}{c^i\bar c_i}$ of the function $H$ are uncharged and dimensionless.
We assume that the function $H$ has a regular power expansion with respect to its arguments,
\be
H(x,y)=\sum_{m,n=0}^\infty
\alpha_{m,n} x^m y^n\,,
\label{H}
\ee
with undefined coefficients $\alpha_{m,n}$.
The reality of the action (\ref{G}) with respect to the tilde-conjugation
implies the symmetry of this function, $H(x,y)=H(y,x)\,$, whence $\alpha_{m,n}=\alpha_{n,m}\,$.

Reordering the summation in (\ref{H}), it is convenient to
represent (\ref{G}) as
\be
\Gamma=\sum_{n=0}^\infty\Gamma_n\,,\qquad
\Gamma_n=\frac\alpha{8}\int d\zeta(^{33}_{11})du
\frac{(\bar \omega^{12}\omega_{23})^2}{(c^i\bar c_i)^2}
\sum_{p=0}^{n}
\alpha_{p,n-p}
\left(\frac{\bar\omega^{12}c^3}{c^i\bar c_i}\right)^p
\left(\frac{\omega_{23}\bar c_1}{c^i\bar c_i}\right)^{n-p}.
\label{Gn}
\ee
The invariance of the action (\ref{Gn}) under the transformations
(\ref{scale}) can be secured order by order, i.e., the
non-vanishing terms from $\delta_{\rm sc} \Gamma_{n}$ are required to be canceled by similar terms
from $\delta_{\rm sc} \Gamma_{n+1}$, and so forth. To simplify the derivation, we put $c^i\bar c_i=1$ and $\alpha=32$;
these constants will be restored in the final expression.

Consider two lowest terms in the series (\ref{Gn}),
\bea
\Gamma_0&=&\alpha_{0,0}\int
d\zeta(^{33}_{11})du(\bar\omega^{12}\omega_{23})^2\,,\nn\\
\Gamma_1&=&\alpha_{0,1}\int d\zeta(_{11}^{33})du
(\bar\omega^{12}\omega_{23})^2(\bar\omega^{12}c^3+\omega_{23}\bar
c_1)\,.
\label{59}
\eea
The superconformal variation of $\Gamma_0$ reads
\be
\delta_{\rm sc}\Gamma_0=2\alpha_{0,0}(A+\bar A)\int d\zeta(^{33}_{11})du
(\bar\omega^{12}\omega_{23})^2\,.
\label{var-59}
\ee
Note that the terms with
$\bar\omega^{12}\bar\omega^{12}\omega_{23}$ and
$\bar\omega^{12}\omega_{23}\omega_{23}$ vanish on shell because of
the relations (\ref{prop}).

The superconformal variation of $\Gamma_1$ reads
\be
\delta_{\rm sc}\Gamma_1=3\alpha_{0,1}\int d\zeta(^{33}_{11})du
\left[(\bar\omega^{12}\omega_{23})^2(\bar A c^3\bar c_3+A c^1 \bar c_1)+
O(\omega^5)\right].
\label{60}
\ee
Using the identities
\be
c^1=D^1_2 c^2= D^1_3 c^3\,,\qquad
\bar c_3 =-D^1_3 \bar c_1=-D^2_3 \bar c_2\,,
\ee
which follow from the definitions (\ref{cccc}), one can write
\bea
c^1\bar c_1&=&\frac13(c^1\bar c_1+\bar c_1 D^1_2 c^2 +\bar c_1 D^1_3
c^3)\,,\nn\\
c^3\bar c_3&=&\frac13(c^3\bar c_3-c^3 D^1_3 \bar c_1- c^3 D^2_3 \bar c_2)\,.
\eea
We substitute these expressions into (\ref{60}) and integrate by
parts with respect to the harmonic derivatives $D^1_2$, $D^2_3$ and $D^1_3$,
\be
\delta_{\rm sc}\Gamma_1=\alpha_{0,1}\int d\zeta(^{33}_{11})du
\left[(\bar A+A)(\bar\omega^{12}\omega_{23})^2+
O(\omega^5)\right].
\label{61}
\ee
Here we made also use of the identity $c^1\bar c_1+c^2\bar c_2+c^3\bar c_3=c^i\bar c_i=1$.
Comparing (\ref{61}) with (\ref{var-59}), we observe that the terms
with four superfield strengths are canceled out under the condition
\be
\alpha_{0,1}=-2\alpha_{0,0}\,.
\ee

Let us now consider the $n$-th term in the series (\ref{Gn}),
\be
\Gamma_n=\int d\zeta(^{33}_{11})du
(\bar \omega^{12}\omega_{23})^2
\sum_{p=0}^{n}
\alpha_{p,n-p}
(\bar\omega^{12}c^3)^p
(\omega_{23}\bar c_1)^{n-p}\,,
\label{Gn_}
\ee
and compute its variation under (\ref{scale}),
\bea
\delta_{\rm sc}\Gamma_n&=&
\int d\zeta(^{33}_{11})du
(\bar \omega^{12}\omega_{23})^2
\sum_{p=0}^{n}
\alpha_{p,n-p}[(p+2)\bar A+(n-p+2)A]
(\bar\omega^{12}c^3)^p
(\omega_{23}\bar c_1)^{n-p}
\nn\\&&
+\int d\zeta(^{33}_{11})du
(\bar \omega^{12}\omega_{23})^2
\sum_{p=1}^{n}
\alpha_{p,n-p}(p+2)\bar A
(\bar\omega^{12}c^3)^{p-1}
(\omega_{23}\bar c_1)^{n-p}c^3\bar c_3
\nn\\&&
+\int d\zeta(^{33}_{11})du
(\bar \omega^{12}\omega_{23})^2
\sum_{p=0}^{n-1}
\alpha_{p,n-p}(n-p+2) A
(\bar\omega^{12}c^3)^p
(\omega_{23}\bar c_1)^{n-p-1}c^1\bar c_1\,.
\label{47}
\eea
In the second line of (\ref{47}) we apply the identity
\be
\bar c_3(c^3)^p(\bar c_1)^{n-p}=
\left(\frac{p}{n+2}\bar c_3-\frac{n-p+1}{n+2}D^1_3\bar c_1
-\frac1{n+2}D^2_3 \bar c_2\right)(c^3)^p(\bar c_1)^{n-p}\,.
\ee
Upon integrating by parts with respect to the harmonic derivatives $D^1_3$ and
$D^2_3$, this expression is replaced by
\be
\frac p{n+2}(\bar c_1)^{n-p}(c^3)^{p-1}\,.
\ee
Similarly, in the last line of (\ref{47}) we apply the identity
\be
c^1(\bar c_1)^{n-p}(c^3)^p=
\left(\frac{n-p}{n+2}c^1+\frac1{n+2}D^1_2 c^2+\frac{p+1}{n+2}D^1_3 c^3\right)
(c_1)^{n-p}(c^3)^p
\ee
and again integrate by parts with respect to the harmonic derivatives. As a result,
the expression $c^1(\bar c_1)^{n-p}(c^3)^p$ in (\ref{47}) produces the term
\be
\frac{n-p}{n+2}(c^3)^p(\bar c_1)^{n-p-1}\,.
\ee

Taking all this into account, the variation (\ref{47}) can be written as
\bea
\delta_{\rm sc}\Gamma_n&=&
\int d\zeta(^{33}_{11})du
(\bar \omega^{12}\omega_{23})^2
\sum_{p=0}^{n}
\alpha_{p,n-p}[(p+2)\bar A+(n-p+2)A]
(\bar\omega^{12}c^3)^p
(\omega_{23}\bar c_1)^{n-p}
\nn\\&&
+\int d\zeta(^{33}_{11})du
(\bar \omega^{12}\omega_{23})^2
\sum_{p=1}^{n}
\alpha_{p,n-p}\frac{p(p+2)}{n+2}\bar A
(\bar\omega^{12}c^3)^{p-1}
(\omega_{23}\bar c_1)^{n-p}
\label{48_}\\&&
+\int d\zeta(^{33}_{11})du
(\bar \omega^{12}\omega_{23})^2
\sum_{p=0}^{n-1}
\alpha_{p,n-p}\frac{(n-p)(n-p+2)}{n+2} A
(\bar\omega^{12}c^3)^p
(\omega_{23}\bar c_1)^{n-p-1}\,.
\nn
\eea
We observe that the terms in the last two lines in (\ref{48_})
cancel similar terms in the first line of
$\delta_{\rm sc}\Gamma_{n-1}$, provided that the coefficients $\alpha_{ij}$ obey the
following two equations
\bea
\alpha_{p,n-p}\,\dfrac{(n-p+2)(n-p)}{n+2}
+\alpha_{p+1,n-p-1}\,\dfrac{(p+3)(p+1)}{n+2}&=&-(n+3)\alpha_{p,n-p-1}
\,,\\
\alpha_{p,n-p}\,\dfrac{(n-p+2)(n-p)}{n+2}
-\alpha_{p+1,n-p-1}\,\dfrac{(p+3)(p+1)}{n+2}&=&-(n-2p-1)\alpha_{p,n-p-1}\,.
\nn
\eea
As a consequence, any two adjacent coefficients are related as
\be
\frac{\alpha_{p,j}}{\alpha_{p,j-1}}
=-\frac{(j+1)(p+j+2)}{(j+2)j}\,.
\ee
The solution of this equation reads
\be
\alpha_{m,n}=(-1)^{m+n}\frac{(m+n+2)!}{(n+2)n!(m+2)m!}\,.
\label{alpha}
\ee

With these coefficients, the series (\ref{H}) can be summed up
to the function
\be
H(x,y)=
  \frac{\ln (1 + x + y)}{x^2y^2}+
\frac{1}
   {xy( 1 + x +
       y ) } -
  \frac{\ln (1 + x)}{x^2 y^2} -
  \frac{\ln (1 + y)}{x^2y^2} \,.
\label{HH}
\ee
We point out that this function is regular at the origin,
\be
\lim_{x,y\to0}H(x,y)=\frac12\,.
\ee
Hence the action (\ref{G}) with this function is well-defined and
the harmonic integral does not encounter any singularities.

The contributions from the last two terms in (\ref{HH}) to the
action (\ref{G}) vanish on shell due to the properties
(\ref{prop}).\footnote{The properties (\ref{prop}) are valid essentially on shell.
Therefore the last two terms in (\ref{HH}) can be neglected only
on the mass shell although they can be important for the off-shell completion of the action.} Therefore, the on-shell effective
action can be rewritten in the following explicit form
\be
\Gamma=\frac\alpha{8}\int d\zeta(^{33}_{11})du\left[\frac{(c^i\bar c_i)^2}{c^3c^3 \bar c_1 \bar c_1}
\ln\left(1+\frac{\bar\omega^{12}c^3}{c^i\bar c_i}
+\frac{\omega_{23}\bar c_1}{c^i\bar c_i}\right)
+\frac{(c^i\bar c_i)\bar\omega^{12}\omega_{23}}{c^3\bar c_1(c^i\bar c_i
+\bar\omega^{12}c^3+\omega_{23}\bar c_1)}
\right].
\label{Gconf}
\ee
Although the charged objects $c^3$ and $\bar c_1$ appear in the
denominators, they do not lead to the divergent harmonic
integrals. It can be explicitly checked  that upon passing
to the component form of the action (\ref{Gconf}), all dangerous
terms with divergent harmonic integrals vanish after performing the integration
over the Grassmann variables.

\subsubsection{Complete $\cN=3$ superconformal symmetry}

In the previous section we found the low-energy effective action (\ref{Gconf}) by imposing
the requirements of scale and $\gamma_5$-invariance only. In this section we demonstrate that this action is invariant
under the full $SU(2,2|3)$ superconformal group.
For this purpose we have to consider the transformations (\ref{dW})
which include all parameters of the superconformal
transformations. The corresponding variations (\ref{scale}) of the
shifted superfield strengths $\bar \omega^{12}$ and $\omega_{23}$
read
\bea
\delta_{\rm sc} \bar \omega^{12}&=&A\bar\omega^{12}+A\bar c_3+\lambda^2_3
\bar c_2+\lambda^1_3 \bar c_1\,,\nn\\
\delta_{\rm sc} \omega_{23}&=&\bar A \omega_{23}+\bar A c^1
-\lambda^1_2 c^2-\lambda^1_3 c^3\,,
\eea
where $A$ and $\bar A$ are given in (\ref{A}) and $\lambda^I_J$
are defined in (\ref{lambda}). The variation of the action (\ref{G})
under these transformations is as follows
\bea
\delta_{\rm sc} \Gamma&=&\frac\alpha{8}\int d\zeta(^{33}_{11})du\, (\bar\omega^{12}\omega_{23})^2
[\frac{2}x H(x,y)+H'_x(x,y)][Ax+Ac^3\bar c_3+\lambda^2_3 c^3\bar c_2
+\lambda^1_3 c^3 \bar c_1]
\nn\\
&&+\,\frac\alpha{8}\int d\zeta(^{33}_{11})du\, (\bar\omega^{12}\omega_{23})^2
[\frac{2}y H(x,y)+H'_y(x,y)][\bar Ay+\bar Ac^1\bar c_1-\lambda^1_2 c^2\bar
c_1 -\lambda^1_3 c^3 \bar c_1]\,.\nn\\
\label{var1}
\eea
For simplicity, we set here $c^i\bar c_i=1\,$, so $x=\bar\omega^{12}c^3$, $y=\omega_{23}\bar c_1$. The
first and second lines in (\ref{var1}) are tilde-conjugated to each other.

Given the explicit form (\ref{HH}) of the function $H(x,y)$, it is
easy to check that it solves the differential equations
\bea
\frac2xH(x,y)+H'_x(x,y)&=&\frac{1}{x(1+x)(1+x+y)^2}\,,\nn\\
\frac2yH(x,y)+H'_y(x,y)&=&\frac{1}{y(1+y)(1+x+y)^2}\,.
\label{relH}
\eea
Taking them into account, we are going to show that the integrand in (\ref{var1}) is a total harmonic derivative,
so the variation (\ref{var1}) vanishes.

To this end, we introduce the auxiliary functions $f(x,y)$ and $\tilde{f}(x,y)$:
\bea
f(x,y)&=&\frac{1}{y(y+1)(x+y+1)}+\frac{\ln(1+x+y)}{xy^2}
-\frac{\ln(1+x)}{xy^2}-\frac{\ln(1+y)}{xy^2}\,,\label{f}\\
\tilde f(x,y)&=&f(y,x)=
\frac{1}{x(x+1)(x+y+1)}+\frac{\ln(1+x+y)}{yx^2}
-\frac{\ln(1+y)}{yx^2}-\frac{\ln(1+x)}{yx^2}\,.\nn
\eea
They possess the following properties
\begin{subequations}
\label{props}
\bea
xf'_x+f&=&-\frac1{(1+x)(1+x+y)^2}=-(xH'_x+2H)\,,
\label{prop0.1}\\
xf'_x+yf'_y+3f&=&\frac1{x(1+x)(1+x+y)^2}-\frac1{x(1+y)^2}
=(H'_x+\frac 2xH)+\ldots\,,\label{prop1.1}\\
y\tilde f'_y+\tilde f&=&-\frac1{(1+y)(1+x+y)^2}=-(yH'_y+2H)\,,\\
y\tilde f'_y+x\tilde f'_x +3\tilde f&=&
\frac1{y(1+y)(1+x+y)^2}-\frac1{y(1+x)^2}=(H'_y+\frac 2x
H)+\ldots\,.~~~~~~~~~~~
\label{prop1}
\eea
\end{subequations}
Here dots stand for the terms integrals of which over the
analytic superspace with the weight
$(\bar\omega^{12}\omega_{23})^2$ are on-shell vanishing due to the relations (\ref{prop}).
Up to these terms, the
equations (\ref{props}) allow one to deduce the relations
\bea
-D^2_3 (f(x,y)c^3 \bar c_2 A)- D^1_3(f(x,y)c^3 \bar c_1 A)
&=&(H'_x+\frac 2x H)(Ax+Ac^3\bar c_3)
\nn\\&&
-\,f(x,y)c^3 \bar c_2
\lambda^2_3
-f(x,y)c^3\bar c_1 \lambda^1_3\,,\nn\\
D^1_2(\tilde f(x,y)c^2\bar c_1\bar A)
+D^1_3(\tilde f(x,y)c^3\bar c_1\bar A)
&=&(H'_y+\frac2yH)(\bar A y+\bar A c^1\bar c_1)
\nn\\&&
+\,\tilde f(x,y)c^2\bar c_1\lambda^1_2
+\tilde f(x,y)c^3\bar c_1\lambda^1_3\,.
~~~~~~~~~~
\label{rel1}
\eea
Here we made use of the obvious identities for the superfield parameters $\lambda^I_J$
\be
\lambda^1_2=D^1_2\bar A\,,\quad
\lambda^2_3=D^2_3A\,,\quad
\lambda^1_3=D^1_3A=D^1_3\bar A\,,
\ee
as well as  of the convention $c^i\bar c_i=1\,$.

Next, we introduce the functions
\begin{subequations}
\label{g}
\bea
g(x,y)&=&\frac1{y(1+y)^2(1+x+y)}-\frac1{y(x+1)}\,,\\
\tilde g(x,y)&=&g(y,x)=\frac1{x(1+x)^2(1+x+y)}-\frac1{x(y+1)}\,,
\eea
\end{subequations}
with the properties
\begin{subequations}
\bea
xg'_x+g&=&\frac1{y(1+y)(1+x+y)^2}-\frac1{y(1+x)^2}
=H'_y+\frac2yH+\ldots\,,~~~~~~\\
y\tilde g'_y+\tilde
g&=&\frac1{x(1+x)(1+x+y)^2}-\frac1{x(1+y)^2}
=H'_x+\frac 2xH+\ldots\,,\\
g(x,y)-\tilde g(x,y)&=&(H'_x+\frac 2x H)-(H'_y+\frac 2y H)\,.
\eea
\end{subequations}
Here, as in (\ref{prop1.1}) and (\ref{prop1}), the dots stand for the terms vanishing on shell after integration over the
analytic superspace with the weight $(\bar
\omega^{12}\omega_{23})^2$. Up to these terms, we obtain the relation
\bea
-D^1_2(\lambda^2_3\tilde g(x,y)c^3\bar c_1)-D^2_3(\lambda^1_2 g(x,y)c^3\bar c_1)
&=&(H'_x+\frac2xH)\lambda^2_3 c^3\bar c_2
-(H'_y+\frac 2yH)\lambda^1_2c^2\bar c_1
\nn\\&&+\,[(H'_x+\frac2xH)-(H'_y+\frac 2yH)]
\lambda^1_3 c^3\bar c_1\,.
\nn\\
\label{rel2}
\eea

Finally, we introduce the functions
\begin{subequations}
\label{h}
\bea
h(x,y)&=&-\frac1{(1+x)y}+\frac{\ln(1+x)}{xy^2}+\frac{\ln(1+y)}{xy^2}
-\frac{\ln(1+x+y)}{xy^2}\,,
\\
\tilde h(x,y)&=&h(y,x)=
-\frac1{(1+y)x}+\frac{\ln(1+y)}{yx^2}+\frac{\ln(1+x)}{yx^2}
-\frac{\ln(1+x+y)}{yx^2}\,,~~~~~~~~
\eea
\end{subequations}
with the properties
\bea
&h(x,y)+yh'_y(x,y)=f(x,y)\,,\qquad
\tilde h(x,y)+x\tilde h'_x(x,y)=\tilde f(x,y)\,,&\\
&h-\tilde h=\tilde f-f\,.&
\eea
These properties allow us to derive one more useful relation
\be
-D^1_2(\lambda^2_3 h(x,y)c^3\bar c_1)-D^2_3(\lambda^1_2 \tilde h(x,y)c^3\bar c_1)
=f\lambda^2_3 c^3\bar c_2
-\tilde f\lambda^1_2c^2\bar c_1
+(f-\tilde f)\lambda^1_3 c^3\bar c_1\,.
\label{rel3}
\ee

Now, taking into account the relations (\ref{rel1}), (\ref{rel2}) and (\ref{rel3}), we observe that the variation (\ref{var1}) can
be represented as a linear combination of harmonic
derivatives acting on the quantities composed of the functions (\ref{f}), (\ref{g}) and (\ref{h}),
\bea
\delta_{\rm sc} \Gamma&=&\frac\alpha{8}\int d\zeta(^{33}_{11})du\, (\bar\omega^{12}\omega_{23})^2
\bigg\{
D^1_2(\tilde f c^2 \bar c_1 \bar A)
- D^2_3( f c^3 \bar c_2 A)
+D^1_3(\tilde f c^3 \bar c_1\bar A- f c^3 \bar c_1 A)
\nn\\&&
-\,D^1_2[(\tilde g+h)\lambda^2_3 c^3\bar c_1]-D^2_3[(g+\tilde h)\lambda^1_2 c^3\bar c_1]
\bigg\}.\label{344}
\eea
The variation (\ref{344}) vanishes as an integral of total
harmonic derivative. This proves the invariance of the action
(\ref{Gconf}) under the full $SU(2,2|3)$ superconformal group.\footnote{Note that (\ref{Gconf}) is $SU(2,2|3)$
invariant for any $c^i\neq 0$, without any restriction on the norm $c^i\bar c_i\,$ which was set equal to $1$
in the above consideration merely for convenience.}

\subsubsection{Independence of the choice of vacua}
By construction, the effective action (\ref{G}) with the function $H$ given in (\ref{HH}) is well defined only
on the Coulomb branch of $\cN=3$ SYM theory. This is manifested in the explicit presence of non-zero vacuum
constants $c^i$ and $\bar c_i$ in the Lagrangian in (\ref{G}). However, the
action itself should be independent of any particular choice of these
constants, except for the point $c^i=0$ at which the effective action
is singular.

Let us rewrite (\ref{G}) in terms of the original
(non-shifted) superfield strengths $\bar W^{12}$ and $W_{23}$
\be
\Gamma[\bar W^{12},W_{23};c^i,\bar c_i]=\frac\alpha{8}\int  d\zeta(^{33}_{11})du\frac{(\bar W^{12}-\bar c_3)^2
(W_{23}-c^1)^2}{(c^i\bar c_i)^2}
H\left(c^3\frac{\bar W^{12}-\bar c_3}{c^i \bar c_i},
\bar c_1\frac{W_{23}-c^1}{c^i\bar c_i}\right).
\label{GG}
\ee
In the previous subsection we proved that this action is invariant
under the full $SU(2,2|3)$ superconformal group. Taking into account that the analytic integration measure is $SU(2,2|3)$ invariant by itself,
the property of superconformal invariance of the action can be written in the finite form as
\be
\Gamma[\bar W^{12},W_{23};c^i,\bar c_i] = \Gamma'[\bar W^{12}{}',W_{23}{}';c^i,\bar c_i] = \Gamma[\bar W^{12}{}',W_{23}{}';c^i,\bar c_i]\,.
\ee
In particular,
consider scale and $\gamma_5$ transformations of the superfield
strength in the finite form,
\be
\bar W^{12}\to e^{\bar A} \bar W^{12}\,,\qquad
W_{23}\to e^A W_{23}\,,
\label{scale-finite}
\ee
where $A=-a+2ib$. The transformation of the action (\ref{GG})
under (\ref{scale-finite}) can be represented as
\bea
\Gamma[\bar W^{12},W_{23};c^i,\bar c_i] &=& \Gamma[e^{\bar A} \bar W^{12},e^A W_{23};c^i,\bar c_i] \nonumber \\
&=&\frac\alpha{8}
\int  d\zeta(^{33}_{11})du\frac{(\bar W^{12}-e^{-\bar A}\bar c_3)^2
(W_{23}-e^{-A}c^1)^2}{(e^{-A-\bar A}c^i\bar c_i)^2}
\nn\\&&\times
\,H\left(e^{-A}c^3\frac{\bar W^{12}-e^{-\bar A}\bar c_3}{e^{-A-\bar A}c^i \bar c_i},
e^{-\bar A}\bar c_1\frac{W_{23}-e^{-A}c^1}{e^{-A-\bar A}c^i\bar c_i}\right).
\eea
Here the $A$-dependence is absorbed into the vev constants,
$c^i\to e^{-A}c^i$, $\bar c_i\to e^{-\bar A}\bar c_i$. Hence, the
superconformal invariance of the action (\ref{GG}) implies its
independence of the complex rescalings of the vev constants,
\be
\Gamma[\bar W^{12},W_{23};c^i,\bar c_i]
=\Gamma[e^{\bar A} \bar W^{12},e^A W_{23};c^i,\bar c_i]
=\Gamma[ \bar W^{12},W_{23};e^{-\bar A}c^i,e^{-A}\bar c_i]\,.
\ee

In a similar way, one can prove that the action
(\ref{GG}) is independent of the parameters of finite $SU(3)$ rotations of the vev constants,
\be
\Gamma[\bar W^{12},W_{23};c^i,\bar c_i]
=\Gamma[ \bar W^{12},W_{23};\Lambda^i_j c^j,\bar \Lambda_i^j\bar
c_j]\,,
\ee
where $\Lambda^i_j$ are $SU(3)$ matrices. As a result, the action
(\ref{GG}) is independent of any particular choice of the vacuum
$c^i$, $c^i\ne0\,$. Indeed, let us assume, without loss of generality,
that $c^3\neq 0\,$. Then, using the coset $SU(3)/[U(1)\times SU(2)]$ transformations with a constant $SU(2)$ doublet as parameters, one can
cast $c^i$ in the form
$c^i = (0,0,c^3)$. The constant $c^3$ can be made real by exploiting the residual $U(1)$ transformation
(a combination of the $\gamma_5$ transformations
and those of $U(1)$ from the denominator of $SU(3)/[U(1)\times SU(2)]$). Finally, it can be rescaled to any non-zero value,
keeping in mind that the action is independent of the rescalings of the vev constants.

\subsection{Component structure}
\subsubsection{$F^4/X^4$ term}
To derive this term from the effective action (\ref{G}), it suffices to consider only constant Maxwell and scalar fields,
omitting all other components in (\ref{W-comp-N3}),
\be
\hat{\bar\omega}^{12}=u^1_i \phi^i+4i\theta_2^\alpha\theta_3^\beta
F_{\alpha\beta}\,,\qquad
\hat\omega_{23}=\bar u_3^i
\bar\phi_i+4i\bar\theta^{1\dot\alpha}\bar\theta^{2\dot\beta}\bar
F_{\dot\alpha\dot\beta}\,.
\ee
Substituting these superfields into (\ref{G}), we integrate over the Grassmann variables and obtain
\be
\Gamma_{F^4/X^4}=\frac\alpha{2}\int d^4x du\, F^2\bar F^2
\sum_{m,n=0}^\infty \frac{(m+1)(n+1)(m+n+2)!(-1)^{m+n}}{m!n!}
(\bar \phi_3 c^3)^m (\phi^1 \bar c_1)^n\,.
\label{G-comp}
\ee
Here we used the series expansion (\ref{H}) for the function
$H$ with the coefficients given by (\ref{alpha}). In this
subsection we assume $c^i\bar c_i=1$ for simplicity
and use the notation
$F^2=F^{\alpha\beta}F_{\alpha\beta}$\,,
$\bar F^2=\bar F^{\dot\alpha\dot\beta}\bar
F_{\dot\alpha\dot\beta}$\,.

In (\ref{G-comp}), we have to calculate the harmonic integrals. According to \cite{GIOS-book},
the definition of harmonic integration over the $SU(3)$ harmonic variables is
\be
\int du\,1=1\,,\qquad
\int du(\mbox{non-singlet $SU(3)$ irreducible representation})=0\,.
\ee
{}From this definition one can derive the following simple relations
\be
\int du\, u^1_i \bar u_1^j=\int du\, u^3_i \bar
u_3^j=\frac13\delta_i^j\,,\quad
\int du\, u^1_i\bar u_1^ju^1_k\bar u_1^l
  =\frac1{6}\delta_i^{(j}\delta_k^{l)}\,,
\quad\mbox{etc.}
\ee
All these integrals appear as particular cases of the general formula
\bea
\int du\,
u^1_{i_1} \bar u_1^{i'_1}\ldots
u^1_{i_n} \bar u_1^{i'_n}
u^3_{j_1} \bar u_3^{j'_1}\ldots
u^3_{j_m} \bar u_3^{j'_m}&=&\sum_{k=0}^m
\frac{2m!(-1)^k}{(m+1)(k+n+2)(k+n+1)k!(m-k)!}\nn\\&&
\times\,
\delta_{i_1}^{(i'_1}\ldots \delta_{i_n}^{i'_n}
\delta_{(j_1}^{\{j'_1}\ldots \delta_{j_k}^{j'_k)}\ldots
\delta_{j_m)}^{j'_m\}}\,.
\eea
Here both $(\ldots)$ and $\{\ldots\}$ denote symmetrization of
the indices.
Contracting this expression with vev constants $c^i$, $\bar c_i$
and with the scalar fields $\phi^i$, $\bar\phi_i\,$, we find
\bea
\int du(\phi^1\bar c_1)^n (c^3\bar\phi_3)^m
&=&\sum_{k=0}^m
\frac{2m!(-1)^k}{(m+1)(k+n+2)(k+n+1)k!(m-k)!}\\&&
\times\phi^{(i_1}\ldots \phi^{i_n}c^{j_1}\ldots c^{j_k)}\ldots c^{j_m}
\bar c_{i_1}\ldots \bar c_{i_n}\bar \phi_{j_1}\ldots \bar \phi_{j_k}
\ldots \bar \phi_{j_m}\,.
\nn
\eea
After some combinatorics, this expression can be rewritten
in the following useful form
\bea
\int du(\phi^1\bar c_1)^n (c^3\bar\phi_3)^m
&=&
\sum_{k=0}^{{\rm min}(m,n)} \frac{2n!m!(m+n-k+1)!(-1)^k}{k!(n-k)!(m-k)!(m+n+2)!(n+1)(m+1)}
\nn\\&&\times\,
(\phi^i\bar\phi_i)^k
(\phi^i\bar c_i)^{n-k}(c^i \bar\phi_i)^{m-k}\,.
\label{hint}
\eea

Now we represent (\ref{G-comp}) as a sum of two
terms,
\be
\Gamma_{F^4/X^4}=\frac\alpha2\int d^4x\, F^2\bar F^2(T_1+T_2)\,,
\ee
where
\bea
T_1&=&\int du\,
\sum_{n=0}^\infty \sum_{m=0}^n \frac{(m+1)(n+1)(m+n+2)!(-1)^{m+n}}{m!n!}
(\phi^1 \bar c_1)^n (\bar \phi_3 c^3)^m \,,\nn\\
T_2&=&\int du\,
\sum_{n=0}^\infty \sum_{m=n+1}^\infty \frac{(m+1)(n+1)(m+n+2)!(-1)^{m+n}}{m!n!}
(\phi^1 \bar c_1)^n(\bar \phi_3 c^3)^m \,.~~~~~~~~~
\eea
The reason for this separation is that the monomials with $m\leq n$ are in $T_1\,$, while those with $m>n$
are in $T_2$. Therefore, for each of these two terms we can apply the
equation (\ref{hint}) for the harmonic integrals,
\bea
T_1&=&2\sum_{n=0}^\infty \sum_{m=0}^n \sum_{l=0}^m
\frac{(m+n-l+1)!(-1)^{m+n+l}}{l!(n-l)!(m-l)!}
(\phi^i\bar\phi_i)^l (\phi^i\bar
c_i)^{n-l}(c^i\bar\phi_i)^{m-l}\,,\nn\\
T_2&=&2\sum_{n=0}^\infty \sum_{m=n+1}^\infty \sum_{l=0}^n
\frac{(m+n-l+1)!(-1)^{m+n+l}}{l!(n-l)!(m-l)!}
(\phi^i\bar\phi_i)^l (\phi^i\bar
c_i)^{n-l}(c^i\bar\phi_i)^{m-l}\,.
\eea
Changing the order of summation, these
terms can be rewritten as
\bea
T_1&=&2\sum_{l,m=0}^\infty \sum_{n=m}^\infty
\frac{(n+m+l+1)!(-1)^{m+n+l}}{l!m!n!}(\phi^i\bar\phi_i)^l
(\phi^i\bar\phi_i)^n (c^i\bar\phi_i)^m\,,\nn\\
T_2&=&2\sum_{l,m=0}^\infty \sum_{n=0}^{m-1}
\frac{(n+m+l+1)!(-1)^{m+n+l}}{l!m!n!}(\phi^i\bar\phi_i)^l
(\phi^i\bar\phi_i)^n (c^i\bar\phi_i)^m\,.
\eea
Putting these two expressions together, we find
\bea
T_1+T_2&=&2\sum_{m,n,k=0}^\infty
\frac{(-1)^{m+n+k}(m+n+k+1)!}{m!n!k!}
(c^i\bar\phi_i)^{m}(\bar c_i\phi^i)^n(\bar\phi_i\phi^i)^{k}
\nn\\&
=&\frac2{(1+c^i\bar\phi_i+\bar c_i \phi^i+\phi^i\bar\phi_i)^2}
=\frac2{(\varphi^i\bar\varphi_i)^2}\,.
\eea
As a result, the $F^4/X^4$ term in the effective action reads
\be
\Gamma_{F^4/X^4}=\alpha\int d^4x\frac{F^2\bar
F^2}{(\varphi^i\bar\varphi_i)^2}\,. \label{F4X}
\ee
This expression is explicitly scale and $U(3)$ invariant, as
expected.

It is a highly non-trivial and remarkable phenomenon that the vev constants $c^i$ and the
shifted scalars $\phi^i$ have combined into the initial scalar
fields $\varphi^i$, (\ref{vev-shift}), after doing the Grassmann and harmonic integrals which
is a rather involved procedure in its own.
This confirms the independence of the action (\ref{G}) of any particular choice of
the vacua, the fact that has been proved in the previous section.

Note that (\ref{F4X}) also respects hidden $SO(6)\simeq SU(4)$ invariance, with the $SU(4)/U(3)$ transformations acting as
\be
\delta \varphi^i = \varepsilon^{ikl}\lambda_k\bar\varphi_l\,, \quad
\delta \bar\varphi_i = \varepsilon_{ikl}\bar\lambda^k\varphi^l\,,
\ee
where $\lambda_i$ comprise 6 corresponding group parameters. This is an indication that the superfield effective action (\ref{G}), besides
the superconformal $SU(2,2|3)$ symmetry, enjoys on shell the $SU(4)$ symmetry, and hence, the superconformal $PSU(2,2|4)$ symmetry
as a closure of these two symmetries.

\subsubsection{Wess-Zumino term}

To single out  the Wess-Zumino term, it is enough to keep only scalar fields in the superfields
(\ref{W-comp-N3}),
\bea
\hat\omega_{23}&=&\phi^1
+i\theta_2^\alpha\bar\theta^{2\dot\alpha}\partial_{\alpha\dot\alpha}\phi^1
-2i\theta_2^\alpha\bar\theta^{1\dot\alpha}\partial_{\alpha\dot\alpha}\phi^2
-2i\theta_3^\alpha\bar\theta^{1\dot\alpha}\partial_{\alpha\dot\alpha}\phi^3
+2\theta_2^\alpha\theta_3^\beta\bar\theta^{1\dot\alpha}\bar\theta^{2\dot\beta}
\partial_{\alpha\dot\alpha}\partial_{\beta\dot\beta}\phi^3\,,
\nn\\
\hat{\bar \omega}{}^{12}&=&\bar\phi_3
-i\theta_2^\alpha\bar\theta^{2\dot\alpha}\partial_{\alpha\dot\alpha}\bar\phi_3
+2i\theta_3^\alpha\bar\theta^{1\dot\alpha}\partial_{\alpha\dot\alpha}\bar\phi_1
+2i\theta_3^\alpha\bar\theta^{2\dot\alpha}\partial_{\alpha\dot\alpha}\bar\phi_2
+2\bar\theta^{1\dot\alpha}\bar\theta^{2\dot\beta}
  \theta_2^\alpha\theta_3^\beta
  \partial_{\alpha\dot\alpha}\partial_{\beta\dot\beta}\bar\varphi_1
  \,.\nn\\&&
\eea
We substitute these superfields into the action (\ref{G}) and
integrate there over the Grassmann variables, keeping only those terms which contain four derivatives contracted with
the antisymmetric $\varepsilon$-symbol,
\bea
\Gamma_{\rm WZ}&=&-\frac{i\alpha}{8}\varepsilon^{mnpq}\int d^4x du
[\partial_m\phi^2 \partial_n\bar\phi_3\partial_p\bar\phi_2\partial_q\phi^3
+\partial_m\bar\phi_2\partial_n\bar\phi_1\partial_p\phi^2\partial_q\phi^1]
\nn\\&&\times
\sum_{i,j=0}^\infty (-1)^{i+j}\frac{(i+j+2)!(i+1)(j+1)}{i!j!}
(c^3\bar\phi_3)^i(\bar c_1\phi^1)^j
\,.
\label{WZ-series}
\eea
To compare this expression with the standard expression (\ref{WZterm}) for Wess-Zumino term,\footnote{To be precise, we compare
(\ref{WZ-series}) with the Wess-Zumino action in the
four-dimensional form (\ref{SWZ0}).} it is necessary to compute the harmonic integrals and to
sum up the series. Unfortunately, it is very difficult to find the
explicit expression for the integral
\be
\int du\,
u^1_{i_1} \bar u_1^{i'_1}\ldots
u^1_{i_n} \bar u_1^{i'_n}
u^3_{j_1} \bar u_3^{j'_1}\ldots
u^3_{j_m} \bar u_3^{j'_m}u^2_k \bar u_2^{k'}
\ee
in terms of (anti)symmetrized irreducible combinations of
the delta-symbols. Therefore, here we restrict our consideration only to the lowest terms in (\ref{WZ-series}), namely,
\be
\Gamma_{\rm WZ}=\frac32i\alpha\varepsilon^{mnpq}\int d^4x du
[\partial_m\phi^2 \partial_n\bar\phi_3\partial_p\bar\phi_2\partial_q\phi^3
+\partial_m\bar\phi_2\partial_n\bar\phi_1\partial_p\phi^2\partial_q\phi^1]
(c^3\bar\phi_3+\bar c_1\phi^1)
+O(\phi^6)
\,.
\label{WZ-leading}
\ee
The corresponding harmonic integral is quite easy to do,
\be
\int du\,u^1_iu^2_ju^3_k \bar u_1^{i'} \bar u_2^{j'} \bar u_3^{k'}
=\frac1{36}\varepsilon_{ijk}\varepsilon^{i'j'k'}
+\frac1{60}\delta_i^{(i'}\delta_j^{j'}\delta_k^{k')}
+\frac1{18}\delta_i^{(i'}\delta_j^{[j')}\delta_k^{k']}
+\frac1{18}\delta_i^{[i'}\delta_j^{(j']}\delta_k^{k')}
\,.
\label{25}
\ee
Then it is straightforward to see that only the first term in the r.h.s. of (\ref{25}) contributes to (\ref{WZ-leading}),
while all other terms  either vanish after contracting the indices, or form total derivatives. As the result,
eq.\ (\ref{WZ-leading}) can be rewritten as
\bea
\Gamma_{\rm WZ}&=&\frac{i\alpha}{24}\varepsilon^{mnpq}\int d^4x\,
\varepsilon_{ijk}\varepsilon^{i'j'k'}
[c^i\partial_m\phi^j \partial_n \phi^k
\bar\phi_{i'} \partial_p\bar\phi_{j'}\partial_q\bar\phi_{k'}
\nn\\&&\qquad\qquad
-\,\phi^i \partial_m\phi^j\partial_n\phi^k
\bar c_{i'}\partial_p\bar \phi_{j'}\partial_q\bar\phi_{k'}]
+O(\phi^6)\,.
\label{label1}
\eea

To compare (\ref{label1}) with (\ref{SWZ0}), we represent the latter as a series expansion over the vevs
\bea
\label{21}
\Gamma_{\rm WZ}&=&\frac{i}{16\pi^2}
\varepsilon^{mnpq}\varepsilon_{ijk}\varepsilon^{i'j'k'}
\int d^4x
(c^i+\phi^i) \partial_m \phi^j \partial_n \phi^k
(\bar c_{i'} +\bar \phi_{i'}) \partial_p \bar \phi_{j'} \partial_q \bar \phi_{k'}
\\&&\times
\sum_{l=1}^\infty \frac{(-1)^{l+1}}{l}[(\phi^i\bar c_i)^l-(c^i\bar \phi_i)^l]
\frac12\sum_{m,n,k=0}^\infty \frac{(m+n+k+2)!}{m!n!k!}
(c^i\bar\phi_i)^m(\bar c_i\phi^i)^n (\phi^i\bar\phi_i)^k\,.
\nn
\eea
Here the fields $\phi^i$ are related with $\varphi^i$ as in
(\ref{vev-shift}) and we assumed that
$c^i \bar c_i=1\,$.

Let us single out, in the series (\ref{21}), the terms with minimal numbers of fields $\phi^i$ and $\bar\phi_i$. These terms
correspond to the choice $m=n=0$ and $l=1$ in the second line in (\ref{21})
\be
\Gamma_{\rm WZ}=\frac{i}{16\pi^2}
\varepsilon^{mnpq}\varepsilon_{ijk}\varepsilon^{i'j'k'}
\int d^4x
(\phi^l\bar c_l-c^l\bar \phi_l)
c^i \partial_m \phi^j \partial_n \phi^k
\bar c_{i'}  \partial_p \bar \phi_{j'} \partial_q \bar
\phi_{k'}+O(\phi^6)\,.
\label{22}
\ee

Up to total derivatives, the following identity holds
\bea
&&\varepsilon^{mnpq}\varepsilon_{ijk}\varepsilon^{i'j'k'}
(\phi^l\bar c_l-c^l\bar \phi_l)
c^i \partial_m \phi^j \partial_n \phi^k
\bar c_{i'}  \partial_p \bar \phi_{j'} \partial_q \bar
\phi_{k'}
\\&=&
\frac13\varepsilon^{mnpq}\varepsilon_{ijk}\varepsilon^{i'j'k'}
(\phi^i \partial_m \phi^j \partial_n \phi^k
\bar c_{i'}  \partial_p \bar \phi_{j'} \partial_q \bar
\phi_{k'}
-c^i \partial_m \phi^j \partial_n \phi^k
\bar \phi_{i'}  \partial_p \bar \phi_{j'} \partial_q \bar
\phi_{k'})
+\mbox{tot. deriv.}\nn
\eea
This identity allows us to bring the action (\ref{22}) to the form
\bea
\Gamma_{\rm WZ}&=&\frac{i}{48\pi^2}
\varepsilon^{mnpq}\varepsilon_{ijk}\varepsilon^{i'j'k'}
\int d^4x(
\phi^i \partial_m \phi^j \partial_n \phi^k
\bar c_{i'}  \partial_p \bar \phi_{j'} \partial_q \bar
\phi_{k'}
\nn\\&&\qquad\qquad
-c^i \partial_m \phi^j \partial_n \phi^k
\bar \phi_{i'}  \partial_p \bar \phi_{j'} \partial_q \bar
\phi_{k'})+
O(\phi^6)\,.
\label{25_}
\eea
This expression coincides with (\ref{label1}) under the choice
\be
\alpha=-\frac1{2\pi^2}\,.
\ee
This proves that the action (\ref{G}) contains the Wess-Zumino
term (\ref{SWZ0}).

\section[Low-energy effective action in $\cN=4$ $USp(4)$ harmonic superspace]{Low-energy effective action in $\cN=4$ $USp(4)$
\break harmonic superspace}
The $USp(4)$ harmonic variables were introduced for the first time in \cite{IKNO}.
Later they were used in \cite{BLS08} to formulate a superparticle model in $\cN=4$ harmonic superspace%
\footnote{Note that the relativistic particle models in the $\cN=2$ and $\cN=3$ harmonic superspaces were studied in \cite{ASB,ABS} and \cite{BS08}, respectively.}
and to study the $\cN=4$ SYM theory with central charge \cite{BP13}. The underlying harmonic superspace proved very efficient
for the construction of the $\cN=4$ SYM low-energy effective action, as was shown in \cite{BISZ}. In this section we review the basic results of the latter work.

\subsection{$\cN=4$ $USp(4)$ harmonic superspace}

The standard $\cN=4$ superspace is parametrized by the coordinates (\ref{zM}), where the indices $i,j=1,2,3,4$ correspond
to the $SU(4)$ R-symmetry group. The covariant spinor derivatives $D^i_\alpha$ and $\bar D_{i\dot\alpha}$ in this superspace
have the form (\ref{D-explicit}) and obey the commutation relations (\ref{D-algebra}). The basic idea of the
$USp(4)$ harmonic superspace is to abandon the manifest $SU(4)$ symmetry and keep only the explicit invariance under $USp(4)\subset SU(4)$.
Then, we extend the standard $\cN=4$ superspace by the harmonic coordinates $u^I{}_i=(u^1{}_i,u^2{}_i,u^3{}_i,u^4{}_i)$
which form the $USp(4)$ matrices
\be
u u^\dag ={\mathbbm 1}_4\,,\qquad
u\Omega u^{\rm T} = \Omega\,.
\label{USp4-u-def}
\ee
Here $\Omega$ is a constant antisymmetric matrix, $\Omega^{\rm T}=-\Omega$. The canonical choice of this matrix is
\be
\Omega = \left(
\begin{array}{cccc}
0 & 1 & 0 & 0\\
-1& 0 & 0 & 0\\
0 & 0 & 0 & 1\\
0 & 0 & -1& 0
\end{array}
\right),
\ee
though other forms are also possible. Being an invariant tensor of the group $USp(4)\,$, $\Omega^{ik}$
can be used to raise and lower the $USp(4)$ indices, e.g.,
\be
u^{Ii} = \Omega^{ij} u^I{}_j\,,\qquad
u^I{}_i = \Omega_{ij} u^{Ij}\,,
\ee
where $\Omega^{ij}$ is the inverse of $\Omega_{ij}$,
\be
\Omega^{ij}\Omega_{jk} = \delta^i_k\,.
\ee

The  group $USp(4)$ contains two independent $U(1)$ subgroups. These subgroups can be chosen
in such a way that the harmonic variables have the following $U(1)$ charge assignment
\be
u^1{}_i = u^{(+,0)}_i\,,\quad
u^2{}_i = u^{(-,0)}_i\,,\quad
u^3{}_i = u^{(0,+)}_i\,,\quad
u^4{}_i = u^{(0,-)}_i\,.
\label{u-charges}
\ee
With these notations, the defining harmonic constraints  (\ref{USp4-u-def}) take the form of the orthogonality conditions
\bea
&&u^{(+,0)i}u^{(-,0)}_i=u^{(0,+)i}u^{(0,-)}_i=1\,,\nn\\
&&u^{(+,0)}_i u^{(0,+)i}=u^{(+,0)}_i u^{(0,-)i}=
u^{(0,+)}_i u^{(-,0)i}=u^{(-,0)}_i u^{(0,-)i}=0\,,
\eea
and the completeness relations
\be
u^{(+,0)}_i u^{(-,0)}_j-u^{(+,0)}_j u^{(-,0)}_i+
u^{(0,+)}_i u^{(0,-)}_j-u^{(0,+)}_j u^{(0,-)}_i=\Omega_{ij}\,.
\ee
Thereby the harmonics can be used to define the $U(1)\times U(1)$ projections of all objects with $USp(4)$ indices. In particular,
for Grassmann coordinates $\theta_{i\alpha}$, $\bar\theta^i_{\dot\alpha}$ and covariant spinor derivatives $D^i_\alpha$, $\bar D_{i\dot\alpha}$
we have
\be
\theta^I_\alpha=-u^{I i}\theta_{i\alpha}\,,\quad
\bar\theta^I_{\dot\alpha}=u^I{}_i\bar\theta^i_{\dot\alpha}\,,\quad
D^I_\alpha=u^I_i D^i_\alpha\,,\quad
\bar D^I_{\dot\alpha}=-u^{I i}\bar D_{i\dot\alpha}\,.
\ee
Among the anticommutators of the derivatives $D^I_\alpha$ and $\bar D^I_{\dot\alpha}$, only the following ones are non-trivial
\bea
\{D^{(+,0)}_\alpha , \bar D^{(-,0)}_{\dot\alpha} \}
 = - \{ D^{(-,0)}_\alpha , \bar D^{(+,0)}_{\dot\alpha} \}
  &=& 2i \sigma^m_{\alpha\dot\alpha} \partial_m\,,\nn\\
\{ D^{(0,+)}_\alpha , \bar D^{(0,-)}_{\dot\alpha} \}
 = - \{ D^{(0,-)}_\alpha, \bar D^{(0,+)}_{\dot\alpha} \}
 &=& 2i \sigma^m_{\alpha\dot\alpha} \partial_m\,.
\eea

Associated with the harmonic variables are the $USp(4)$-covariant harmonic derivatives defined as
\begin{align}
&D^{(\pm\pm,0)}=u^{(\pm,0)}_i\dfrac{\partial}{\partial
u^{(\mp,0)}_i}\,,&&
D^{(0,\pm\pm)}=u^{(0,\pm)}_i\dfrac{\partial}{\partial
u^{(0,\mp)}_i}\,,\nn\\
&D^{(\pm,\pm)}=u^{(\pm,0)}_i\dfrac\partial{\partial u^{(0,\mp)}_i}
+u^{(0,\pm)}_i\dfrac\partial{\partial u^{(\mp,0)}_i}\,,&&
D^{(\pm,\mp)}=u^{(\pm,0)}_i\dfrac\partial{\partial u^{(0,\pm)}_i}
-u^{(0,\mp)}_i\dfrac\partial{\partial u^{(\mp,0)}_i}\,,\nn\\
&S_1=u^{(+,0)}_i\dfrac\partial{\partial u^{(+,0)}_i}
 -u^{(-,0)}_i\dfrac\partial{\partial u^{(-,0)}_i}\,,&&
S_2=u^{(0,+)}_i\dfrac\partial{\partial u^{(0,+)}_i}
 -u^{(0,-)}_i\dfrac\partial{\partial u^{(0,-)}_i}\,.
\label{D-harm-usp}
\end{align}
It is easy to check that they obey the commutation relations of the  Lie algebra $usp(4)$. In particular,
the operators $S_1$ and $S_2$ are the generators of the two $U(1)$ subgroups, and they count the corresponding harmonic $U(1)$ charges
\be
[S_1,D^{(s_1,s_2)}]=s_1 D^{(s_1,s_2)}\,,\qquad
[S_2,D^{(s_1,s_2)}]=s_2 D^{(s_1,s_2)}\,,\qquad
[S_1,S_2]=0 \ .
\label{u1u1}
\ee
They appear on the right-hand sides of the appropriate commutators
\be
\label{su2su2}
[D^{(++,0)},D^{(--,0)}]=S_1\,, \quad
[D^{(0,++)},D^{(0,--)}]=S_2\,.
\ee
It is also easy to check that any operators from the set $\{D^{(++,0)} , D^{(--,0)}, S_1\}$ commute with those
from the set $\{D^{(0,++)} , D^{(0,--)}, S_2\}\,$. Thus, these sets form two independent mutually commuting $su(2)$ subalgebras
in the full $usp(4)$ algebra  of the harmonic derivatives.

The harmonic variables and the matrix $\Omega$ reveal the following complex conjugation properties
\be
\overline{(u_i^{(\pm,0)})}=\mp u^{(\mp,0)i}\,, \quad
\overline{(u_i^{(0,\pm)})}=\mp u^{(0,\mp)i}\,, \quad
\overline{(\Omega_{i j})}=-\Omega^{i j} \,.
\ee
As was already mentioned earlier, the conventional complex conjugation is not too useful in the harmonic superspace, since
it does not allow to ensure reality for the analytic  subspaces of the full superspace. In the harmonic superspace approach,
it is customary to use the generalized $\,\widetilde{\phantom{m}}\,$-conjugation
which, in the present case,  is defined to act on the harmonics by the rules
\be
\label{tilde-u-usp}
\widetilde{u^{(\pm,0)}_i}=u^{(0,\pm)i}\,,\quad
\widetilde{u^{(0,\pm)}_i}=u^{(\pm,0)i}\,,\quad
\widetilde{u^{(\pm,0)i}}=-u^{(0,\pm)}_i\,,\quad
\widetilde{u^{(0,\pm)i}}=-u^{(\pm,0)}_i\,.
\ee
The transformations of the Grassmann variables and covariant spinor derivatives under this conjugation read
\bea
&&\widetilde{\theta^{(\pm,0)}_\alpha}=\bar\theta^{(0,\pm)}_{\dot\alpha}\,,
\quad
\widetilde{\theta^{(0,\pm)}_\alpha}=\bar\theta^{(\pm,0)}_{\dot\alpha}\,,
\quad
\widetilde{\bar\theta^{(0,\pm)}_{\dot\alpha}}=-\theta^{(\pm,0)}_\alpha\,,
\quad
\widetilde{\bar\theta^{(\pm,0)}_{\dot\alpha}}=-\theta^{(0,\pm)}_\alpha\,,
\nn\\
&&\widetilde{D^{(\pm,0)}_\alpha}=-\bar
D^{(0,\pm)}_{\dot\alpha}\,,\quad
\widetilde{D^{(0,\pm)}_\alpha}=-\bar
D^{(\pm,0)}_{\dot\alpha}\,,\quad
\widetilde{\bar
D^{(\pm,0)}_{\dot\alpha}}=D^{(0,\pm)}_\alpha\,,\quad
\widetilde{\bar D^{(0,\pm)}_{\dot\alpha}}=D^{(\pm,0)}_\alpha\,.~~~~~~
\label{tilde-D-usp}
\eea

$\cN=4$ $USp(4)$ harmonic superspace with the coordinates $\{x^m,\theta^I_{\alpha},\bar\theta^I_{\dot\alpha},u^I{}_i\}$
contains several analytic subspaces with 8 (out of the total 16) Grassmann coordinates. One of these subspaces is parametrized
by the set of coordinates
\be
\{\zeta,u\}=\{(x^m_{A},\theta^{(+,0)}_\alpha,\theta^{(-,0)}_\alpha,
\bar\theta^{(0,+)}_{\dot\alpha},\bar\theta^{(0,-)}_{\dot\alpha}),u^I{}_i\}\,,
\label{ancor}
\ee
where
\be
x^m_{A}=x^m-i\theta^{(0,-)}\sigma^m\bar\theta^{(0,+)}
+i\theta^{(0,+)}\sigma^m \bar\theta^{(0,-)}
-i\theta^{(+,0)}\sigma^m\bar\theta^{(-,0)}
+i\theta^{(-,0)}\sigma^m \bar\theta^{(+,0)}\,.
\label{N4anal}
\ee
In the analytic basis $(\zeta, u, \theta^{(0,\mp)\alpha}, \bar\theta^{(\mp,0)\dot\alpha})$, the following Grassmann derivatives become short,
\be
D^{(0,\pm)}_{\alpha}
=\pm\frac\partial{\partial\theta^{(0,\mp)\alpha}}\,,\qquad
\bar D^{(\pm,0)}_{\dot\alpha}=\pm\frac\partial{
\partial\bar\theta^{(\mp,0)\dot\alpha}}\ .
\label{Dshort}
\ee

The harmonic derivatives (\ref{D-harm-usp}) in the analytic basis acquire the form
\bea
D^{(\pm,\pm)}&=&u^{(\pm,0)}_i\frac\partial{\partial u^{(0,\mp)}_i}
+u^{(0,\pm)}_i\frac\partial{\partial u^{(\mp,0)}_i}
\pm2i(\theta^{(0,\pm)}\sigma^m \bar\theta^{(\pm,0)}
-\theta^{(\pm,0)}\sigma^m \bar\theta^{(0,\pm)})\frac\partial{\partial x_A^m}
\nn\\&& + \theta^{(0,\pm)}_\alpha \frac\partial{\partial\theta^{(\mp,0)}_\alpha}
+ \theta^{(\pm,0)}_\alpha \frac\partial{\partial\theta^{(0,\mp)}_\alpha}
+\bar\theta^{(0,\pm)}_{\dot\alpha} \frac\partial{\partial\bar\theta^{(\mp,0)}_{\dot\alpha}}
+ \bar\theta^{(\pm,0)}_{\dot\alpha} \frac\partial{\partial\bar\theta^{(0,\mp)}_{\dot\alpha}}\,,\nn\\
D^{(\pm\pm,0)}&=&u^{(\pm,0)}_i\frac{\partial}{\partial
u^{(\mp,0)}_i}+ \theta^{(\pm,0)}_\alpha \frac\partial{\partial \theta^{(\mp,0)}_\alpha}
+ \bar\theta^{(\pm,0)}_{\dot\alpha} \frac\partial{\partial \bar\theta^{(\mp,0)}_{\dot\alpha}}\,,\nn\\
D^{(0,\pm\pm)}&=&u^{(0,\pm)}_i\frac{\partial}{\partial
u^{(0,\mp)}_i}+ \theta^{(0,\pm)}_\alpha \frac\partial{\partial \theta^{(0,\mp)}_\alpha}
+ \bar\theta^{(0,\pm)}_{\dot\alpha} \frac\partial{\partial \bar\theta^{(0,\mp)}_{\dot\alpha}}\,,\nn\\
D^{(\pm,\mp)}&=&u^{(\pm,0)}_i\frac\partial{\partial u^{(0,\pm)}_i}
-u^{(0,\mp)}_i\frac\partial{\partial u^{(\mp,0)}_i}
\pm2i(\theta^{(\pm,0)}\sigma^m \bar\theta^{(0,\mp)}
 - \theta^{(0,\mp)}\sigma^m \bar\theta^{(\pm,0)})
 \frac\partial{\partial x^m_A}\nn\\
&&+\theta^{(\pm,0)}_\alpha \frac\partial{\partial\theta^{(0,\pm)}_\alpha}
+\bar\theta^{(\pm,0)}_{\dot\alpha} \frac\partial{\partial\bar\theta^{(0,\pm)}_{\dot\alpha}}
-\theta^{(0,\mp)}_\alpha \frac\partial{\partial\theta^{(\mp,0)}_\alpha}
-\bar\theta^{(0,\mp)}_{\dot\alpha} \frac\partial{\partial\bar\theta^{(\mp,0)}_{\dot\alpha}}\,.
\label{6.19}
\eea
It is interesting to note that the operators $D^{(\pm\pm,0)}$ and $D^{(0,\pm\pm)}$ in the analytic basis
do not involve terms with the $x_A^m$ derivatives.

Note also that the analytic subspace (\ref{ancor}) is closed under the $\,\widetilde{\phantom{m}}\,$-conjugation
defined in (\ref{tilde-u-usp}) and (\ref{tilde-D-usp}).

\subsection{$\cN=4$ SYM constraints in the $USp(4)$
harmonic superspace}

Within the standard geometric approach, the gauge theory is introduced through adding gauge connections to the superspace derivatives,
as in eq.\ (\ref{A-connections}).
The $\cN=4$ SYM constraints have the same form as in the $\cN=3$ case (\ref{N3constraints}), but the indices $i,j$ take now the values $1,2,3,4$.
In the abelian case, these constraints imply the following Bianchi identities
\begin{subequations}
\label{N4-constraints}
\bea
&&D_\alpha^i W^{j k}+D_\alpha^j W^{i k}=0\,,\\
&&\bar D_{i\dot\alpha}W^{j k}=\frac{1}{3}(\delta_i^j\bar D_{l\dot\alpha}W^{l k}-\delta_i^k\bar D_{l\dot\alpha}W^{l j})\,.
\label{N4-bianchi}
\eea
\end{subequations}
Besides this, the $\cN=4$ superfield strengths $W^{ij}=-W^{ji}$ should be subject to the reality condition
which is a superfield counterpart of (\ref{phi-reality}):
\be
\overline{W^{i j}} = \bar W_{ij} =\frac12\varepsilon_{i j k l}W^{k l}\,.
\label{N4-reality}
\ee

The constraints (\ref{N4-constraints}) and (\ref{N4-reality}) can be rewritten in $\cN=4$ harmonic superspaces based on different cosets of the $SU(4)$ group \cite{Bandos88,HH94,AFSZ}. The aim of the present subsection is to rewrite them in the $USp(4)$ harmonic superspace introduced in the previous subsection.

Given the $\cN=4$ superfield strength $W^{ij}\,$, we can project it on the harmonics:
\be
W^{IJ} = u^I{}_i u^J{}_j W^{ij}\,.
\ee
Recall that the harmonic variables have the $U(1)$-charge assignment indicated in eq.\ (\ref{u-charges}).
Then, the corresponding charges of $W^{IJ}$ are
\bea
&& W^{12} = W \,,\quad
W^{13} = W^{(+,+)}\,,\quad
W^{14} = W^{(+,-)}\,,\nn\\
&& W^{23} = W^{(-,+)}\,,\quad
W^{24} = W^{(-,-)}\,,\quad
W^{34} = {\cal W}\,,
\label{W-charges}
\eea
where $W$ and $\cal W$ are two different uncharged projections
\be
S_1 W = S_2 W =0\,,\qquad
S_1 {\cal W} = S_2 {\cal W} =0\,.
\ee

Let us examine the superfield ${\cal W}=u^{(0,+)}_i u^{(0,-)}_j W^{ij}$. By construction, this superfield obeys
the following equations with the harmonic derivatives (\ref{D-harm-usp})
\begin{subequations}
\label{W-short-usp}
\bea
&&D^{(++,0)}{\cal W}=D^{(--,0)}{\cal W}=D^{(0,++)}{\cal W}=D^{(0,--)}{\cal W}=0\,,\label{6.25a}\\
&&(D^{(+,+)})^2{\cal W}=0\,.
\label{6.25b}
\eea
\end{subequations}
The equations (\ref{N4-bianchi}) imply certain analyticity properties for $\cal W$
\be
D^{(0,+)}_\alpha{\cal W}=D^{(0,-)}_\alpha{\cal W}
=\bar D^{(+,0)}_{\dot\alpha}{\cal W}
=\bar D^{(-,0)}_{\dot\alpha}{\cal W}=0\,.
\label{W-analyt-usp}
\ee
Eq.\ (\ref{N4-reality}) means that $\cal W$ is real under the $\,\widetilde{\phantom{m}}\,$-conjugation (\ref{tilde-u-usp}):
\be
\widetilde{\cal W}={\cal W}\,.
\label{W-real-usp}
\ee
In a similar way one can find the equations for all other superfield strengths (\ref{W-charges}), see \cite{BLS08} for details.

It is instructive to consider the equations (\ref{W-short-usp}) and
(\ref{W-analyt-usp}) in the analytic basis. As follows from (\ref{Dshort}), the constraints (\ref{W-analyt-usp})
are automatically solved by an arbitrary real analytic $\cal W$
\be
{\cal W} = {\cal W}(\zeta,u)\,.
\ee
The equations (\ref{6.25a}) are not dynamical, since the harmonic derivatives $D^{(\pm\pm,0)}$ and $D^{(0,\pm\pm)}$ in the analytic coordinates
do not contain $\partial/{\partial x_A^m}$, see (\ref{6.19}). These equations serve to eliminate auxiliary fields in the component field
expansion of $\cal W$, but they do not impose any constraint on the physical components. Only eq.\ (\ref{6.25b}) is dynamical: It leads to
the standard free equations of motion for physical components in $\cal W$. The solution of the total set of equations
(\ref{W-short-usp})--(\ref{W-real-usp}) is given by the following component field expansions:
\begin{subequations}
\label{calW}
\bea
\cal W&=&  {\cal W}_{\rm bos}+{\cal W}_{\rm ferm}\,,\\
{\cal W}_{\rm bos} &=&\varphi+f^{ij}(u^{(+,0)}_{[i}u^{(-,0)}_{j]}-u^{(0,+)}_{[i}u^{(0,-)}_{j]})
\nn\\&& +\frac1{\sqrt2} (\theta^{(+,0)}_\alpha\theta^{(-,0)}_\beta
\sigma^{m\alpha}{}_{\dot\alpha}\sigma^{n\beta\dot\alpha}
-\bar\theta^{(0,+)}_{\dot\alpha}\bar\theta^{(0,-)}_{\dot\beta}
\sigma^{m\dot\alpha}{}_\alpha \sigma^{n\alpha\dot\beta})F_{mn}
\nn\\&& -4i\theta^{(+,0)}_\alpha\bar\theta^{(0,+)}_{\dot\alpha}
\partial^{\alpha\dot\alpha}f^{ij}u^{(-,0)}_{[i}u^{(0,-)}_{j]}
-4i\theta^{(-,0)}_\alpha\bar\theta^{(0,-)}_{\dot\alpha}
\partial^{\alpha\dot\alpha}f^{ij}u^{(+,0)}_{[i}u^{(0,+)}_{j]}
\nn\\&& +4i\theta^{(+,0)}_\alpha\bar\theta^{(0,-)}_{\dot\alpha}
\partial^{\alpha\dot\alpha}f^{ij}u^{(-,0)}_{[i}u^{(0,+)}_{j]}
+4i\theta^{(-,0)}_\alpha\bar\theta^{(0,+)}_{\dot\alpha}
\partial^{\alpha\dot\alpha}f^{ij}u^{(+,0)}_{[i}u^{(0,-)}_{j]}
\nn\\&& +4\theta^{(+,0)}_\alpha\theta^{(-,0)}_\beta
\bar\theta^{(0,+)}_{\dot\alpha}\bar\theta^{(0,-)}_{\dot\beta}
\partial^{\alpha\dot\alpha}\partial^{\beta\dot\beta}
[\varphi-f^{ij}(u^{(+,0)}_{[i}u^{(-,0)}_{j]}-u^{(0,+)}_{[i}u^{(0,-)}_{j]})] \,,\label{x15_}\\
{\cal W}_{\rm ferm}&=&
i\theta^{(+,0)\alpha}\psi^i_\alpha u^{(-,0)}_i
-i\theta^{(-,0)\alpha}\psi^i_\alpha u^{(+,0)}_i
+i\bar\theta^{(0,+)}_{\dot\alpha}\bar\psi^{i\dot\alpha} u^{(0,-)}_i
-i\bar\theta^{(0,-)}_{\dot\alpha}\bar\psi^{i\dot\alpha} u^{(0,+)}_i
\nn\\&&
-2\theta^{(+,0)\alpha}\theta^{(-,0)\beta}\bar\theta^{(0,+)\dot\alpha}
\partial_{(\alpha\dot\alpha}\psi^i_{\beta)} u^{(0,-)}_i
+2\theta^{(+,0)\alpha}\theta^{(-,0)\beta}\bar\theta^{(0,-)\dot\alpha}
\partial_{(\alpha\dot\alpha}\psi^i_{\beta)} u^{(0,+)}_i
\nn\\&&
-2\theta^{(+,0)\alpha}\bar\theta^{(0,+)\dot\beta}\bar\theta^{(0,-)\dot\alpha}
\partial_{\alpha(\dot\alpha}\bar\psi^i_{\dot\beta)} u^{(-,0)}_i
+2\theta^{(-,0)\alpha}\bar\theta^{(0,+)\dot\beta}\bar\theta^{(0,-)\dot\alpha}
\partial_{\alpha(\dot\alpha}\bar\psi^i_{\dot\beta)} u^{(+,0)}_i\,.
\nn\\
\eea
\end{subequations}
Here, the component fields satisfy the free equations of motion
\begin{align}
&\square \varphi =0&\quad&\mbox{1 real scalar},\nn\\
&\square f^{ij}=0,\quad (f^{ij}\Omega_{ij}=0)&&
\mbox{5 real scalars},\nn\\
&\sigma^{m\alpha}{}_{\dot\alpha}\partial_m\psi^i_\alpha=0,\quad
\sigma^m{}_\alpha{}^{\dot\alpha}\partial_m\bar\psi^i_{\dot\alpha}=0&&
\mbox{4 Weyl spinors},\nn\\
&\partial^m F_{mn}=0&& \mbox{1 Maxwell field} \,.
\end{align}
All component fields in (\ref{calW}) depend on $x^m_A$\
defined in  (\ref{N4anal}).
These fields are subject to the reality conditions
\be
\overline{\varphi} = \varphi\,,\quad
\overline{ f^{ij} } = \bar f_{ij}= f_{ij}
\,,\quad
\overline{\psi^i_\alpha} = \bar\psi_{i\dot\alpha}\,,\quad
\overline{F_{mn}} = F_{mn}\,.
\ee

Recall that the group $USp(4)$ is locally isomorphic to $SO(5)$. For computational reasons, it is useful to express ${\cal W}_{\rm bos}$
in terms of $SO(5)$ harmonic variables. Recall also that the representation ${\bf 5}$ of $USp(4)\simeq SO(5)$ is given by the antisymmetric
$\Omega$-traceless ${\bf 4}\times{\bf 4}$ matrix. The corresponding
Clebsch-Gordan coefficients are gamma matrices $\gamma_a^{i j}$, with
$a=1,2,3,4,5$ of $SO(5)$ and $i=1,2,3,4$ of $USp(4)$, such that
\bea
&& \gamma_a^{ij}=-\gamma_a^{j\,i}\,, \quad \Omega_{i j}\gamma_a^{i j}=0\,, \quad
\gamma_{a i j}\gamma_b^{j k}+\gamma_{b i j}\gamma_a^{j k}=2\delta_{ab}\delta_i^k\,, \quad
\overline{(\gamma_a^{i j})}=-\gamma_{a i j} \,, \nn\\[3pt]
&& \gamma_a^{ij}\gamma_{b\,ij}=-4\delta_{ab}\,,\quad
\gamma_{a i j}\gamma_a^{k l}=-2(\delta_i^k\delta_j^l-\delta_i^l\delta_j^k)
-\Omega_{ij}\Omega^{kl}\,. \label{gamma65}
\eea
Using the bilinear combinations of $USp(4)/[U(1)\times U(1)]$ harmonics appearing in (\ref{x15_}), we define
\bea
&&v^{(-,-)}_a = \gamma_a^{ij}u^{(-,0)}_{[i}u^{(0,-)}_{j]}, \quad
v^{(+,+)}_a = \gamma_a^{ij}u^{(+,0)}_{[i}u^{(0,+)}_{j]}
\,,\nn\\
&&v^{(-,+)}_a = \gamma_a^{ij}u^{(-,0)}_{[i}u^{(0,+)}_{j]}, \quad
v^{(+,-)}_a = \gamma_a^{ij}u^{(+,0)}_{[i}u^{(0,-)}_{j]}
\,, \nn\\
&&v^{(0,0)}_a = \gamma^{ij}_a(u_{[i}^{(+,0)}u^{(-,0)}_{j]}-u_{[i}^{(0,+)}u^{(0,-)}_{j]})
\,.
\label{so5u1u1h}
\eea
These objects have definite $U(1)\times U(1)$ charges, but  they do not form an $SO(5)$ matrix on their own because their non-zero products are
\be
v_a^{(-,-)}v_a^{(+,+)}=-2\,, \quad
v_a^{(-,+)}v_a^{(+,-)}=+2\,, \quad
v_a^{(0,0)}v_a^{(0,0)}=-4 \,.
\ee
The correct definition of $SO(5)$ harmonics $v_b^a$ is provided by the formulas
\bea
&& v_a^1=\frac{1}{2}(v_a^{(-,-)}-v_a^{(+,+)})\,, \quad
v_a^2=\frac{i}{2}(v_a^{(-,-)}+v_a^{(+,+)}) \nn\\
&& v_a^3=\frac{i}{2}(v_a^{(-,+)}-v_a^{(+,-)})\,, \quad
v_a^4=\frac{1}{2}(v_a^{(-,+)}+v_a^{(+,-)}), \quad
v_a^5=-\frac{i}{2}v_a^{(0,0)} \,.
\label{cor-harm}
\eea
These harmonics are real, $\overline{(v_a^b)}=v_a^b$, and obey the needed $SO(5)$ relations
\be
v^a_c v^b_c=\delta^{ab}\,,\qquad
\varepsilon^{abcde}v^1_a v^2_b v^3_c v^4_d v^5_e=1\,.
\ee
The integration over $SO(5)$ harmonic variables is defined by
\be
\int dv\,1=1\,,\qquad \int dv\,(\mbox{non-singlet $SO(5)$ irrep})=0\ .
\label{SO5hint}
\ee
Two basic harmonic integrals are
\be
\int dv \, v^5_a v^5_b=\frac15\delta_{ab}\,,\qquad
\int dv\, v^1_a v^2_b v^3_c v^4_d v^5_e=\frac1{5!}\varepsilon_{abcde}\ .
\label{SO5hint1}
\ee
A small amount  of combinatorics yields the following generalization of
these integrals
\bea
\label{206}
\int dv\, v^5_{a_1}\ldots v^5_{a_{k}}
&=&\left\{
\begin{array}{ll}\displaystyle
\frac{3}{(2n+1)(2n+3)}\delta_{(a_1 a_2}\ldots
\delta_{a_{k-1}a_{k})}\,,\ \ & k=2n \nn\\
0\,, &k=2n+1
\end{array}
\right.
\nn\\
\int dv\, v^1_a v^2_b v^3_c v^4_d v^5_e v^5_{e_1}\ldots
v^5_{e_{k}}&=&
\left\{
\begin{array}{ll}\displaystyle
\frac{\varepsilon_{abcd(e}\delta_{e_1e_2}\ldots
\delta_{e_{k-1}e_{k})}}{8(5+2n)(2n+3)}\,,\ \ & k=2n \\
0&k=2n+1\,.
\end{array}
\right.
\eea

The gamma matrices defined in (\ref{gamma65}) can also be used to relate the scalars $f^{i j}$ to the $SO(5)$ vector $X_a$,
\be
f^{ij}=\frac12\gamma^{ij}_a X_a\,,\quad
X_a=\gamma_{aij}f^{ij}\,,\quad
f^{ij}f_{ij}=-X_a X_a\,.
\ee
The sixth scalar $\varphi$ is $SO(5)$ singlet,  $\varphi=X_6\,$.

Taking into account the above redefinition of the scalars, we rewrite the bosonic part of the superfield strength (\ref{x15_})
in terms of $SO(5)$ harmonic variables as
\bea
{\cal W}_{\rm bos}&=&\varphi+i X_a v^5_a
+\frac1{\sqrt2} (\theta^{(+,0)}_\alpha\theta^{(-,0)}_\beta
\sigma^{m\alpha}{}_{\dot\alpha}\sigma^{n\beta\dot\alpha}
-\bar\theta^{(0,+)}_{\dot\alpha}\bar\theta^{(0,-)}_{\dot\beta}
\sigma^{m\dot\alpha}{}_\alpha \sigma^{n\alpha\dot\beta})F_{mn}
\nn\\
&& -2i\theta^{(+,0)}_\alpha\bar\theta^{(0,+)}_{\dot\alpha}
\partial^{\alpha\dot\alpha}X_a (v^1_a-iv^2_a)
+2i\theta^{(-,0)}_\alpha\bar\theta^{(0,-)}_{\dot\alpha}
\partial^{\alpha\dot\alpha}X_a (v^1_a+iv^2_a)
\nn\\
&& +2i\theta^{(+,0)}_\alpha\bar\theta^{(0,-)}_{\dot\alpha}
\partial^{\alpha\dot\alpha}X_a (v^4_a-iv^3_a)
+2i\theta^{(-,0)}_\alpha\bar\theta^{(0,+)}_{\dot\alpha}
\partial^{\alpha\dot\alpha}X_a (v^4_a+iv^3_a)
\nn\\
&& +4\theta^{(+,0)}_\alpha\theta^{(-,0)}_\beta
\bar\theta^{(0,+)}_{\dot\alpha}\bar\theta^{(0,-)}_{\dot\beta}
\partial^{\alpha\dot\alpha}\partial^{\beta\dot\beta}
[\varphi-iX_a v^5_a]\,.
\label{x15.2}
\eea
We will use this form of the superfield strength in subsection \ref{subsect6.4} for studying the bosonic component structure of the
low-energy effective action in $\cN=4$ SYM theory.

\subsection{Scale invariant low-energy effective action}
\label{sect6.3}

In general, the four-derivative part of the $\cN=4$ SYM low-energy effective action can be represented
by the following functional in the $\cN=4$ superspace
\be
\Gamma=\int d\zeta du\, {\cal H}({\cal W})\,,
\label{GH4_}
\ee
where $d\zeta$ is the measure of integration over the analytic subspace with the coordinates (\ref{ancor}).
We assume that this measure is defined so that
\be
d\zeta = d^4x_A d^8\theta\,,\qquad
\int d^8\theta (\theta^{(+,0)})^2(\theta^{(-,0)})^2
(\bar\theta^{(0,+)})^2 (\bar\theta^{(0,-)})^2=1\,,
\label{6.43}
\ee
and the integration over the harmonic variables is defined by the standard rules
\be
\int du\,1=1\,,\qquad
\int du(\mbox{non-singlet $USp(4)$ irreducible representation})=0\,.
\ee
As is seen from (\ref{6.43}), the analytic measure is uncharged and dimensionless. Effectively, it contains  eight covariant spinor derivatives
which produce four space-time derivatives on the component fields. Hence, all the space-time derivatives in (\ref{GH4_}) are already hidden
in the superspace measure and the function ${\cal H}({\cal W})$ should contain neither space-time, nor covariant spinor derivatives
of the superfield strength $\cal W$.
This is very similar to the situation with the effective action in the $\cN=2$ and $\cN=3$ harmonic superspaces
considered in the previous sections.

Now we implement the requirement of scale invariance of the effective action $\Gamma$. The function ${\cal H}({\cal W})$
should be dimensionless, since the analytic measure (\ref{6.43}) has the dimension zero, but the superfield strength $\cal W$
has the dimension one. Thus, we are led to
introduce a parameter $\Lambda$, such that ${\cal W}/\Lambda$ is
dimensionless, and to choose
\be
{\cal H}({\cal W},\Lambda)={\cal H}({\cal W}/\Lambda)\,.
\ee
Since the dependence on $\Lambda$ must disappear upon doing the integration over
superspace, the function ${\cal H}$ is uniquely determined  to be
\be
{\cal H}=c\ln\frac{\cal W}{\Lambda}\,,
\label{log}
\ee
where $c$ is some constant coefficient. Rescaling $\cal W$ amounts to shifting $\cal H$ by a constant,
which yields zero under the $d\zeta$ integral.

We conclude that the four-derivative part of the SYM effective
action on the Coulomb branch in $\cN=4$ $USp(4)$ harmonic superspace has
the following simple unique form
\be
\Gamma=c\int d\zeta du\, \ln\frac{\cal W}\Lambda\,.
\label{G4}
\ee
We will show that this action contains the $F^4/X^4$ term (\ref{F4X4}), as well as the Wess-Zumino term (\ref{WZso5}).
This will allow us to fix the coefficient $c$.

\subsection{Component structure}
\label{subsect6.4}
\subsubsection{$F^4/X^4$ term}
In order to identify the $F^4/X^4$ term (\ref{F4X4})
it is sufficient to consider the bosonic part of the superfield strength, ${\cal W}_{\rm bos}$ (\ref{x15_}).
Recall that it can be rewritten through
the $SO(5)$ harmonic variables,
see (\ref{x15.2}). Hence, for deriving the $F^4/X^4$ term we substitute (\ref{x15.2}) into (\ref{G4})
and replace the integration measure $du$ by $dv\,$,
\be
\Gamma_{F^4/X^4} = c \int d\zeta dv \ln\frac{{\cal W}_{\rm bos}}\Lambda\,.
\label{S4-v}
\ee
Moreover, it suffices to consider ${\cal W}_{\rm bos}$ with \emph{constant} scalar fields
$\varphi$ and $X_a$. Then only the first line in (\ref{x15.2}) survives.
Substituting this simplified expression for ${\cal W}_{\rm bos}$ into the action (\ref{S4-v})
and integrating there over $\theta$'s by the rule (\ref{6.43}), we find
\be
\Gamma_{F^4/X^4}=\frac14\int d^4x dv\,{\cal H}^{(4)}(\varphi+i X_a v^5_a)
\Big[F_{mn}F^{nk}F_{kl}F^{lm}-\frac14(F_{pq}F^{pq})^2\Big],
\label{SS4_}
\ee
where ${\cal H}^{(n)}$ stands for the $n$'th derivative of ${\cal H}$ with respect to its argument.
To compute the harmonic integral, we expand ${\cal H}^{(4)}$ in the Taylor series,
\be
{\cal H}^{(4)}(\varphi+i X_a v^5_a)=\sum_{n=0}^\infty \frac1{n!} {\cal H}^{(4+n)}(\varphi)
(i X_a v^5_a)^n\,.
\label{ser}
\ee
Applying (\ref{206}) to each term in this series, we obtain
\be
\Gamma_{F^4/X^4}=\frac14\int d^4x\,
\Big[F_{mn}F^{nk}F_{kl}F^{lm}-\frac14(F_{pq}F^{pq})^2\Big]
\sum_{n=0}^\infty \frac{3 (-X_a X_a)^n}{(2n+1)!(2n+3)}
{\cal H}^{(4+2n)}(\varphi)\,.
\label{S4.1}
\ee
For the function $\cal H$ defined in (\ref{log}), we obtain
\be
{\cal H}^{(n)}(\varphi)=c\frac{(-1)^{n-1}(n-1)!}{\varphi^n}\,.
\label{Hderiv}
\ee
Substituting this expression  into (\ref{S4.1}) and summing up the series, we find
\be
\Gamma_{F^4/X^4}=-\frac32 c\int d^4x\,
\frac{F_{mn}F^{nk}F_{kl}F^{lm}-\frac14(F_{pq}F^{pq})^2}
{(\varphi^2+X_a X_a)^2}\,.
\label{SF4}
\ee
This precisely matches with (\ref{F4X4}), provided that we identify $\varphi=X_6$ and set
\be
c=-\frac1{96\pi^2}\,.
\label{k}
\ee
Thus, the superfield action (\ref{G4}) contains the $F^4/X^4$ term (\ref{F4X4}).

\subsubsection{Wess-Zumino term}

Recall that the Wess-Zumino term (\ref{WZso5}) depends only on the scalar fields and their derivatives.
Hence, for singling out this term in the component field representation of (\ref{G4}) it is enough
to use the same superfield expression (\ref{S4-v}),
but in the superfield (\ref{x15.2}) we now need to keep only the scalar fields. Then, performing integration
over $\theta$'s by the rule (\ref{6.43}), we find
\bea
\Gamma &=&\int d^4xdv\,
{\cal H}^{(4)}(\varphi+i X_e v^5_e)
\partial^{\alpha\dot\alpha} X_a
\partial^{\beta\dot\beta}X_b
\partial_{\alpha\dot\beta}X_c
\partial_{\beta\dot\alpha}X_d\nn\\
&&\hspace{100pt}\times
(v^1_a-iv^2_a)(v^1_b+iv^2_b)
(v^3_c+iv^4_c)(v^3_d-iv^4_d)\nn\\
&-& \int d^4x dv\,{\cal H}^{(3)}(\varphi+i X_e v^5_e)
 \partial^{\alpha\dot\alpha}X_a \partial^{\beta\dot\beta}X_b
 \partial_{\alpha\dot\alpha}\partial_{\beta\dot\beta}(\varphi-i X_c v^5_c)
\nn\\&&\hspace{100pt}\times
 (v^1_a-iv^2_a)(v^1_b+iv^2_b)\nn\\
&-&\int d^4x dv\,{\cal H}^{(3)}(\varphi+i X_e v^5_e)
 \partial^{\alpha\dot\beta}X_a \partial^{\beta\dot\alpha}X_b
 \partial_{\alpha\dot\alpha}\partial_{\beta\dot\beta}(\varphi-i X_c v^5_c)
 \nn\\&&\hspace{100pt}\times
 (v^3_a+i v^4_a)(v^3_b-i v^4_b)\nn\\
&+&\frac12\int d^4x dv\, {\cal H}^{(2)}(\varphi+i X_e v^5_e)
\partial^{\alpha\dot\alpha}\partial^{\beta\dot\beta}
(\varphi-i X_a v^5_a)
\partial_{\alpha\dot\alpha}\partial_{\beta\dot\beta}
(\varphi-i X_b v^5_b)+\ldots\,,\qquad
\label{xxx}
\eea
where ellipsis stand for other component fields in (\ref{G4}).

To extract the Wess-Zumino term from (\ref{xxx}), we point out that the
Levi-Civita tensor $\varepsilon^{mnpq}$ can arise only from the cyclic contraction of the spinor indices on four $x$-derivatives $\partial$'s,
recall (\ref{trace-4sigma}).
In addition, if two $\partial$'s act on the same object, no contribution to the Wess-Zumino term appears,
since $\varepsilon^{m n p q}\partial_m\partial_n$ vanishes.
Therefore, only the first integral in (\ref{xxx}) can contribute, and we find
\be
\Gamma_{\rm WZ}=8i\varepsilon^{mnpq}\int d^4xdv\,{\cal H}^{(4)}(\varphi+i X_e v^5_e)
\partial_m X_a
\partial_n X_b
\partial_p X_c
\partial_q X_d
v^1_a v^2_b v^3_c v^4_d\,.
\label{SWZ}
\ee
Once again, using the power series  expansion (\ref{ser}) and computing the harmonic integral
for each term in the series with the help of (\ref{206}), we obtain
\be
\Gamma_{\rm WZ}=-
\varepsilon^{mnpq}\varepsilon^{abcde}\int d^4x\,
X_a
\partial_m X_b
\partial_n X_c
\partial_p X_d
\partial_q X_e
\sum_{n=0}^\infty\frac{(-X_fX_f)^n {\cal H}^{(2n+5)}(\varphi)}{
(2n+5)(2n+3)(2n+1)!}\,.
\label{SWZ.1}
\ee
Substituting (\ref{Hderiv}) into (\ref{SWZ.1}) and summing up the series, we eventually find
\be
\Gamma_{\rm WZ}=-\frac85 c\,
\varepsilon^{mnpq}\varepsilon^{abcde}\int d^4x\,
\frac{g\left(\sqrt{\frac{X_f X_f}{\varphi^2}}\right)}{\varphi^5}
X_a
\partial_m X_b
\partial_n X_c
\partial_p X_d
\partial_q X_e
\,,
\label{SWZ.2}
\ee
where
\be
g(z)=\frac5{8 z^5}\left[
      3\arctan z-
          \frac{z
           \left( 3 +
          5z^2 \right) }
         {{\left( 1 + z^2
           \right) }^2}
 \right].
\ee
This perfectly matches with (\ref{WZso5}), (\ref{gsum}) for $N=1\,$, provided that we once again identify
$\varphi=X_6$ and take $c$ as in (\ref{k}).


\section{Low-energy effective action in $\cN=4$ $SU(2)\times SU(2)$ harmonic superspace}
In section \ref{SectWZ} we discussed various forms of the Wess-Zumino term in the $\cN=4$ SYM
effective action and showed that there exist four different representations of this term which are associated
with four maximal subgroups of $SU(4)$ listed in (\ref{msubs}). In the previous sections we presented
three different superspace formulations of the $\cN=4$ SYM low-energy effective action which correspond
to three different forms of the Wess-Zumino term. Namely, the $\cN=2$ harmonic superspace gives
the Wess-Zumino term in the $SO(4)\times SO(2)$ covariant form, the $\cN=3$ harmonic superspace corresponds
to the $SU(3)\times U(1)$ covariant form of the Wess-Zumino term, while the $\cN=4$ superspace with $USp(4)$ harmonic
variables gives rise to the Wess-Zumino term with manifest $SO(5)$. The last option in the list (\ref{msubs})
is the group $SO(3)\times SO(3)$ which is locally isomorphic to $SU(2)\times SU(2)$. In the present section
we will show that this case is naturally reproduced within the formulation of the low-energy effective action in
the $\cN=4$ superspace equipped with $SU(2)\times SU(2)$ harmonic variables. This formulation
was developed in \cite{BelSam2}.

\subsection{$\cN=4$ bi-harmonic superspace}

In the present section we will consider the $\cN=4$ harmonic superspace which is based on the harmonic variables
for the  maximal subgroup $SU(2)\times SU(2)$ of $SU(4)$. In \cite{BelSam2} it was christened the bi-harmonic $\cN=4$ superspace,
by analogy with the earlier works, where this kind of harmonic variables appeared \cite{IS1,Ivanov:1995jb,Ivanov:1995yp,BelIv,IS2,BIS,IN}.

The basic idea is to give up the manifest $SU(4)$ symmetry of $\cN=4$ SYM theory and use a superspace formulation which keeps
manifest only the maximal $SU(2)\times SU(2)$ subgroup of $SU(4)$ and employs two independent sets of $SU(2)$ harmonic variables
for this subgroup.\footnote{In principle, it is possible to define also another type
of bi-harmonic ${\cal N}=4$ superspace by reducing $SU(4)$ to its $SU(2)\times SU(2) \times U(1)$ subgroup
and harmonizing both $SU(2)$ groups in this product. The ${\cal N}=4$ SYM effective action in such a superspace is expected to
be equivalent to its ${\cal N}=2$ superspace formulation considered in sect. 4.}
In this section, we change our conventions for the indices: The $SU(4)$
indices will be denoted by capital letters $I,J,K,\ldots\,,$ while the indices of the two $SU(2)$'s will be represented
by $i,j,k,\ldots$ and $\tilde i,\tilde j,\tilde k,\ldots$, respectively. Then, every $SU(4)$ index $I$ is
replaced by a pair of $SU(2)$ indices $(i,\tilde i)$
\be
I=(i,\tilde i)=\{(1,1),(1,2),(2,1),(2,2)\}\,.
\label{I}
\ee
For instance, the Grassmann variables $\theta_I^\alpha$ and $\bar \theta^{I\dot\alpha}=\overline{\theta_I^\alpha}$
are now labeled as $\theta_{i \tilde i}^\alpha$ and $\bar\theta^{i\tilde i\,\dot\alpha}=\overline{\theta_{i\tilde i}^\alpha}$,
respectively. The $SU(2)$ indices are raised and lowered  by the standard rules, e.g.
\be
\theta^{i\tilde i\,\alpha}=\varepsilon^{ij}\varepsilon^{\tilde i\tilde j}\theta_{j\tilde j}^\alpha\qquad
(\varepsilon^{12}=-1)\,.
\ee
In these new notations, the covariant spinor derivatives are represented as
\be
D^{i\tilde i}_\alpha = \frac\partial{\partial\theta^\alpha_{i\tilde i}}
+ i \bar\theta^{i\tilde i\,\dot\alpha}\sigma^m_{\alpha\dot\alpha}
\partial_m\,,\qquad
\bar D_{i\tilde i\,\dot\alpha} = -\frac\partial{\partial\bar\theta^{i\tilde i\,\dot\alpha}}
- i\theta^\alpha_{i\tilde i}\sigma^m_{\alpha\dot\alpha}
\partial_m\,.
\label{D-double}
\ee
They obey the anti-commutation relation
\be
\{ D^{i\tilde i}_\alpha , \bar D_{j\tilde j\,\dot\alpha} \}
 = - 2i\delta^i_j \delta^{\tilde i}_{\tilde j} \sigma^m_{\alpha\dot\alpha}
 \partial_m\,.
\ee

Now we introduce two sets of $SU(2)$ harmonic variables, $u^\pm_i$ and $v^\pm_{\tilde i}$, with the defining properties
\be
u^{+i}u^-_i=1\,,\quad v^{+\tilde i} v^-_{\tilde i}=1\,,\quad
u^{+i}u^+_i=u^{-i}u^-_i=0\,,\quad
v^{+\tilde i}v^+_{\tilde i}=v^{-\tilde i}v^-_{\tilde i}=0\,.
\label{u-harm}
\ee
Respectively, we have two sets of the covariant harmonic derivatives
\bea
D^{(2,0)}=u^+_i\frac\partial{\partial u^-_i}\,,\quad
D^{(-2,0)}=u^-_i\frac\partial{\partial u^+_i}\,,\quad
S_1=[D^{(2,0)},D^{(-2,0)}]=u^+_i\frac\partial{\partial u^+_i}
-u^-_i\frac\partial{\partial u^-_i}\,,\nn\\
D^{(0,2)}=v^+_{\tilde i}\frac\partial{\partial v^-_{\tilde i}}\,,\quad
D^{(0,-2)}=v^-_{\tilde i}\frac\partial{\partial v^+_{\tilde i}}\,,\quad
S_2=[D^{(0,2)},D^{(0,-2)}]=v^+_{\tilde i}\frac\partial{\partial v^+_{\tilde i}}
-v^-_{\tilde i}\frac\partial{\partial v^-_{\tilde i}}\,,\hspace{7pt}
\label{Dh}
\eea
which generate two mutually commuting $su(2)$ algebras.
The operators $S_1$ and $S_2$ form $u(1)$
subalgebras in these two $su(2)$'s and count the $U(1)$ charges of
other operators:
\be
[S_1,D^{(s_1,s_2)}]=s_1 D^{(s_1,s_2)}\,,\qquad
[S_2,D^{(s_1,s_2)}]=s_2 D^{(s_1,s_2)}\,.
\ee

Having the harmonic variables $u^\pm_i$ and $v^\pm_{\tilde i}$, one can
define the harmonic projections of all objects with $SU(2)\times SU(2)$ indices.
In particular, the Grassmann variables are projected as
\bea
\theta^{(1,1)}_\alpha=u^+_i v^+_{\tilde i} \theta^{i\tilde i}_\alpha\,,\quad
\theta^{(1,-1)}_\alpha=u^+_i v^-_{\tilde i} \theta^{i\tilde i}_\alpha\,,\quad
\theta^{(-1,1)}_\alpha=u^-_i v^+_{\tilde i} \theta^{i\tilde i}_\alpha\,,\quad
\theta^{(-1,-1)}_\alpha=u^-_i v^-_{\tilde i} \theta^{i\tilde i}_\alpha\,,
\nn\\
\bar\theta^{(1,1)}_{\dot\alpha}=u^+_i v^+_{\tilde i} \bar\theta^{i\tilde i}_{\dot\alpha}\,,\quad
\bar\theta^{(1,-1)}_{\dot\alpha}=u^+_i v^-_{\tilde i} \bar\theta^{i\tilde i}_{\dot\alpha}\,,\quad
\bar\theta^{(-1,1)}_{\dot\alpha}=u^-_i v^+_{\tilde i} \bar\theta^{i\tilde i}_{\dot\alpha}\,,\quad
\bar\theta^{(-1,-1)}_{\dot\alpha}=u^-_i v^-_{\tilde i} \bar\theta^{i\tilde i}_{\dot\alpha}\,.
\label{7.8}
\eea
Here, the  superscripts stand for the $U(1)$ charges.

In what follows, to simplify the subsequent expressions, we will label the $U(1)$ charges by the boldface
capital index {\boldmath$I$}={\bf 1},{\bf 2},{\bf 3},{\bf 4}, so that
\newcommand{\1}{{\bf1}}
\newcommand{\2}{{\bf2}}
\newcommand{\3}{{\bf3}}
\newcommand{\4}{{\bf4}}
\bea
&&
\theta^{\1}_\alpha\equiv \theta^{(1,1)}_\alpha\,, \quad
\theta^{\2}_\alpha\equiv \theta^{(1,-1)}_\alpha\,, \quad
\theta^{\3}_\alpha\equiv \theta^{(-1,1)}_\alpha\,, \quad
\theta^{\4}_\alpha\equiv \theta^{(-1,-1)}_\alpha\,,\nn\\
&&
\bar\theta^{\1}_{\dot\alpha}\equiv \bar\theta^{(-1,-1)}_{\dot\alpha}\,,\quad
\bar\theta^{\2}_{\dot\alpha}\equiv \bar\theta^{(-1,1)}_{\dot\alpha}\,, \quad
\bar\theta^{\3}_{\dot\alpha}\equiv \bar\theta^{(1,-1)}_{\dot\alpha}\,, \quad
\bar\theta^{\4}_{\dot\alpha}\equiv \bar\theta^{(1,1)}_{\dot\alpha}\,.
\eea
In this new notation, the harmonic projections of the covariant spinor derivatives (\ref{D-double}) are written as
\begin{align}
&D_\alpha^\1= \frac{\partial}{\partial\theta^{\1\alpha}}+i\bar\theta^{\1\dot\alpha}\partial_{\alpha\dot\alpha}\,, 
&&\bar D_{\dot\alpha}^\1=-\frac{\partial}{\partial\bar\theta^{\1\dot\alpha}}-i\theta^{\1\alpha}\partial_{\alpha\dot\alpha}\,, \nn\\
&D_\alpha^\2=-\frac{\partial}{\partial\theta^{\2\alpha}}+i\bar\theta^{\2\dot\alpha}\partial_{\alpha\dot\alpha}\,, 
&&\bar D_{\dot\alpha}^\2= \frac{\partial}{\partial\bar\theta^{\2\dot\alpha}}-i\theta^{\2\alpha}\partial_{\alpha\dot\alpha}\,, \nn\\
&D_\alpha^\3=-\frac{\partial}{\partial\theta^{\3\alpha}}+i\bar\theta^{\3\dot\alpha}\partial_{\alpha\dot\alpha}\,, 
&&\bar D_{\dot\alpha}^\3= \frac{\partial}{\partial\bar\theta^{\3\dot\alpha}}-i\theta^{\3\alpha}\partial_{\alpha\dot\alpha}\,, \nn\\
&D_\alpha^\4= \frac{\partial}{\partial\theta^{\4\alpha}}+i\bar\theta^{\4\dot\alpha}\partial_{\alpha\dot\alpha}\,, 
&&\bar D_{\dot\alpha}^\4=-\frac{\partial}{\partial\bar\theta^{\4\dot\alpha}}-i\theta^{\4\alpha}\partial_{\alpha\dot\alpha}\,.
\label{D}
\end{align}
The non-vanishing anticommutation relations between these projections are
\be
\{D^\1_\alpha,\bar D^\1_{\dot\alpha} \}=
\{D^\4_\alpha,\bar D^\4_{\dot\alpha}
\}=-2i\partial_{\alpha\dot\alpha}\,,\qquad
\{D^\2_\alpha,\bar D^\2_{\dot\alpha} \}=
\{D^\3_\alpha,\bar D^\3_{\dot\alpha}
\}=2i\partial_{\alpha\dot\alpha}\,.
\ee

In order to be able to define real structures in harmonic superspaces, one needs the proper definition
of the generalized conjugation.
Recall that in the $\cN=2$ harmonic superspace such a conjugation is given by the involution (\ref{tilde-u})
which is a generalization of the standard complex conjugation. In the $\cN=4$ bi-harmonic superspace considered
here the analogous operation can be defined in different ways. We postulate that the $\,\widetilde{\phantom{m}}\,$-conjugation
acts on the $u$-harmonics by the same rules (\ref{tilde-u}), but on the $v$-harmonics it is realized as the conventional complex conjugation,
\bea
&&
\widetilde{u^\pm_i }=u^{\pm i}\,,\qquad
\widetilde{u^{\pm i}}=-u^{\pm}_i\,,\nn\\
&&
\widetilde{ v^{+\tilde i} }= v^-_{\tilde i}\,,\quad
\widetilde{ v^{+}_{\tilde i} }= -v^{-\tilde i}\,,\quad
\widetilde{ v^{-\tilde i} } = - v^+_{\tilde i} \,,\quad
\widetilde{ v^-_{\tilde i} } = v^{+\tilde i}\,.
\label{tilde1}
\eea
Assuming that all the harmonic-independent objects behave under this conjugation in the same way as under the complex conjugation,
we can specify the $\,\widetilde{\phantom{m}}\,$-conjugation properties of Grassmann variables
(\ref{7.8})
\bea
&&
\widetilde{\theta^\1_\alpha}=-\bar\theta^{\3}_{\dot\alpha}\,,\quad
\widetilde{\theta^\2_\alpha}=\bar\theta^{\4}_{\dot\alpha}\,,\quad
\widetilde{\theta^\3_\alpha}=-\bar\theta^{\1}_{\dot\alpha}\,,\quad
\widetilde{\theta^\4_\alpha}=\bar\theta^{\2}_{\dot\alpha}\,,
\nn\\&&
\widetilde{\bar\theta^\1_{\dot\alpha}}=\theta^\3_\alpha\,,\quad
\widetilde{\bar\theta^\2_{\dot\alpha}}=-\theta^\4_\alpha\,,\quad
\widetilde{\bar\theta^\3_{\dot\alpha}}=\theta^\1_\alpha\,,\quad
\widetilde{\bar\theta^\4_{\dot\alpha}}=-\theta^\2_\alpha\,.
\label{tilde2}
\eea

By definition, the full $\cN=4$ bi-harmonic superspace is parametrized by the coordinates
\be
\{x^m, \theta^{\bm{I}\alpha},\bar\theta^{\bm{I}\dot\alpha},u,v \}\,.
\ee
This superspace has several analytic subspaces, each involving eight Grassmann variables out of sixteen ones. Every analytic subspace  is closed under
the full supersymmetry. All these subspaces were considered in detail in \cite{BelSam2}. Here we will need only one of them,
parametrized by the coordinates
\be
\{ \zeta, u ,v \} = \{ (x^m_A, \theta^\1_\alpha, \bar\theta^\2_{\dot\alpha} ,\bar\theta^\3_{\dot\alpha}, \theta^\4_\alpha),u,v  \}\,,
\label{anal-biharm}
\ee
where
\be
x_{A}^m=x^m
+i\theta^\1\sigma^m\bar\theta^\1
+i\theta^\2\sigma^m\bar\theta^\2
+i\theta^\3\sigma^m\bar\theta^\3
+i\theta^\4\sigma^m\bar\theta^\4\,.
\label{anal-coord-biharm}
\ee
It is straightforward to check that this subspace is closed under the $\,\widetilde{\phantom{m}}\,$-conjugation (\ref{tilde1}), (\ref{tilde2}).

In the analytic basis involving (\ref{anal-biharm}) as the coordinate subset,  half of the covariant spinor derivatives (\ref{D}) become short:
\bea
&&D^{\2}_\alpha=-\frac\partial{\partial\theta^{\2\alpha}}
 \,,\qquad
D^{\1}_\alpha=\frac\partial{\partial\theta^{\1\alpha}}
 +2i\bar\theta^{\1\dot\alpha}\partial_{\alpha\dot\alpha}\,,\nn\\
&&D^\3_\alpha=-\frac\partial{\partial\theta^{\3\alpha}}
 \,,\qquad
D^{\4}_\alpha=\frac\partial{\partial\theta^{\4\alpha}}
+2i\bar\theta^{\4\dot\alpha}\partial_{\alpha\dot\alpha} \,,\nn\\
&&\bar D^{\1}_{\dot\alpha}=-\frac\partial{\partial\bar\theta^{\1\dot\alpha}}
\,,\qquad
\bar D^{\2}_{\dot\alpha}=\frac\partial{\partial\bar\theta^{\2\dot\alpha}}
-2i\theta^{\2\alpha}\partial_{\alpha\dot\alpha}\,,\nn\\
&&\bar D^{\4}_{\dot\alpha}=-\frac\partial{\partial\bar\theta^{\4\dot\alpha}}\,,\qquad
\bar D^{\3}_{\dot\alpha}=\frac\partial{\partial\bar\theta^{\3\dot\alpha}}
-2i\theta^{\3\alpha}\partial_{\alpha\dot\alpha}
\,.
\label{D-short-biharm}
\eea

A superfield $\Phi_A$ is called analytic if it is annihilated by the following covariant spinor derivatives
\be
D^{\2}_\alpha \Phi_A = D^\3_\alpha \Phi_A
=\bar D^{\1}_{\dot\alpha} \Phi_A
=\bar D^{\4}_{\dot\alpha} \Phi_A =0\,.
\ee
The general solution of these constraints is given by
\be
\Phi_A = \Phi_A(\zeta,u,v)\,.
\ee

For completeness and for the further use, we give the expressions
of the covariant harmonic derivatives (\ref{Dh}) in the analytic basis
\bea
D^{(2,0)}&=&u^+_i\frac\partial{\partial u^-_i}
+2i\theta^{\2\alpha}\bar\theta^{\4\dot\alpha}\partial_{\alpha\dot\alpha}
+2i\theta^{\1\alpha}\bar\theta^{\3\dot\alpha}\partial_{\alpha\dot\alpha}
+\theta^{\1}_\alpha\frac\partial{\partial\theta^{\3}_\alpha}
+\theta^{\2}_\alpha\frac\partial{\partial\theta^{\4}_\alpha}
+\bar\theta^{\4}_{\dot\alpha}\frac\partial{\partial\bar\theta^{\2}_{\dot\alpha}}
+\bar\theta^{\3}_{\dot\alpha}\frac\partial{\partial\bar\theta^{\1}_{\dot\alpha}}
\,,
\nn\\
D^{(-2,0)}&=&u^-_i\frac\partial{\partial u^+_i}
+2i\theta^{\4\alpha}\bar\theta^{\2\dot\alpha}\partial_{\alpha\dot\alpha}
+2i\theta^{\3\alpha}\bar\theta^{\1\dot\alpha}\partial_{\alpha\dot\alpha}
+\theta^{\3}_\alpha\frac\partial{\partial\theta^{\1}_\alpha}
+\theta^{\4}_\alpha\frac\partial{\partial\theta^{\2}_\alpha}
+\bar\theta^{\2}_{\dot\alpha}\frac\partial{\partial\bar\theta^{\4}_{\dot\alpha}}
+\bar\theta^{\2}_{\dot\alpha}\frac\partial{\partial\bar\theta^{\3}_{\dot\alpha}}\,,
\nn\\
D^{(0,2)}&=&v^+_{\tilde i}\frac\partial{\partial v^-_{\tilde i}}
+2i\theta^{\1\alpha} \bar\theta^{\2\dot\alpha}
\partial_{\alpha\dot\alpha}
+2i\theta^{\3\alpha}\bar\theta^{\4\dot\alpha}\partial_{\alpha\dot\alpha}
+\theta^{\1}_\alpha\frac\partial{\partial\theta^{\2}_\alpha}
+\theta^{\3}_\alpha\frac\partial{\partial\theta^{\4}_\alpha}
+\bar\theta^{\4}_{\dot\alpha}\frac\partial{\partial\bar\theta^{\3}_{\dot\alpha}}
+\bar\theta^{\2}_{\dot\alpha}\frac\partial{\partial\bar\theta^{\2}_{\dot\alpha}}\,,
\nn\\
D^{(0,-2)}&=&v^-_{\tilde i}\frac\partial{\partial v^+_{\tilde i}}
+2i\theta^{\4\alpha} \bar\theta^{\3\dot\alpha}
\partial_{\alpha\dot\alpha}
+2i\theta^{\2\alpha}\bar\theta^{\1\dot\alpha}\partial_{\alpha\dot\alpha}
+\theta^{\2}_\alpha\frac\partial{\partial\theta^{\1}_\alpha}
+\theta^{\4}_\alpha\frac\partial{\partial\theta^{\3}_\alpha}
+\bar\theta^{\3}_{\dot\alpha}\frac\partial{\partial\bar\theta^{\4}_{\dot\alpha}}
+\bar\theta^{\1}_{\dot\alpha}\frac\partial{\partial\bar\theta^{\2}_{\dot\alpha}}\,,
\nn\\
\label{HA1}
\eea
where $\partial_{\alpha\dot\alpha} = \sigma^m_{\alpha\dot\alpha}
\frac\partial{\partial x^m_A}$.

\subsection{$\cN=4$ SYM constraints in bi-harmonic superspace}

Recall that the $\cN=4$ SYM constraints are given in the abelian case by eqs. (\ref{N4-constraints}) and (\ref{N4-reality}).
With employing the notations of the present section, they are rewritten as
\begin{subequations}
\bea
&& D^I_\alpha W^{JK}+D^J_\alpha W^{IK}=0\,,\label{c2}\\
&&\bar D_{I\dot\alpha} W^{JK}=\frac13(\delta_I^J\bar D_{L\dot\alpha} W^{LK}
 -\delta_I^K\bar D_{L\dot\alpha}W^{LJ})\,,
\label{c1}\\
&&\overline{W^{IJ}} \equiv \bar W_{IJ}=\frac12\varepsilon_{IJKL}W^{KL}\,.
\label{c3}
\eea
\end{subequations}
Here $W^{IJ}=-W^{JI}$ is the $\cN=4$ superfield strength with $SU(4)$ indices. Representing the $SU(4)$ indices
 as pairs of the $SU(2)$ ones, like in (\ref{I}), we find
\be
W^{IJ} \equiv W^{i\tilde i,j\tilde j}=\varepsilon^{ij}W^{\tilde i\tilde j}+\varepsilon^{\tilde i\tilde j}W^{ij}\,,
\ee
so that the superfield strength $W^{IJ}$ is now split into a pair of \emph{symmetric} $SU(2)$ tensors:
$W^{\tilde i\tilde j}=W^{\tilde j\tilde i}$ and $W^{ij}=W^{ji}$. The constraints (\ref{c2})--(\ref{c3})
can be readily rewritten in terms of  these tensors. In particular, using the identity
\be
\varepsilon_{IJKL} \equiv
\varepsilon_{i\tilde i,j\tilde j,k\tilde k,l\tilde l}=
\varepsilon_{il}\varepsilon_{jk}\varepsilon_{\tilde i\tilde j}\varepsilon_{\tilde k\tilde l}
-\varepsilon_{ij}\varepsilon_{kl}\varepsilon_{\tilde i\tilde l}\varepsilon_{\tilde j \tilde k}\,,
\ee
we find that the reality condition  (\ref{c3}) is equivalent to the following reality properties
\be
\overline{W^{ij}} \equiv \bar W_{ij}=W_{ij}\,,\qquad
\overline{W^{\tilde i\tilde j}} \equiv \bar W_{\tilde i\tilde j}=-W_{\tilde i\tilde j}\,.
\label{Wconj}
\ee
It is also straightforward to rewrite the constraints (\ref{c2}) and (\ref{c1}) in terms of the newly introduced  superfield strengths
\begin{subequations}
\label{cc}
\bea
&&D^{\tilde i(i}_\alpha W^{jk)}=0\,,\quad
D^{i(\tilde i}_\alpha W^{\tilde j\tilde k)}=0\,,\quad
D^{k\tilde i}_\alpha W^i_k+D^{i\tilde k}_\alpha W^{\tilde i}_{\tilde k}=0\,,
\label{cc2}\\
&&\bar D^{\tilde i(i}_{\dot\alpha} W^{jk)}=0\,,\quad
\bar D^{i(\tilde i}_{\dot\alpha }W^{\tilde j\tilde k)}=0\,,\quad
\bar D^{k\tilde i}_{\dot\alpha} W^i_k-\bar D^{i\tilde k}_{\dot\alpha} W^{\tilde i}_{\tilde k}=0\,.
\label{cc1}
\eea
\end{subequations}
It should be stressed that the equations (\ref{Wconj}), (\ref{cc2}) and (\ref{cc1}) are
equivalent to the $\cN=4$ SYM constraints (\ref{c3}), (\ref{c2}) and (\ref{c1}).

Now we introduce the harmonic projections of the superfields $W^{ij}$ and $W^{\tilde i\tilde j}$:
\begin{align}
&
W=u^+_i u^-_j W^{ij}-v^+_{\tilde i} v^-_{\tilde j} W^{\tilde i\tilde j}\,,\qquad&&
W'=u^+_i u^-_j W^{ij}+v^+_{\tilde i} v^-_{\tilde j} W^{\tilde i\tilde j}\,,\label{35}\\
&W^{(2,0)}=u^+_i u^+_j W^{ij}\,,\qquad&&
W^{(-2,0)}=u^-_i u^-_j W^{ij}\,,\label{36}\\
&W^{(0,2)}=v^+_{\tilde i} v^+_{\tilde j} W^{\tilde i\tilde j}\,,\qquad&&
W^{(0,-2)}=v^-_{\tilde i} v^-_{\tilde j} W^{\tilde i\tilde j}\,.
\label{37}
\end{align}
According to the conjugation rules (\ref{tilde1}) and (\ref{Wconj}), these harmonic projections obey the reality properties:
\be
\widetilde{W}={W}\,,\quad
\widetilde{{W}'}={W}'\,,\quad
\widetilde{W^{(\pm 2,0)}}=W^{(\pm 2,0)}\,,\quad
\widetilde{W^{(0,\pm2)}}=- W^{(0,\mp2)}\,.
\label{conj}
\ee

For the goals of the present subsection, we need to consider only one of these superfields, $W$;
the remaining ones were studied in \cite{BelSam2}. In order to find the differential constraints for this basic superfield,
we are led to consider contractions of the equations (\ref{cc}) with various combinations of harmonic variables.
As a result, we derive the set of the first-order differential constraints on $W$
\be
\bar D^\1_{\dot\alpha} W =D^\2_\alpha W= D^\3_\alpha W=
\bar D^\4_{\dot\alpha} W=0\,.
\label{Wan}
\ee
These equations are easily recognized as the analyticity conditions, since the covariant spinor derivatives appearing in (\ref{Wan})
become short in the analytic basis, see (\ref{D-short-biharm}). Thus the general solution of (\ref{Wan}) is
an arbitrary analytic superfield
\be
W=W(x^m_A,\theta^\1_\alpha,\bar\theta^\2_{\dot\alpha},\bar\theta^\3_{\dot\alpha},
\theta^\4_\alpha,u,v)\,.
\label{3.32}
\ee

It is obvious that there remain many auxiliary fields in $W$ which
should be removed by the other constraints also following from (\ref{cc}):
\be
(D^\1)^2 W =(\bar D^\2)^2 W=(\bar D^\3)^2 W=(D^\4)^2 W=(D^\1 D^\4)W=(\bar D^\2\bar D^\3) W=0\,.
\label{Wlin}
\ee
These second-order constraints eliminate the unphysical components in $W$, but do not imply any dynamical
equations for the physical components. The equations of motion for the physical components follow from the relations
\be
D^{(2,0)}D^{(2,0)}{ W}=D^{(0,2)}D^{(0,2)}{ W}=
D^{(2,0)}D^{(0,2)}{ W}=0\,.
\label{Wharm}
\ee
In the central basis, these constraints are satisfied for the superfields (\ref{35})  by construction. However, they
become non-trivial dynamical equations in the analytic basis, in which the harmonic derivatives involve the space-time derivatives,
see (\ref{HA1}).

The constraints (\ref{conj}), (\ref{Wan}), (\ref{Wlin}) and  (\ref{Wharm}) completely specify the superfield $W$:
\begin{subequations}
\label{W-biharm}
\bea
W&=&W_{\rm bos}+W_{\rm ferm}\,,\\
W_{\rm bos}&=&
\omega+u^+_i u^-_j \phi^{ij}+v^+_{\tilde i} v^-_{\tilde j} i \varphi^{\tilde i\tilde j}\nn\\&&
+\frac1{\sqrt2}(\theta^\1_\alpha \theta^\4_\beta
 \sigma^{m\alpha}{}_{\dot\alpha}\sigma^{n\beta\dot\alpha}
+\bar\theta^\3_{\dot\alpha}\bar\theta^\2_{\dot\beta}\sigma^{m\dot\alpha}{}_\alpha
\sigma^{n\alpha\dot\beta})F_{mn}\nn\\&&
+2\theta^{\1\alpha}\bar\theta^{\2\dot\alpha}
 \partial_{\alpha\dot\alpha}\varphi^{\tilde i\tilde j} v^-_{\tilde i} v^-_{\tilde j}
+2\theta^{\4\alpha}\bar\theta^{\3\dot\alpha}
 \partial_{\alpha\dot\alpha}\varphi^{\tilde i\tilde j} v^+_{\tilde i} v^+_{\tilde j}
\nn\\&&
-2i \theta^{\4\alpha}\bar\theta^{\2\dot\alpha}
 \partial_{\alpha\dot\alpha} \phi^{ij}u^+_i u^+_j
-2i \theta^{\1\alpha}\bar\theta^{\3\dot\alpha}
 \partial_{\alpha\dot\alpha}\phi^{ij} u^-_i u^-_j
 \nn\\&&
+4\theta^{\1\alpha}\theta^{\4\beta}\bar\theta^{\3\dot\alpha}
\bar\theta^{\2\dot\beta}\partial_{\alpha\dot\alpha}\partial_{\beta\dot\beta}
(u^+_i u^-_j \phi^{ij}-v^+_{\tilde i} v^-_{\tilde j} i \varphi^{\tilde i\tilde j})\,,
\label{W-bos-biharm}\\
W_{\rm ferm}&=&
\theta^{\1\alpha}\psi_\alpha^{i\tilde i}u^-_i v^-_{\tilde i}
-\theta^{\4\alpha}\psi_\alpha^{i\tilde i}u^+_i v^+_{\tilde i}
+\bar\theta^{\2}_{\dot\alpha}\bar\psi^{\dot\alpha\, i\tilde i}u^+_i
v^-_{\tilde i}
-\bar\theta^{\3}_{\dot\alpha}\bar\psi^{\dot\alpha\, i\tilde i}
u^-_i v^+_{\tilde i}
\nn\\&&
+2i\theta^{\4\alpha}\theta^{\1\beta}\bar\theta^{\3\dot\beta}
 \partial_{\beta\dot\beta}\psi^{i\tilde i}_\alpha u^-_i v^+_{\tilde i}
-2i\theta^{\1\alpha}\theta^{\4\beta}\bar\theta^{\2\dot\beta}
\partial_{\beta\dot\beta}\psi^{i\tilde i}_\alpha u^+_i v^-_{\tilde i}
\nn\\&&
+2i\bar\theta^{\3\dot\alpha}\theta^{\4\beta}\bar\theta^{\2\dot\beta}
\partial_{\beta\dot\beta}\bar\psi^{i\tilde i}_{\dot\alpha} u^+_i v^+_{\tilde i}
-2i\bar\theta^{\2\dot\alpha}\theta^{\1\beta}\bar\theta^{\3\dot\beta}
\partial_{\beta\dot\beta}\bar\psi^{i\tilde i}_{\dot\alpha} u^-_i v^-_{\tilde i}\,.
\eea
\end{subequations}
Here, all the component fields depend on $x^m_A\,$,  $\phi^{ij}=\phi^{(ij)}$ and $\varphi^{\tilde i\tilde j}=\varphi^{(\tilde i\tilde j)}$
are two triplets of scalar fields, $\psi^{i\tilde i}_\alpha$ are four Weyl spinors and $F_{mn}$ is the Maxwell field strength.
These fields obey the classical free equations of motion
\be
\square \phi^{ij}=\square \varphi^{\tilde i\tilde j}=0\,,\quad
\partial^{\alpha\dot\alpha}\psi^{i\tilde i}_\alpha=0\,,\quad
\partial^m F_{mn}=0\,.
\label{eom}
\ee
No any auxiliary field component is present in $W$ as they all have been eliminated by the constraints (\ref{Wan}),
(\ref{Wlin}) and (\ref{Wharm}).

The component field expansion (\ref{W-bos-biharm}) starts with an arbitrary constant $\omega$. This constant would have never appeared,
had we started with the component form of $W^{IJ}$ that solves (\ref{c2})--(\ref{c3}), defined $W$ by the rule (\ref{35})
and then passed to analytic coordinates. Instead, here we postulated $W$ to be \emph{defined} by the constraints (\ref{conj}), (\ref{Wan}),
(\ref{Wlin}), and (\ref{Wharm}). Finally, these constraints proved quite sufficient to properly restrict the component degrees of freedom,
except for the residual appearance of an extra constant parameter $\omega$.

We set $\omega$ equal to zero by requiring that $W$ transforms linearly under the scale transformations with a constant parameter $\lambda$,
\be
\delta W=\lambda\, W \quad \Rightarrow \quad \omega=0\,.
\ee
This requirement is particularly natural for the purposes of the next subsection, where we will construct the superconformal effective action of
$\cN=4$ SYM theory in the bi-harmonic $\cN=4$ superspace.

Note that the bosonic part of the superfield strength (\ref{W-bos-biharm}) involves only a few harmonic monomials,
 $u^+_i u^+_j$, $u^-_i u^-_j$, $u^+_{(i}u^-_{j)}$,
and $v^+_{\tilde i} v^+_{\tilde j}$, $v^-_{\tilde i} v^-_{\tilde j}$, $v^+_{(\tilde i}v^-_{\tilde j)}$.
For the computational reasons, it is convenient to rewrite
these $SU(2)$ monomials in terms of the $SO(3)$ harmonics
$U^1_{a'}$, $U^2_{a'}$, $U^3_{a'}$ and $V^1_a$, $V^2_a$, $V^3_a$,
\bea
&&
V^1_a= i \gamma_a^{\tilde i\tilde j} v^+_{\tilde i} v^-_{\tilde j}\,,\quad
V^2_a=\frac12\gamma_a^{\tilde i\tilde j}(v^+_{\tilde i} v^+_{\tilde j} + v^-_{\tilde i} v^-_{\tilde j})
\,,\quad
V^3_a=\frac i2\gamma_a^{\tilde i\tilde j}(v^+_{\tilde i} v^+_{\tilde j} - v^-_{\tilde i} v^-_{\tilde j})\,,
\nn\\&&
U^1_{a'}= i \gamma_{a'}^{ij} u^+_i u^-_j\,,\quad
U^2_{a'}=\frac12\gamma_{a'}^{ij}(u^+_i u^+_j + u^-_i u^-_j)
\,,\quad
U^3_{a'}=\frac i2\gamma_{a'}^{ij}(u^+_i u^+_j - u^-_i u^-_j)\,,~~~~~~~~~~~~
\label{B2}
\eea
where $\gamma^a_{\tilde i\tilde j}$, $\gamma^{a'}_{ij}$ are two copies of $SO(3)$ gamma-matrices with the defining properties
\be
\gamma^a_{\tilde i\tilde j}\gamma^{b\,\tilde j\tilde k}+\gamma^b_{\tilde i\tilde j}\gamma^{a\,\tilde j\tilde k}=
2\delta^{ab}\delta_{\tilde i}^{\tilde k}\,,\qquad
\gamma^{a'}_{ij}\gamma^{{b'}\,jk}+\gamma^{b'}_{ij}\gamma^{{a'}\,jk}=
2\delta^{{a'}{b'}}\delta_i^k\,.
\label{gamma}
\ee
Using (\ref{u-harm}), (\ref{cc}) and (\ref{gamma}), it is straightforward to
check that the objects (\ref{B2}) are real under  the usual complex
conjugation and obey the standard properties of $SO(3)$ matrices,
\be
\begin{array}{lll}
U^{a'}_{b'} U^{c'}_{b'}=\delta^{{a'}{c'}}\,,\quad&
\varepsilon^{a'b'c'}U^1_{a'}U^2_{b'}U^3_{c'}=1\,,\quad&
\overline{U^{b'}_{a'}}=U^{b'}_{a'}\,,\\
V^a_b V^c_b=\delta^{ac}\,,&
\varepsilon^{abc}V^1_{a}V^2_{b}V^3_{c}=1\,,&
\overline{V^a_b}=V^a_b\,.
\end{array}
\label{SO3harm}
\ee

In terms of the $SO(3)$-harmonics (\ref{B2}), the bosonic part (\ref{W-bos-biharm})
of the superfield strength $W$ can be rewritten as
\bea
W_{\rm bos}&=& \varphi^a V^1_a-i \phi^{a'} U^1_{a'}
+\frac1{\sqrt2}(\theta^\1_\alpha \theta^\4_\beta
 \sigma^{m\alpha}{}_{\dot\alpha}\sigma^{n\beta\dot\alpha}
+\bar\theta^\3_{\dot\alpha}\bar\theta^\2_{\dot\beta}\sigma^{m\dot\alpha}{}_\alpha
\sigma^{n\alpha\dot\beta})F_{mn} \nn\\&&
+2\theta^{\1\alpha}\bar\theta^{\2\dot\alpha}
 \partial_{\alpha\dot\alpha}\varphi^a (V^2_a+iV^3_a)
+2\theta^{\4\alpha}\bar\theta^{\3\dot\alpha}
 \partial_{\alpha\dot\alpha}\varphi^a (V^2_a-i V^3_a)
\nn\\&&
-2i \theta^{\4\alpha}\bar\theta^{\2\dot\alpha}
 \partial_{\alpha\dot\alpha} \phi^{a'}(U^2_{a'}-iU^3_{a'})
-2i \theta^{\1\alpha}\bar\theta^{\3\dot\alpha}
 \partial_{\alpha\dot\alpha}\phi^{a'}(U^2_{a'}+i U^3_{a'})
 \nn\\&&
-4\theta^{\1\alpha}\theta^{\4\beta}\bar\theta^{\3\dot\alpha}
\bar\theta^{\2\dot\beta}\partial_{\alpha\dot\alpha}\partial_{\beta\dot\beta}
(V^1_a\varphi^a+i U^1_{a'}\phi^{a'})
\,,
\label{Wbos-SO3}
\eea
where we have defined the $SO(3)$ triplets of the scalars:
\be
\varphi^a=\frac12\gamma^a_{\tilde i\tilde j}\varphi^{\tilde i\tilde j}\,,\qquad
\phi^{a'}=\frac12\gamma^{a'}_{ij}\phi^{ij}\,.
\ee

\subsection{Scale invariant low-energy effective action}

We will look for the low-energy effective action in the form of a functional of $W$
\be
\Gamma=\int d\zeta du dv\,H(W)\,,
\label{G-biharm}
\ee
where $H(W)$ is some function of $W$ without derivatives.  The integration goes over the analytic superspace (\ref{anal-biharm})
with the analytic measure defined as
\be
d\zeta = d^4x d^8\theta\,,\qquad
\int d^8\theta
\,(\theta^\1)^2(\theta^\4)^2(\bar\theta^\2)^2(\bar\theta^\3)^2=1\,.
\label{measure-biharm}
\ee
The integration over harmonic variables $du$ and $dv$ is defined by the same rule (\ref{int-du}). We point out that the
function $H(W)$ must have zero $U(1)$ charges, since the integration
measure $d\zeta$ of the analytic superspace (\ref{anal-biharm}) is uncharged.

Note that the integration measure (\ref{measure-biharm}) amounts to eight spinor covariant
derivatives, or, equivalently, to four space-time ones on the component fields.  Therefore, we expect
that the action (\ref{G-biharm}) with some appropriate $H(W)$ describes the
four-derivative term in the $\cN=4$ low-energy effective action, and that
this term is the leading one in the derivative expansion. We will now
determine the function $H$ by requiring scale invariance of
the action (\ref{G-biharm}), in exactly the same way as we proceeded in sect.\ \ref{sect6.3}.

As the measure $d\zeta$ is dimensionless, the function $H(W)$ must also be dimensionless.
Recalling that the mass dimension of $W$ is one, we are forced to introduce a parameter $\Lambda$
such that $W/\Lambda$ is dimensionless, and choose $H=H(W/\Lambda)$. However, the dependence on $\Lambda$
should disappear after doing the integral over Grassmann variables. This requirement uniquely fixes the form of the
function $H$,
\be
H=c\ln\frac W\Lambda\,,
\label{H-log}
\ee
with some coefficient $c$. The corresponding low-energy effective action
\be
\Gamma=c\int d\zeta du dv\,\ln\frac W\Lambda
\label{G-log-biharm}
\ee
is scale invariant. Indeed, rescaling $W$ shifts the integrand in (\ref{G-log-biharm}) by a constant,
which gives a zero contribution under the Grassmann integral. Thus, the requirement of scale invariance fixes the form
of the low-energy effective action. Surprisingly, this form is very similar to (\ref{G4}).

\subsection{Component structure}
\subsubsection{$F^4/X^4$ term}
To find the $F^4/X^4$ term in the component field expansion of the low-energy effective action (\ref{G-log-biharm}),
it suffices to substitute in it the bosonic part of the superfield strength $W$ in the form (\ref{Wbos-SO3}),
\be
\Gamma_{F4/X^4} = c\int d\zeta dU dV \ln \frac{W_{\rm bos}}\Lambda\,,
\label{G-F4}
\ee
where we have replaced the integration over the $SU(2)$ harmonics by that over the $SO(3)$ harmonics.
Moreover, we can neglect all terms with derivatives of the scalar fields in (\ref{Wbos-SO3}),
since they do not contribute to the $F^4/X^4$ term,
\be
W_{\rm bos}\, \Rightarrow \, \varphi^a V^1_a-i \phi^{a'} U^1_{a'}
+\frac1{\sqrt2}(\theta^\1_\alpha \theta^\4_\beta
 \sigma^{m\alpha}{}_{\dot\alpha}\sigma^{n\beta\dot\alpha}
+\bar\theta^\3_{\dot\alpha}\bar\theta^\2_{\dot\beta}\sigma^{m\dot\alpha}{}_\alpha
\sigma^{n\alpha\dot\beta})F_{mn}\,.
\label{WW}
\ee
Substituting (\ref{WW}) into (\ref{G-F4}) and integrating there over the Grassmann variables by the rules (\ref{measure-biharm}),
we find
\be
\Gamma_{F^4/X^4}=\frac14\int d^4x dU dV\,H^{(4)}(\varphi^a V^1_a-i\phi^{a'} U^1_{a'})
\left[F_{mn}F^{nk}F_{kl}F^{lm}-\frac14(F_{pq}F^{pq})^2\right].
\label{GF}
\ee
Here we have applied the standard identity  (\ref{trace-4sigma}) for the trace of four sigma-matrices.
Choosing now the function $H$ as in (\ref{H-log}), we expand it in the Taylor series over $i\phi^{a'}U^1_{a'}$,
\bea
H^{(4)}(\varphi^a V^1_a-i\phi^{a'} U^1_{a'})
&=&\sum_{n=0}^\infty \frac1{n!}H^{(n+4)}(\varphi^a V^1_a) (-i\phi^{a'}U^1_{a'})^n
\nn\\&=&
c\sum_{n=0}^\infty \frac{(-1)^{n+1}(n+3)!}{n!}
\frac{(-i\phi^{a'} U^1_{a'})^n}{(\varphi^a V^1_a)^{n+4}}\,.
\label{ser1}
\eea
Here $H^{(n)}$ stands for the $n$'th derivative of the function
$H$ with respect to its argument. Next, we substitute this series into (\ref{GF}) and compute the harmonic integral over $dU$, using the rules
\be
\int dU\,1=1\,,\qquad
\int dU(\mbox{non-singlet $SO(3)$ irreducible representation})=0\,.
\label{SO3-int}
\ee
As a result, we obtain
\bea
\Gamma_{F^4/X^4}&=&-\frac c4\int d^4x dV\,[F_{mn}F^{nk}F_{kl}F^{lm}-\frac14(F_{pq}F^{pq})^2]\,
\sum_{n=0}^\infty (2n+2)(2n+3)
\frac{(-\phi^{a'}\phi^{a'})^{n}}{(\varphi^a V^1_a)^{2n+4}}
\nn\\&=&
\frac c2\int d^4x dV\,[F_{mn}F^{nk}F_{kl}F^{lm}-\frac14(F_{pq}F^{pq})^2]\,
\frac{\phi^{a'}\phi^{a'}-3(\varphi^a V^1_a)^2}{[\phi^{b'}\phi^{b'}+(\varphi^b V^1_b)^2]^3}
\,.
\label{4.9}
\eea
It is notable that the series
in the first line in (\ref{4.9}) is summed up into the concise analytical expression given in the second line.
This allows us to expand the expression in the second line in (\ref{4.9}) in a series over another argument, $\varphi^a V^1_a$,
and compute the harmonic integral over $dV$, using the same rules (\ref{SO3-int}):
\bea
\Gamma_{F^4/X^4}&=&\frac c2\int d^4x dV
\frac{F_{mn}F^{nk}F_{kl}F^{lm}-\frac14(F_{pq}F^{pq})^2}{(\phi^{b'}\phi^{b'})^2}
\sum_{n=0}^\infty (-1)^n(2n+1)(n+1)
\frac{(\varphi^a V^1_a)^{2n}}{(\phi^{a'}\phi^{a'})^n}
\nn\\&=&
\frac c2\int d^4x \frac{F_{mn}F^{nk}F_{kl}F^{lm}-\frac14(F_{pq}F^{pq})^2}{(\phi^{b'}\phi^{b'})^2}
\sum_{n=0}^\infty (-1)^n(n+1)\left(
\frac{\varphi^a \varphi^a}{\phi^{a'}\phi^{a'}}
\right)^n.
\eea
This series can be easily re-summed, and we obtain the following result
\be
\Gamma_{F^4/X^4}=\frac c2\int d^4x
\frac{F_{mn}F^{nk}F_{kl}F^{lm}-\frac14(F_{pq}F^{pq})^2}{(\phi^{a'}\phi^{a'}
+\varphi^a\varphi^a)^2}
=\frac c2\int d^4x
\frac{F_{mn}F^{nk}F_{kl}F^{lm}-\frac14(F_{pq}F^{pq})^2}{(X_A X_A)^2}\,.
\ee
Note that the scalar fields in the denominator appear in
the right $SO(6)$-invariant form, and we end up exactly with the $F^4/X^4$-term in the form (\ref{F4X4}),  under the choice
\be
c=\frac1{32\pi^2}\,.
\label{c-biharm}
\ee

\subsubsection{Wess-Zumino term}
\label{SectWZ-biharm}

To derive the Wess-Zumino term, we can start from the same superfield expression (\ref{G-F4}). However, in the expansion (\ref{Wbos-SO3})
we have to omit the Maxwell field strength  and keep all terms with derivatives of scalars:
\bea
W_{\rm bos}&=&V^1_a\varphi^a-i U^1_{a'} \phi^{a'}
\nn\\&&
+2\theta^{\1\alpha}\bar\theta^{\2\dot\alpha}
 \partial_{\alpha\dot\alpha}\varphi^a (V^2_a+iV^3_a)
+2\theta^{\4\alpha}\bar\theta^{\3\dot\alpha}
 \partial_{\alpha\dot\alpha}\varphi^a (V^2_a-i V^3_a)
\nn\\&&
-2i \theta^{\4\alpha}\bar\theta^{\2\dot\alpha}
 \partial_{\alpha\dot\alpha} \phi^{a'}(U^2_{a'}-iU^3_{a'})
-2i \theta^{\1\alpha}\bar\theta^{\3\dot\alpha}
 \partial_{\alpha\dot\alpha}\phi^{a'}(U^2_{a'}+i U^3_{a'})
 \nn\\&&
-4\theta^{\1\alpha}\theta^{\4\beta}\bar\theta^{\3\dot\alpha}
\bar\theta^{\2\dot\beta}\partial_{\alpha\dot\alpha}\partial_{\beta\dot\beta}
(V^1_a\varphi^a+i U^1_{a'}\phi^{a'})\,.
\label{Wscal}
\eea
The terms in the last line do not contribute to the Wess-Zumino term, as they contain two space-time derivatives acting on the same scalar.
Substituting the remaining terms into (\ref{G-F4}) and computing the integral over the Grassmann variables, we find
\bea
\Gamma_{\rm WZ}&=&\int d^4x dU dV\,H^{(4)}(V^1_d\varphi^d-i U^1_{d'} \phi^{d'})
\partial_{\alpha\dot\alpha}\varphi^a
\partial_{\beta\dot\beta}\varphi^b
\partial^{\beta\dot\alpha}\phi^{a'}
\partial^{\alpha\dot\beta}\phi^{b'}
\nn\\&&\times
(V^2_a+iV^3_a)(V^2_b-iV^3_b)(U^2_{a'}-iU^3_{a'})(U^2_{b'}+iU^3_{b'})\,.
\eea
Re-expressing $\partial_{\alpha\dot\alpha}$ as  $\partial_{\alpha\dot\alpha}=\sigma^m_{\alpha\dot\alpha}\partial_m$,
we apply the trace formula (\ref{trace-4sigma})  for the sigma-matrices and single out the term with the antisymmetric $\varepsilon$-tensor,
\be
\Gamma_{\rm WZ}=-8i\varepsilon^{mnpq}\int d^4x dU dV\,H^{(4)}(V^1_d\varphi^d-i U^1_{d'} \phi^{d'})
\partial_m\varphi^a
\partial_n\varphi^b
\partial_p\phi^{a'}
\partial_q\phi^{b'}
V^2_a V^3_b U^2_{a'} U^3_{b'}\,.
\label{4.15}
\ee
Substituting the power series expansion (\ref{ser1}) into (\ref{4.15}) and computing the integral over the $U$-harmonics
by the rules (\ref{SO3-int}), we obtain
\bea
\label{4.16}
\Gamma_{\rm WZ}&=&-16c\,\varepsilon^{mnpq}\varepsilon_{{a'}{b'}{c'}}
\int d^4x dV\,
\sum_{n=0}^\infty (n+1)(n+2)(-1)^n\frac{(\phi^{d'}\phi^{d'})^n}{(V^1_d\varphi^d)^{2n+5}}
\\&&\times
\phi^{a'}
\partial_p\phi^{b'}
\partial_q\phi^{c'}
\partial_m\varphi^a
\partial_n\varphi^b
V^2_a V^3_b
\nn\\
&&\hspace{-30pt}=
-32c\,\varepsilon^{mnpq}\varepsilon_{{a'}{b'}{c'}}
\int d^4x dV\,
\frac{V^1_c\varphi^c}{[\phi^{d'}\phi^{d'}+(V^1_d\varphi^d)^2]^3}
\phi^{a'}
\partial_p\phi^{b'}
\partial_q\phi^{c'}
\partial_m\varphi^a
\partial_n\varphi^b
V^2_a V^3_b
\,.\nn
\eea
Next, we expand the integrand in a series over $V^1_d\varphi^d$
and perform the integration over the $V$-harmonics in a similar way,
\bea
\Gamma_{\rm WZ}&=&
-8c\,\varepsilon^{mnpq}
\int d^4x\,
\frac{1}{(\phi^{d'}\phi^{d'})^3}
\sum_{n=0}^\infty \frac{(-1)^n(n+2)(n+1)}{2n+3}
\left(
\frac{\varphi^d\varphi^d}{\phi^{d'}\phi^{d'}}
\right)^n
\nn\\&&\times
(\varepsilon_{{a'}{b'}{c'}}
\phi^{a'}
\partial_p\phi^{b'}
\partial_q\phi^{c'})
(\varepsilon_{abc}
\varphi^a
\partial_m\varphi^b
\partial_n\varphi^c)\,.
\eea
The series can be summed up, and we obtain the following result
\be
\label{4.17_}
\Gamma_{\rm WZ}=
-2c\,\varepsilon^{mnpq}
\int d^4x\,\frac{h(z)}{(\phi^{d'}\phi^{d'})^3}
(\varepsilon_{{a'}{b'}{c'}}
\phi^{a'}
\partial_p\phi^{b'}
\partial_q\phi^{c'})
(\varepsilon_{abc}
\varphi^a
\partial_m\varphi^b
\partial_n\varphi^c)\,,
\ee
where
\be
h(z)=\frac{z^2-1}{z^2(z^2+1)^2}+\frac{\arctan z}{z^3}\,,\qquad
z^2=\frac{\varphi^a\varphi^a}{\phi^{a'} \phi^{a'}}\,.
\ee

Let us now introduce the normalized scalars,
\be
Y^a=\frac{\varphi^a}{\sqrt{\varphi^b\varphi^b+\phi^{b'}\phi^{b'}}}\,,\qquad
Y^{a'}=\frac{\phi^{a'}}{\sqrt{\varphi^b\varphi^b+\phi^{b'}\phi^{b'}}}\,,
\ee
which lie on the unit five-sphere, $Y^a Y^a+Y^{a'} Y^{a'}=1$.  In terms of these scalars, the action (\ref{4.17_}) is rewritten as
\be
\Gamma_{\rm WZ}
=-2c\,\varepsilon^{mnpq}
\int d^4x\,g(z)
(\varepsilon_{abc} Y^a
\partial_p Y^b
\partial_q Y^c)
(\varepsilon_{{a'}{b'}{c'}} Y^{a'}
\partial_m Y^{b'}
\partial_n Y^{c'})\,,
\label{4.18}
\ee
where
\be
g(z)=\frac{z^4-1}{z^2}+\frac{(z^2+1)^3}{z^3}\arctan z\,,\qquad
z^2=\frac{Y^a Y^a}{Y^{a'} Y^{a'}}\,.
\ee
Comparing (\ref{4.18}) with (\ref{WZ-biharm}), we observe the perfect agreement between the two  expressions,
provided that the coefficient $c$ is chosen as in (\ref{c-biharm}).

Thus in this section we demonstrated that the superfield functional (\ref{G-biharm}) does contain, in its component structure,
the $F^4/X^4$ and Wess-Zumino terms as the necessary ingredients of the $\cN=4$ SYM low-energy effective action. In principle,
it is possible to explicitly compute  all other component terms in the action (\ref{G-biharm}) needed to complete these
selected bosonic terms  to the full $\cN=4$ supersymmetry invariants.

\section{Concluding remarks}
The present review was devoted  to the problem of constructing the low-energy effective action in $\cN=4$ SYM theory, based upon
the powerful off- and on-shell superfield methods of extended supersymmetry. The consideration was basically concentrated around the papers \cite{BuIv,BelSam1,BelSam2,BISZ},
in which the four-derivative part of the low-energy effective action in the Coulomb branch was studied.
This part of the effective action represents the leading quantum correction in the theory. Although it was known
for a long time that this contribution to the effective action is one-loop exact \cite{DS,S16,BKO} and does not
receive instanton corrections \cite{Dorey}, only some selected terms in the action were studied before.
In particular, in the papers \cite{Henningson,deWit,Lindstrom,non-hol2,non-hol3} there was considered
that part of the $\cN=4$ SYM effective action, which refers to the $\cN=2$ vector multiplet.
The derivation of the completely $\cN=4$ supersymmetric extension of these results appeared a quite non-trivial problem. It
was resolved in \cite{BuIv,BelSam1,BelSam2,BISZ}, with making use of different harmonic superspace approaches. It turned out
that the corresponding superfield effective action can be restored solely on the symmetry ground, by requiring it to enjoy the ${\cal N}=4$
supersymmetry and/or superconformal $PSU(2,2|4)$ symmetry. Although only some part of the underlying supersymmetries can be realized off shell
($\cN=2$ supersymmetry in the $\cN=2$ harmonic approach and $\cN=3$ supersymmetry in the $\cN=3$ harmonic approach), the on-shell realization of
the remaining part proved quite sufficient to fully fix the superfield effective actions.

Dine and Seiberg \cite{DS} argued that the $F^4/X^4$ term in the low-energy effective action
of $\cN=4$ SYM theory is one-loop exact, so that the coefficient in front of this term
is non-renormalized against higher-order quantum loop corrections. The origin of this non-renormalizability
was clarified in \cite{BelSam1}. It is very important to realize that the $\cN=4$ SYM low-energy effective action
contains the Wess-Zumino term \cite{TZ} for six scalar fields of the $\cN=4$ gauge multiplet.
This Wess-Zumino term is obviously one-loop exact because it appears in the Coulomb branch
as the necessary consequence  of the anomaly-matching condition for the $SU(4)$ R-symmetry \cite{Intriligator}. Because this term involves
four space-time derivatives of scalars, it is of the same order as the $F^4/X^4$ term. Thus,
these two terms are related to each other by $\cN=4$ supersymmetry and are, in fact, different components
of the same superfield expression for the four-derivative part of the low-energy effective action \cite{BelSam1}.
This explains the non-renormalizability of the coefficient in the $F^4/X^4$ term.

The presence of the potential anomaly of the $SU(4)$ R-symmetry current in $\cN=4$ SYM theory was explicitly demonstrated in \cite{Sokatchev98}.
Therefore, the effective Lagrangian is invariant under $SU(4)$ only up to the total derivative terms. The $SU(4) \sim SO(6)$ group
has four maximal subgroups: $SO(5)\sim USp(4)$, $SO(4)\times SO(2)\sim SU(2)\times SU(2)\times U(1)$, $SO(3)\times SO(3)\sim SU(2)\times SU(2)$
and $SU(3)\times U(1)$. Only the last of these groups is anomalous, while the others are not. As a consequence,
only the first three groups can appear as the manifest symmetry of the effective action. As we showed in the present paper,
each of these subgroups correspond to a particular superspace description of the $\cN=4$ SYM low-energy effective action.
In particular, the $SU(2)\times SU(2)\times U(1)$ group is manifest in the $\cN=2$ harmonic superspace,
the group $USp(4)$ is manifest in the $\cN=4$ superspace equipped with $USp(4)$ harmonic variables, while the
group $SU(2)\times SU(2)$ corresponds to the $\cN=4$ bi-harmonic superspace. The last option $SU(3)\times U(1)$
is the R-symmetry group of the $\cN=3$ harmonic superspace.

Each of the four superspace approaches considered here has its own specific features.
The $\cN=4$ harmonic superspaces with $USp(4)$ and $SU(2)\times SU(2)$ harmonic variables
provide the most elegant description of the $\cN=4$ SYM low-energy effective action:
the effective Lagrangian is given simply by the logarithm of the uncharged $\cN=4$ superfield strength.
All four-derivative component terms in the low-energy effective action prove to be encapsulated
in this simple superfield expression.

The effective Lagrangian in the $\cN=3$ harmonic superspace is still simple enough
as it is expressed in terms of elementary functions, but it explicitly involves the constants $c^i$ which
correspond to the vevs of the scalars fields $\varphi^i$. These constants break manifest $SU(3)$ symmetry,
although the latter is implicitly realized modulo total derivative terms. This is a manifestation of the
fact that the $SU(3)$ subgroup of the R-symmetry group is anomalous in  $\cN=4$ gauge theory.
An important advantage of the $\cN=3$ harmonic superspace is that, in principle, it provides a way to
realize the maximal number of supersymmetries off the mass shell owing to the existence of an
unconstrained superfield formulation  of the $\cN=3$ SYM classical action in this superspace \cite{GIKOS-N3a,GIKOS-N3}.

The $\cN=2$ harmonic superspace is the most deeply elaborated approach among all the superspace approaches
discussed here. In particular, the quantum perturbation theory is well developed in it \cite{GIOS2,BFM}.
These perturbative methods were applied in \cite{BIP,BBP,BP2005} for direct computations of the low-energy effective
action in $\cN=4$ SYM theory. In principle, this approach opens the ways to study higher-order quantum corrections
to the low-energy effective action in $\cN=4$ SYM theory \cite{BPT2002,Kuz2004}. However, this issue is very subtle and
below we will only briefly comment on it.

Let us dwell on possible generalizations of the results reviewed here.

In the present paper we considered only the gauge group $SU(2)$ spontaneously broken down to $U(1)$.
It is rather trivial to generalize it to an arbitrary simple Lie group $G$ broken down to its maximal abelian subgroup $H$.
For instance, consider the gauge group $G=SU(N)$ spontaneously broken down to $H=[U(1)]^{N-1}$. The $\cN=4$ superfield $W$
in this case is the diagonal $N\times N$ matrix in the Cartan subalgebra of $su(N)\,$,
\be
W={\rm diag}(W^1,W^2,\ldots W^N)\,,\qquad
\sum_{i=1}^N W^i =0\,,
\ee
with all eigenvalues being distinct, $W^i\ne W^j$ for $i\ne j$. Then the effective action (\ref{G4}) generalizes to this case as
\be
\Gamma = -\frac1{96\pi^2} \int d\zeta du \sum_{i< j}^N
\ln \frac{|W^i - W^j|}{\Lambda}\,.
\ee
Here $W^i - W^j$ correspond to root subspaces in the Lie algebra $su(N)$ of the gauge group
and the summation is performed over the positive roots. Taking this into account, one can immediately
write down the low-energy effective action in $\cN=4$ SYM theory for any other simple gauge group.
In the same manner one can generalize all other superfield actions (\ref{Gamma-N2-final}), (\ref{Gconf}) and (\ref{G-log-biharm})
considered in this paper.

Another possible generalization is the study of the next-to-leading terms in the $\cN=4$ SYM low-energy effective action.
Indeed, in this paper we considered only the four-derivative part of the effective action, the typical representative of which
is the $F^4/X^4$ component term. In general, the effective action contains the terms $F^{2n+2}/X^{2n}$, $n\in{\mathbbm N}$,
with all their supersymmetric complements. The interest in these terms is motivated by the AdS/CFT conjecture \cite{CT,Maldacena,BI-review},
which predicts that the $\cN=4$ SYM low-energy effective action is related to the D3-brane action in $AdS_5\times S^5$.
The latter is described by the following action in the bosonic sector
\bea
S_{\rm D3}&=& \frac1{2\pi g_s} \int d^4x \left(
h^{-1}-\sqrt{-\det(g_{mn}+F_{mn})}
\right)\,,\label{D3-action}\\
g_{mn} &=& h^{-1/2}\eta_{mn} + h^{1/2} \partial_m X^I \partial_n X^I\,,\qquad
h= \frac{g_s N}{\pi(X^I X^I)^2}\,,\nn
\eea
where $X^I$ are six coordinates transverse to the world-volume of the D3-brane, $N$ is the number
of D3-branes which create the background $AdS_5\times S^5$ geometry and $g_s$ is the string coupling constant.
Upon the series expansion of the square root of the determinant in (\ref{D3-action}),
one uncovers all terms of the form $F^{2n+2}/X^{2n}$, which are present in the $\cN=4$ SYM effective action as well.
In this expansion, the $F^2$ term is a part of the abelian $\cN=4$ SYM classical action, while the $F^4/X^4$ term
should originate from the low-energy effective action described in the present paper. After the appropriate redefinition
of the constants in (\ref{D3-action}), the coefficients before its $F^2$ and $F^4/X^4$ terms
exactly match those in the $\cN=4$ SYM low-energy effective action.

However, it is hard to match the higher order terms in these actions. This problem is multi-fold.
It is quite obvious that (\ref{D3-action}) cannot exactly match the $\cN=4$ SYM low-energy effective action in the bosonic sector.
Indeed, the D3-brane action (\ref{D3-action}) involves only the first space-time derivatives of physical scalars,
while the $\cN=4$ SYM low-energy effective action in any superfield formulation discussed here inevitably contains
higher-order derivatives of the scalars. Thus, these actions can coincide only upon the appropriate redefinition of fields,
\bea
X'^I &=& X'^I(X^I, \partial_m X^i, \partial_m \partial_n X^I, F_{mn}, \ldots)\,,\nn\\
F'_{mn} &=& F'_{mn}(F_{mn}, X^I, \partial_m X^I, \partial_m\partial_n X^I,\ldots )\,.
\eea
Such a redefinition was worked out to some order in
\cite{GKPR}, but in general, it is still a non-trivial issue which has never been presented in literature in a closed form.
The reason for such a field redefinition was explained in \cite{BIK}: the superconformal group $SU(2,2|4)$ is realized
differently on the fields inherent to the field theory and those appearing in the AdS settings.

The problem of higher-order terms in the low-energy effective action is even more subtle.
Different superspace methods of quantum computations of the coefficient in the $F^6/X^8$ term used in  \cite{BPT2002,Kuz2004} ($\cN=1$)
and \cite{Kuzenko-McArthur-2004} ($\cN=2$) give different results. This mismatch is explained \cite{Kuz2004} by the fact
that in distinct superfield methods different gauges
are applied and it is very difficult to perform higher-loop quantum computations in a gauge-independent way. In \cite{KMT} it was also argued that the higher-order terms can be found by employing the quantum-deformed conformal symmetry.

To understand this issue better, it would be interesting to develop the methods of computations of quantum corrections
to the effective action in the $\cN=3$ harmonic superspace. Although the basic principles of quantum perturbation theory
in this superspace were formulated in \cite{Delduc}, the background field method has never been worked out
in the $\cN=3$ superfield approach. Given the $\cN=3$ superfield background field method, it would be possible to check
the conjecture made in \cite{BISZ} that the $F^6/X^8$ term does not receive quantum corrections beyond one loop and
the correct value of this coefficient appears after elimination of all auxiliary fields in the $\cN=3$ effective action
(\ref{Gconf}) considered together with the classical action (\ref{N3-class-action}) in the abelian case.

It is also tempting to develop alternative superspace methods for studying classical and quantum aspects of the $\cN=4$ SYM theory. For instance, in the recent papers \cite{CS1,CS2} the so-called Lorentz harmonic chiral superspace was proposed for computing certain classes of correlation functions. It would be very interesting to apply this approach to the problem of low-energy effective action in the $\cN=4$ SYM theory.

The relation of the $\cN=4$ SYM low-energy effective action to the D3-brane dynamics discussed above
suggests that a similar correspondence can be established for supersymmetric gauge theories in space-times
of dimension other than four. In particular, the low-energy dynamics of multiple M2-branes in M-theory can be
understood through the three-dimensional superconformal gauge theories with $\cN=6$ and $\cN=8$ supersymmetries,
which are known as the ABJM \cite{ABJM} and BLG \cite{BLG1,BLG2,Gustavsson} theories. In \cite{Schwarz} it is
conjectured that the low-energy effective action in the ABJM theory should describe the effective dynamics
of single M2-brane on the $AdS_4\times S^7$ background, in a similar way as the $\cN=4$ SYM low-energy effective
action is related to the D3-brane. In the three-dimensional case, this conjecture has never been tested.
We expect that the extended superspace methods could be useful for solving this problem.
For the Lagrangians of the ABJM and BLG theories, the 3D $\cN=3$ harmonic superspace \cite{BILPSZ} seems
to provide the highest number of off-shell supersymmetries (see also \cite{KS} for a recent discussion).
It would be interesting to study the superfield low-energy effective action in the ABJM theory.

As the final remark, we point out that the harmonic superspace methods turned out to be very useful
also in the recent studies of effective actions in higher-dimensional supersymmetric models \cite{BP2015-1,BP2015-2,BIS15}.

\vspace{30pt}
\noindent
{\bf Acknowledgments}\nopagebreak\\[3mm]
The authors are very grateful to D.~V.~Belyaev, S.~M.~Kuzenko, 
O.~Lechtenfeld, N.~G.~Pletnev, A.~A.~Tseytlin and \fbox{B.~M.~Zupnik} for fruitful collaboration and numerous discussions.
The work of I.L.B. was supported in part by Ministry of Education and Science of Russian Federation under contract 3.867.2014/K.
The work of I.L.B. and E.A.I. was
partially supported by the RFBR grants No 15-02-06670 and No 16-52-12012. The work of I.B.S. was supported in part by the Australian Research Council, project No. DP140103925.
The work of E.A.I. was also supported by a grant
of the Heisenberg-Landau program. He thanks Emery Sokatchev for useful correspondence.


\end{document}